\pdfoutput=1
\newcommand*{\ATLASLATEXPATH}{latex/}
\documentclass[cernpreprint,texlive=2016,txfonts,UKenglish]{\ATLASLATEXPATH atlasdoc}
\usepackage{\ATLASLATEXPATH atlaspackage}
\usepackage{\ATLASLATEXPATH atlasbiblatex}

\usepackage{\ATLASLATEXPATH atlascontribute}

\usepackage{\ATLASLATEXPATH atlasphysics}
\usepackage{multirow}
\usepackage{graphicx}

\graphicspath{{logos/}{figures/}}

\usepackage{EXOT-2016-10-PAPER-defs}

\AtlasTitle{Search for heavy resonances decaying into a $W$ or $Z$ boson and a Higgs boson in final states with leptons and $b$-jets in 36~fb$^{-1}$ of $\sqrt s = 13$~\TeV\ $pp$ collisions with the ATLAS detector}

\author{The ATLAS Collaboration}

\date{\today}

\PreprintIdNumber{CERN-EP-2017-250}

\AtlasJournal{JHEP}

\AtlasJournalRef{JHEP 03 (2018) 174}

\AtlasDOI{DOI:10.1007/JHEP03(2018)174}

\AtlasRefCode{EXOT-2016-10}
\AtlasAbstract{
  A search is conducted for new resonances decaying into a $W$ or $Z$ boson
  and a 125~\GeV\ Higgs boson in the  $\nu\bar{\nu}b\bar{b}$, $\ell^{\pm}{\nu}b\bar{b}$, and $\ell^+\ell^-b\bar{b}$  final states,
  where $\ell ^{\pm}= e^{\pm}$ or $\mu^{\pm}$, in $pp$ collisions at $\sqrt s = 13$~\TeV.
  The data used correspond to a total integrated luminosity of 36.1~fb$^{-1}$ collected with the ATLAS detector at the Large Hadron Collider during the 2015 and 2016 data-taking periods.
  The search is conducted by examining the reconstructed invariant or transverse mass distributions of $Wh$ and $Zh$ candidates for evidence of a localised excess in the mass range of 220~\GeV\ up to 5~\TeV.
  No significant excess is observed and the results are interpreted in terms of constraints on the production cross-section times branching 
  fraction of heavy $W^\prime$ and $Z^\prime$ resonances in heavy-vector-triplet models and the CP-odd scalar boson $A$ in two-Higgs-doublet models.
  Upper limits are placed at the 95\% confidence level and range between $9.0\times 10^{-4}$~pb and $8.1\times 10^{-1}$~pb depending on the model and mass of the resonance.
}

\hypersetup{pdftitle={ATLAS document},pdfauthor={The ATLAS Collaboration}}

\begin{document}

\maketitle

%\tableofcontents

% List of contributors - print here or after the Bibliography.
%\PrintAtlasContribute{0.30}
%\clearpage

\section{Introduction}
\label{sec:intro}

The ATLAS~\cite{HIGG-2012-27} and the CMS~\cite{CMS-HIG-12-028} collaborations discovered a Higgs boson (\h) with a mass near 125 \GeV\ and properties consistent with the Standard Model (SM) predictions~\cite{HIGG-2013-17,CMS-HIG-14-018,CMS-HIG-14-042}. Two of the most important questions 
that remain are how the Higgs boson mass is protected against large radiative corrections (the naturalness problem~\cite{naturalness1, naturalness2,naturalness3}) 
and whether the Higgs boson is part of an extended scalar sector~\cite{deFlorian:2016spz}, thus making this particle important for searches for new physics beyond the SM.

%The recently discovered Higgs boson (\h) with a mass of 125 \GeV\ by the ATLAS~\cite{:2012gk} and the CMS~\cite{:2012gu} collaborations, with properties consistent with the Standard Model (SM), can be used as a tool to search for physics beyond the SM. Two of the most important questions that remain is how the Higgs mass is protected against large radiative corrections (naturalness problem~\cite{naturalness1, naturalness2,naturalness3}) and whether it is part of an extended scalar sector.

Various models with dynamical electroweak symmetry breaking scenarios attempt to solve the naturalness problem 
by assuming a new strong interaction at a higher energy scale. These models generically predict the existence of new vector resonances that naturally decay 
into a vector boson and a Higgs boson, for example in Minimal Walking Technicolour~\cite{Sannino:2004qp,Foadi:2007ue,Belyaev:2008yj}, 
Little Higgs~\cite{Schmaltz:2005ky}, and composite Higgs models~\cite{Dugan:1984hq,Agashe:2004rs}.  The decays into a vector-boson and Higgs boson final state are 
frequently enhanced in these models.
%Resonance searches are typically not sensitive to all of the free parameters of the underlying theory, thus simplified models~\cite{Alves:2011wf}  can be used to parametrise a broad a broad class of model, wherein only the relevant couplings and mass parameters are retained in the Lagrangian.

Another possible extension of the SM includes the addition of a second Higgs doublet~\cite{Branco:2011iw}. A second Higgs doublet arises in many 
theories beyond the SM, collectively called two-Higgs-doublet models (2HDMs), such as the minimal supersymmetric SM~\cite{Fayet:1976et,Fayet:1977yc,
Farrar:1978xj,Fayet:1979sa,Dimopoulos:1981zb}, axion models~\cite{Kim:1986ax}, %(e.g. Ref. \cite{Kim:1986ax}), 
and baryogenesis models~\cite{Joyce:1994zt}. %(e.g. Ref. \cite{Joyce:1994zt}). 
In 2HDMs with a CP-conserving Higgs potential, the scalar sector of the theory consists of five Higgs bosons: two charged ($H^{\pm}$),  two neutral 
CP-even ($h,H$) and one neutral CP-odd ($A$). 

This paper describes a search for the production of new heavy vector bosons, denoted hereafter by $W^\prime$ and $Z^\prime$,
that decay into a $W$ or a $Z$ boson and an \h\ boson and a search for a heavy CP-odd scalar boson $A$ that decays into a $Z$ and an \h\ boson. 
%The analysed dataset corresponds to 36.1~fb$^{-1}$ of integrated luminosity collected in $pp$ 
%collisions at a centre-of-mass energy $\sqrt s = 13$~\TeV\ with the ATLAS detector at the Large Hadron Collider (LHC). 
The analyses described here target leptonic decays of the vector bosons 
($W^{\pm}\rightarrow \ell^{\pm} \nu$, $Z\rightarrow \ell^+ \ell^-/\nu\bar{\nu}$; $\ell^{\pm} = e^{\pm},\mu^{\pm}$) 
and decays of the \h\ boson into a $b$-quark pair. This results in three search channels: $W^\prime \rightarrow W^{\pm}\h
\rightarrow \ell^{\pm} \nu b\bar{b}$, $Z^\prime/A \rightarrow Z\h \rightarrow \ell^+ \ell^- b\bar{b}$, and $Z^\prime/A \rightarrow Z\h \rightarrow \nu \bar{\nu}
b\bar{b}$.  

Resonance searches are typically not sensitive to all free parameters of the underlying theory, thus simplified models~\cite{Alves:2011wf}  can be used to parameterise a broad class of models, wherein only the relevant couplings and mass parameters are retained in the Lagrangian.  For the interpretation of the results in the context of models with heavy vector triplets (HVT), a simplified model~\cite{Pappadopulo:2014qza, deBlas:2012qp},
based on a phenomenological Lagrangian is used as a benchmark. This model incorporates an SU(2)$_{\mathrm{L}}$ triplet of heavy vector bosons, which allows 
the results to be interpreted in a large class of models. The new heavy vector bosons, ${W}^\prime$ and ${Z}^\prime$, collectively denoted 
by $V'$, couple to the Higgs and gauge bosons via a combination of parameters $g_\text{V}c_\text{H}$ and to the fermions via the combination $(g^2/g_\text{V})~c_\text{F}$, 
where $g$ is the SU(2)$_{\mathrm{L}}$ gauge coupling. The parameter $g_\text{V}$ represents the strength of the new vector-boson interaction, and $c_\text{H}$ and $c_\text{F}$ represent 
corrections to the coupling strength specific to Higgs bosons and fermions, respectively. 
Two benchmark models are used in this analysis. In the first model, referred to as \textit{Model A}, the branching 
fractions to fermions and gauge bosons are comparable, as in some models with an extended gauge symmetry~\cite{Barger:1980ix}.  
For \textit{Model B}, fermionic couplings are suppressed, as in strong dynamical models such as the minimal composite Higgs model~\cite{Contino:2011np}. At low resonance masses 
and large $g_\text{V}$ couplings, the HVT models fail to reproduce the SM parameters, thus this search focuses on high masses, from 500~\GeV\ up to 5~\TeV.

The results from the 
$A \rightarrow Z$\h\ search are interpreted as exclusion limits on the ratio of the 
vacuum expectation values of the two Higgs doublets, $\tan(\beta)$, and  on $\cos(\beta-\alpha)$, where $\alpha$ is the mixing angle between the two CP-even 
Higgs bosons. The exclusion limits are evaluated for the Type I, Type II, Lepton-specific, and Flipped 2HDMs. These differ with respect to which doublets 
couple to the up-type and down-type quarks as well as to the charged leptons~\cite{Branco:2011iw}. Both the production via gluon--gluon fusion and the production 
with associated $b$-quarks (\bbA) are considered in this search. The $A\rightarrow Zh$ decay mode is mostly relevant below the \ttbar\ production threshold and 
the cross-section falls steeply with increasing $A$ boson mass. Therefore, this search starts at the $Zh$ threshold of approximately 220~\GeV\ and goes up to 2~\TeV.

Previous searches in the same final states have been performed by the ATLAS and the CMS collaborations using data at $\sqrt s = 8$~\TeV\ and 13~\TeV. 
The ATLAS searches for ${W}^\prime \rightarrow W\h$ (${Z}^\prime \rightarrow Z\h$) exclude, at 95\% confidence level (CL), 
${W}^\prime$ (${Z}^\prime$) resonances with masses below 1.75~(1.49)~\TeV\ assuming the HVT benchmark \textit{Model A} ($g_\text{V}=1$)
and below 2.22~(1.58)~\TeV\ assuming \textit{Model B} ($g_\text{V}=3$)~\cite{EXOT-2013-23,EXOT-2015-18}.  
Searches by the CMS Collaboration exclude resonances with masses  less than 2.0~\TeV\ at 95\% CL 
assuming the HVT benchmark \textit{Model B} ($g_\text{V}=3$)~\cite{CMS-B2G-16-003}. 
Searches using the fully hadronic final state ($W/Z\h \rightarrow qq^\prime b\bar{b}$) have also been performed by CMS and ATLAS  and exclude \Wprime\ (\Zprime) 
resonances below 3.15~\TeV\ (2.6~\TeV) assuming the HVT benchmark \textit{Model B} ($g_\text{V}=3$)~\cite{CMS-EXO-14-009,cmsVHqqbb2017,Aaboud:2017ahz}.
Previous searches for a CP-odd scalar boson $A$ in the $Z\h$ decay mode are reported in Refs.~\cite{Schael2006,HIGG-2013-31,CMS-HIG-13-021,HIGG-2013-06,CMS-HIG-14-011}.

The search presented in this paper is performed by looking for a localised excess in the distribution of the reconstructed mass of the $\nu \bar{\nu} b\bar{b}$, $\ell^{\pm} \nu b\bar{b}$, and  $\ell^+ \ell^- b\bar{b}$  systems. 
%The mass range covered by the search from 220~\GeV\ to 5~\TeV\ implies that the Higgs boson candidates can be produced in a wide range of  transverse momenta. 
The mass range covered by the search, from 220~\GeV\ to 5~\TeV, probes a wide range of Higgs boson transverse momenta.
Thus, two methods are used to reconstruct Higgs boson candidates. 
At low transverse momenta, the decay products of the Higgs boson are reconstructed as individual jets. At high transverse momenta, the decay products start to merge and are reconstructed as a single jet.
%The signal yield and background normalisations are determined from a binned maximum likelihood fit to the data distribution and are used to set upper limits on the production cross-section times decay branching fraction for each of the $V^\prime$ and $A$ boson models.  
The signal yield and background normalisations are determined from a binned maximum-likelihood fit to the data distribution  for each of the $V^\prime$ and $A$ boson models (${W}^\prime$, ${Z}^\prime$, gluon--gluon fusion $A$,  $bbA$) and are used to set upper limits on the production cross-section times decay branching fraction.  
%A separate fit is performed for the HVT model in the case the $V^\prime$ are degenerate, using the combination all three channels.
A combined fit using all three lepton channels sets bounds on the HVT model in the case where the $V^\prime$ bosons are degenerate in mass.

This paper is structured as follows. Sections~\ref{sec:detector} and~\ref{sec:data_MC_samples} provide a brief description of the ATLAS experiment and the data and simulated event 
samples.  The event reconstruction and selections are discussed in Sections~\ref{sec:object_selection} and~\ref{sec:event_selection}. The background estimation and systematic 
uncertainties are described in Sections~\ref{sec:background} and~\ref{sec:systematics}. 
Finally, Sections~\ref{sec:results} and~\ref{sec:conclusion} detail the statistical analysis and provide a discussion of the results and concluding remarks.

\section{ATLAS detector}
\label{sec:detector}

% Footnote with ATLAS coordinate system
\newcommand{\AtlasCoordFootnote}{%
ATLAS uses a right-handed coordinate system with its origin at the nominal interaction point (IP)
in the centre of the detector and the $z$-axis along the beam pipe.
The $x$-axis points from the IP to the centre of the LHC ring,
and the $y$-axis points upwards.
Cylindrical coordinates $(r,\phi)$ are used in the transverse plane, 
$\phi$ being the azimuthal angle around the $z$-axis.
The pseudorapidity is defined in terms of the polar angle $\theta$ as $\eta = -\ln \tan(\theta/2)$.
Angular distance is measured in units of $\Delta R \equiv \sqrt{(\Delta\eta)^{2} + (\Delta\phi)^{2}}$.}

The ATLAS detector~\cite{PERF-2007-01} at the LHC covers nearly the entire solid angle\footnote{\AtlasCoordFootnote} around the collision point.
It consists of an inner tracking detector (ID) surrounded by a thin superconducting solenoid, electromagnetic and hadronic calorimeters,
and a muon spectrometer incorporating three large superconducting toroid magnets.

The ID is immersed in a \SI{2}{\tesla} axial magnetic field 
and provides charged-particle tracking in the range $|\eta| < 2.5$.
It consists of silicon pixel, silicon microstrip, and transition radiation tracking detectors.
Prior to the data-taking at the increased centre-of-mass energy of 13~\TeV, 
the ID was enhanced by adding a new layer of pixel detectors (the IBL~\cite{ATLAS-TDR-19}) 
inside the existing pixel detector layers in the barrel region (at a radius of approximately 33 mm). 
The upgraded detector typically provides four three-dimensional measurements for tracks originating from the luminous region. 
The silicon microstrip tracker provides four two-dimensional measurement points per track.
The transition radiation tracker enables track reconstruction at large radii up to $|\eta| = 2.0$ and 
provides electron identification information 
based on the number of hits above the threshold for transition radiation.

The calorimeter system covers the pseudorapidity range $|\eta| < 4.9$.
Within the region $|\eta|< 3.2$, electromagnetic calorimetry is provided by barrel and 
endcap high-granularity lead/liquid-argon (LAr) electromagnetic calorimeters.
An additional thin LAr presampler, covering $|\eta| < 1.8$,
is used to correct for energy loss in material upstream of the calorimeters.
Hadronic calorimetry is provided by a steel/scintillator-tile calorimeter,
segmented into three barrel structures within $|\eta| < 1.7$, and two copper/LAr hadronic endcap calorimeters.
The solid-angle coverage is completed with forward copper/LAr and tungsten/LAr calorimeter modules
optimised for electromagnetic and hadronic measurements, respectively.

The muon spectrometer is composed of separate trigger and
high-precision tracking chambers, measuring the deflection of muons in a magnetic field generated by superconducting air-core toroids.
The precision chamber system covers the region $|\eta| < 2.7$ with three layers of monitored drift tubes,
complemented by cathode strip chambers in the forward region, where the particle flux is highest.
The muon trigger system covers the range $|\eta| < 2.4$ with resistive plate chambers in the barrel, and thin-gap chambers in the endcap regions.

A two-level trigger system is used to select interesting events~\cite{TRIG-2016-01}.
The level-1 trigger is implemented in hardware and uses a subset of detector information
to reduce the event rate to a design value of at most \SI{100}{\kHz}.
This is followed by the software-based trigger level, the high-level trigger, which reduces the event rate further to about \SI{1}{\kHz}.

\section{Data and simulated event samples}
\label{sec:data_MC_samples}

The data used in this analysis were recorded with the ATLAS detector during the 2015 and 2016 $pp$-collision runs at $\sqrt{s}=$ 13~\TeV\ 
and correspond to a total integrated luminosity of $36.1$~fb$^{-1}$. The data are required to satisfy 
a number of criteria that ensure that the ATLAS detector was in good operating condition. 
A number of Monte Carlo (MC) simulation samples are used to model the background and signal 
processes for this search.

For the $W'$ and $Z'$ processes, simulated events were generated with \textsc{MadGraph5\_aMC@NLO} (\MGMCatNLO) 2.2.2 \cite{Alwall:2014hca} 
at leading-order (LO) accuracy using the \NNPDF~2.3 \LO parton density function (PDF) set ~\cite{Ball:2011mu}. The parton shower 
and hadronisation were simulated with \PYTHIA~8.186~\cite{Pythia8} using the A14 set~\cite{ATL-PHYS-PUB-2014-021}
of tuned parameters (``tune'') together with the \NNPDF~2.3 \LO PDF set. 
Events were generated 
for a range of resonance masses from 500 to 5000~\GeV, assuming a zero natural width. %The Higgs boson mass is set to 125.5~\gev\ for the HVT signal. 
Higgs boson decays into $b\bar{b}$ and $c\bar{c}$ pairs were simulated, with a relative branching fraction 
$B(\h\rightarrow c\bar{c})/B(\h\rightarrow b\bar{b})=0.05$ fixed to the SM prediction~\cite{Heinemeyer:2013tqa}.

Events for the gluon--gluon fusion production of \A\ bosons were generated at LO accuracy using the same set-up as for the $W'$ and $Z'$ samples.
The $b$-quark associated production of \A\ bosons was simulated with \MGMCatNLO~2.2.3 
using next-to-leading-order (\NLO) matrix elements with massive $b$-quarks and the \CTtenF NLO PDF set~\cite{Lai:2010vv}.
The parton shower and hadronisation were simulated with \PYTHIA~8.210~\cite{Sjostrand:2014zea}. Events were generated for a range of \A\ boson masses from 220 to 2000~\GeV\ assuming a zero natural width.
For the \A\ boson signals, only decays of the Higgs boson into a $b\bar{b}$ pair were generated. 

For the interpretation of the \AZh\ search in the context of 2HDMs, the masses of the $H^{\pm}$ and $H$ bosons are assumed to be equal to the mass of the $A$ boson. 
The cross-sections were calculated using up to next-to-next-to-leading-order (NNLO) QCD corrections for gluon--gluon fusion and $b$-quark associated production in the five-flavour scheme as implemented in \textsc{Sushi}~\cite{Harlander:2012pb,Harlander:2005rq,Harlander:2003ai,Harlander:2002wh}. For the $b$-quark associated production, a cross-section in the four-flavour scheme was also calculated as described in Refs.~\cite{Dawson:2003kb,Dittmaier:2003ej} and the results were combined with the five-flavour scheme calculation following Ref.~\cite{Harlander:2011aa}. The \A\ boson width and the branching fractions for $A\rightarrow Zh$ and $h\rightarrow b\bar{b}$ have been calculated using \textsc{2HDMC}~\cite{Eriksson:2009ws,Haber:2015pua}. The procedure for the calculation of the cross-section and branching fractions as well as the choice of the 2HDM parameters follows Ref.~\cite{deFlorian:2016spz}.
 
The production of \W\ and \Z\ 
bosons in association with jets was simulated with \SHERPA~2.2.1~\cite{Gleisberg:2008ta} using the \NNPDF~3.0 \NNLO 
PDF set~\cite{Ball:2014uwa} for both the matrix element calculation and the dedicated parton-shower tuning developed 
by the \SHERPA authors.  The event generation utilises \Comix~\cite{Gleisberg:2008fv} and \OpenLoops~\cite{Cascioli:2011va}, 
for the matrix element calculation, matched to the \SHERPA parton shower using the \MEandPSatNLO prescription~\cite{Hoeche:2012yf}.
The matrix elements were calculated for up to two additional partons at NLO and for three and four partons at LO in QCD.
%The \NLO electroweak corrections are not taken into account. 
The cross-sections for \W/\Z+jets were calculated at \NNLO accuracy~\cite{Melnikov:2006kv}.

%%%paragraph from HVT paper
The \ttbar\ process was simulated with \POWHEGBOX~v2~\cite{powheg1,powheg2,powheg3} interfaced to \PYTHIA~6.428~\cite{Sjostrand:2006za}.
The \CTten PDF set~\cite{Lai:2010vv} was used in the calculation of the matrix elements, while the parton shower used the Perugia 2012  tune~\cite{Skands:2010ak}
with the \CTEQ PDF set~\cite{Pumplin:2002vw}.  The cross-section was calculated at \NNLO accuracy including the resummation of next-to-next-to-leading logarithmic (NNLL) 
soft gluon terms with \textsc{Top}++2.0~\cite{Beneke:2011mq,Cacciari:2011hy,Baernreuther:2012ws,Czakon:2012zr,Czakon:2012pz,Czakon:2013goa,Czakon:2011xx}.
The predicted transverse momentum spectra of top quarks and the \ttbar\ system were reweighted to the corresponding \NNLO parton-level 
spectra~\cite{Czakon2017}.

%The production of single top quarks in the $t$-channel is simulated with the same setup as the one use for the \ttbar\ process. 
%For the $s$-channel and $Wt$ production of single top quarks, the \POWHEGBOX generator is used with the \CTten\ PDF set. 
The production of single top quarks ($t$-channel, $s$-channel, and $Wt$) was simulated using the \POWHEGBOX\ event generator with the \CTten\ PDF set.   
The shower and hadronisation was simulated using the same event generator and set-up as for the \ttbar\ process.
The cross-section for the $t$- and $s$-channel single-top-quark production was calculated at \NLO accuracy
using \textsc{Hathor}~v2.1~\cite{Aliev:2010zk,Kant:2014oha}, while for the $Wt$ process an approximate \NNLO calculation was used~\cite{Kidonakis:2010ux}.
The top mass is set fixed to 172.5 \GeV\ in the \ttbar\ and single-top-quark samples.

Diboson events ($WW$, $WZ$, $ZZ$) were simulated using the \SHERPA~2.1.1 event generator using the \CTten PDFs.
Matrix elements were calculated for up to one ($ZZ$) or no ($WW$, $WZ$) additional partons at \NLO and up to three additional partons at \LO, and the cross-sections were calculated at \NLO accuracy.

Finally, the SM processes \V\h($\h\rightarrow\bbbar$), \ttbar\h, \ttbar$W$, and \ttbar$Z$ are included in the total background estimation. 
The $q\bar{q}\rightarrow Zh$ and $q\bar{q}\rightarrow Wh$ processes were simulated at \LO with \PYTHIA~8.186 using the \NNPDF~2.3 \LO PDF set
and the A14 tune. The $gg\rightarrow Zh$ process was simulated at \NLO using the \POWHEGBOX~v2 event generator with the \CTten PDF set. The modelling of the shower,
hadronisation and underlying event was provided by \PYTHIA~8.186 using the AZNLO tune~\cite{STDM-2012-23} with the same PDF set as for the matrix element calculation.
%The mass of the Higgs boson is fixed to 125~\GeV\ and the $b\bar{b}$ branching fraction is fixed to 58\%. 
%The $h\rightarrow b\bar{b}$ branching fraction is fixed to 58\%.
The cross-sections for the $Wh$ and $Zh$ processes were taken from Ref.~\cite{deFlorian:2016spz}.
The \ttbar\h\ and \ttbar\V\ samples were generated at NLO accuracy with \MGMCatNLO~2.3.2 interfaced to 
\PYTHIA~8.210. The \NNPDF~3.0 \NLO PDF set was used in the matrix element calculation while
for the parton shower the A14 tune was used with the \NNPDF~2.3 \LO PDF set.

A summary of event generators used for the simulation of signal and background processes is shown in Table~\ref{tab:mc_samples}.
{
\def\arraystretch{1.2}
\begin{table}[t]
\caption{\label{tab:mc_samples} 
Event generators used for the simulation of the signal and background processes. The acronyms ME and PS are used for matrix element and parton shower, respectively.
}

\begin{center}

\begin{footnotesize}

\begin{tabular}{l| c | c| c | c| c}
\hline\hline
\multirow{2}{*}{Process} & \multirow{2}{*}{ME generator} & \multirow{2}{*}{ME PDF} & PS and & \multirow{2}{*}{MC tune} & Cross-section\\
& & & Hadronisation &  & calc. order \\
\hline
\multicolumn{6}{c}{Signal} \\
\hline
$V'$   & \MGMCatNLO~2.2.2 &  \NNPDF~2.3 \LO & \PYTHIA~8.186 & A14 & \LO\\
Gluon--gluon fusion $A$   & \MGMCatNLO~2.2.2 &  \NNPDF~2.3 \LO & \PYTHIA~8.186 & A14 & \NNLO\\
$bbA$ &  \MGMCatNLO~2.2.3 & \CTtenF & \PYTHIA~8.210 & A14 & \NLO\\
\hline
\multicolumn{6}{c}{Top quark} \\
\hline
\ttbar      & \POWHEGBOX~v2 & \CTten & \PYTHIA~6.428 & Perugia 2012 & NNLO+NNLL\\
$s$-channel & \POWHEGBOX & \CTten & \PYTHIA~6.428 & Perugia 2012 & \NLO\\
$t$-channel & \POWHEGBOX & \CTten & \PYTHIA~6.428 & Perugia 2012 & \NLO\\
$Wt$        & \POWHEGBOX & \CTten & \PYTHIA~6.428 & Perugia 2012 & approx. \NNLO\\
\hline
\multicolumn{6}{c}{Vector boson} \\
\hline
$V$ + jets & \SHERPA~2.2.1 & \NNPDF~3.0 \NNLO&  \SHERPA~2.2.1& Default & \NNLO\\
$WW$, $WZ$, $ZZ$    &  \SHERPA~2.1.1 &  \CTten & \SHERPA~2.2.1 & Default & \NLO  \\
\hline
\multicolumn{6}{c}{Other} \\
\hline
\ttbar\h, \ttbar$W$, \ttbar$Z$ & \MGMCatNLO~2.3.2 & \NNPDF~3.0 \NLO & \PYTHIA~8.210 & A14 & \NLO \\
$q\bar{q}\rightarrow Zh$, $Wh$ & \PYTHIA~8.186 & \NNPDF~2.3 \LO& \PYTHIA~8.186 & A14 & NNLO+NLO \\
$gg\rightarrow Zh$ & \POWHEGBOX~v2 & \CTten & \PYTHIA~8.186  & AZNLO & NLO+NLL\\
\hline\hline
\end{tabular}
\end{footnotesize}
\end{center}
\end{table}
}

All simulated event samples include the effect of multiple \pp\ interactions in the same and neighbouring bunch crossings (pile-up) by overlaying 
simulated minimum-bias events on each generated signal or background event. The minimum-bias events were simulated with the single-, double- and
non-diffractive \pp\ processes of \PYTHIA~8.186 using the A2 tune~\cite{ATL-PHYS-PUB-2012-003} and the \MSTW \LO PDF~\cite{Martin:2009iq}.
For all \MADGRAPH and \POWHEG samples, the \EvtGen~v1.2.0 program~\cite{Lange:2001uf} was used for the bottom and charm hadron decays.
The generated samples were processed using the \GEANT-based ATLAS detector simulation~\cite{Agostinelli:2002hh,SOFT-2010-01} and the same 
event reconstruction algorithms were used as for the data.

\section{Event reconstruction}
\label{sec:object_selection}

This search makes use of the reconstruction of multi-particle vertices, the identification and the kinematic properties 
of reconstructed electrons, muons, $\tau$ leptons, jets, and the determination of missing transverse momentum. 

Collision vertices are reconstructed from at least two ID tracks with transverse momentum $\pt > 400$~\MeV. 
%If an event contains more than one vertex candidate, the one with the highest $\sum\pt^{2}$, calculated considering all associated tracks, is selected as the primary vertex.
The primary vertex is selected as the one with the highest $\sum\pt^{2}$, calculated considering all associated tracks.

Electrons are reconstructed from ID tracks that are matched to energy clusters in the electromagnetic calorimeter. The clusters 
are reconstructed using the standard ATLAS sliding-window algorithm, which clusters calorimeter cells within fixed-size
$\eta/\phi$ rectangles~\cite{PERF-2013-05}. Electron candidates are required to satisfy requirements for the electromagnetic shower shape, 
track quality, and track--cluster matching; these requirements are applied using a likelihood-based approach. 
The "Loose" and "Tight" working points defined in Ref.~\cite{PERF-2013-03} are used.

Muons are identified by matching tracks found in the ID to either full tracks or track segments reconstructed in the muon spectrometer ("combined muons"),  
or by stand-alone tracks in the muon spectrometer~\cite{PERF-2015-10}. Muons are required to pass identification requirements based 
on quality requirements applied to the ID and muon spectrometer tracks. The "Loose" and "Medium" identification working points defined 
in Ref.~\cite{PERF-2015-10} are used in this analysis. The "Loose" working point includes muons reconstructed with the muon spectrometer 
alone to extend the acceptance to $|\eta|=2.7$. 

Electron and muon candidates are required to have a minimum \pt\ of 7~\GeV\ and to lie within a region where there is good reconstruction and 
identification efficiency ($|\eta| < 2.7$ for muons and $|\eta|<2.47$ for electrons).  %All candidates are required to pass at least 
%the "Loose" identification requirements and to originate from the primary vertex using requirements on the significance of the transverse impact 
%parameter $d_0$ ($|d_0|/\sigma(d_0)<5.0$ for electrons, $<3.0$ for muons)  and on the longitudinal impact parameter $z_0$ ($|z_0/\sin(\theta)| < 0.5$~mm,  
%where $\theta$ is the polar angle of the lepton). 
The "Loose" lepton identification criterion has to be fulfilled and all candidates have to originate from the primary vertex. The last condition is satisfied by requiring that 
the significance of the transverse impact parameter $|d_0|/\sigma(d_0)$ is less than $5.0$ for electrons ($<3.0$ for muons)  and $|z_0\sin(\theta)|$ 
is less than $0.5$~mm,  where $z_0$ is the longitudinal impact parameter and $\theta$ is the polar angle defined in Section~\ref{sec:detector}.
The lepton candidates are required to be isolated using requirements on the sum of the \pt\ of the tracks lying 
in a cone around the lepton direction whose size, $\Delta R$, decreases as a function of the 
lepton \pt~\cite{miniiso}. The efficiency of the isolation selection is tuned to be larger than 99\% in a sample of $Z\to\ell^+\ell^-$ 
decays~\cite{PERF-2013-05,PERF-2015-10,ATLAS-CONF-2016-024}. The identification and isolation efficiencies of 
both the electrons and muons are calibrated using a tag-and-probe method in $Z\to\ell^+\ell^-$ data events~\cite{PERF-2013-05,PERF-2015-10}.

Two types of calorimeter-based jets, "\smallR" and "\largeR" jets, are used to reconstruct Higgs boson candidates over a wide momentum spectrum.
\SmallR\ jets are reconstructed from noise-suppressed topological clusters in the calorimeter~\cite{ATL-LARG-PUB-2008-002} using the anti-\kt\ \cite{Cacciari:2008gp} 
algorithm implemented in the \textsc{FastJet} package~\cite{Cacciari:2011ma} with a radius parameter $R = 0.4$ and are required to have a $\pT>20$~\GeV\ for 
$|\eta|<2.5$ (central jets) or $\pT>30$~\GeV\ for $2.5<|\eta|<4.5$ (forward jets). 
To reduce the number of \smallR\ jets originating from pile-up interactions, these jets are required to pass the jet vertex tagger~\cite{PERF-2014-03} 
selection, with an efficiency of about 90\%, if they are in the range $\pT<60$~\GeV\ and $|\eta|<2.4$. 

\LargeR\ jets are used to reconstruct Higgs boson candidates with high momenta for which the $b$-quarks are emitted close to 
each other. They are constructed using the anti-\kt\ algorithm with a radius parameter of $R = 1.0$ and are 
trimmed~\cite{Krohn:2009th} to remove the energy of clusters that originate from initial-state radiation, pile-up interactions 
or the underlying event. This is done by reclustering the constituents of the initial jet, using the \kt\ 
algorithm~\cite{Catani:1993hr,Ellis:1993tq}, into smaller $R_{\mathrm{sub}}=0.2$ subjets and then removing 
any subjet that has a \pt\ less than 5\% of the \pt\ of the parent jet~\cite{ATL-PHYS-PUB-2015-033}. The jet mass resolution is improved at 
high momentum using tracking in addition to calorimeter information~\cite{ATLAS-CONF-2016-035}.  \LargeR\ jets are required to have $\pT>250\,\GeV$ and $|\eta|<2.0$.

The momenta of both the \largeR\ and \smallR\ jets are corrected for energy losses in passive material and for the non-compensating response of the calorimeter. 
\SmallR\ jets are also corrected for the average additional energy due to pile-up interactions~\cite{PERF-2011-03,ATL-PHYS-PUB-2015-015}. 

% suggestion from James: A third type of jet, build from tracks, is used in this analysis.. 
A third type of jet, built from tracks (hereafter referred to as a track-jet), is used in this analysis for the identification of $b$-jets from decays of boosted Higgs bosons. The jets are built with the anti-\kt\ 
algorithm with $R = 0.2$ from at least two ID tracks with $\pt > 400$~\MeV\ associated with the primary vertex, or with a longitudinal impact parameter 
$|z_0\sin(\theta)|<3$~mm~\cite{ATLAS-CONF-2016-039}. Track-jets are required to have $\pt > 10$~\GeV\ and 
$|\eta| < 2.5$, and are matched to the \largeR\ jets via ghost-association~\cite{Cacciari:2008gn}.

%\SmallR\ jets and track jets containing $b$-hadrons are identified with the multivariate MV2c10 $b$-tagging algorithm~\cite{Aad:2015ydr,ATL-PHYS-PUB-2016-012} 
% at the $70\%$ efficiency working point as defined in a \ttbar\ MC sample. The algorithm makes use of information about the jet kinematics, the properties 
%of tracks within jets, and the presence of displaced secondary vertices to provide a factor of 380 (120) in the \smallR\ (track) light-quark/gluon-jet rejection and
%a factor of 6 (4) in the \smallR\ (track) $c$-jet rejection.%~\cite{ATL-PHYS-PUB-2017-003}. 
\SmallR\ jets and track-jets containing $b$-hadrons are identified with the multivariate MV2c10 $b$-tagging algorithm~\cite{PERF-2012-04,ATL-PHYS-PUB-2016-012}, 
which  makes use of information about the jet kinematics, the properties of tracks within jets, and the presence of displaced secondary vertices.
The algorithm is used at the $70\%$ efficiency working point and provides a factor of 380 (120) in rejecting \smallR\ jets (track-jets) from gluons and light quarks, 
and a factor of 12 (7) in rejecting \smallR\ jets (track-jets) from c-quarks. Jets satisfying these requirements are referred to as "$b$-tagged jets".
% benchmarks: https://twiki.cern.ch/twiki/bin/view/AtlasProtected/BTaggingBenchmarksRelease20

\begin{figure}[htb]
\centering
\includegraphics[width=0.45\textwidth]{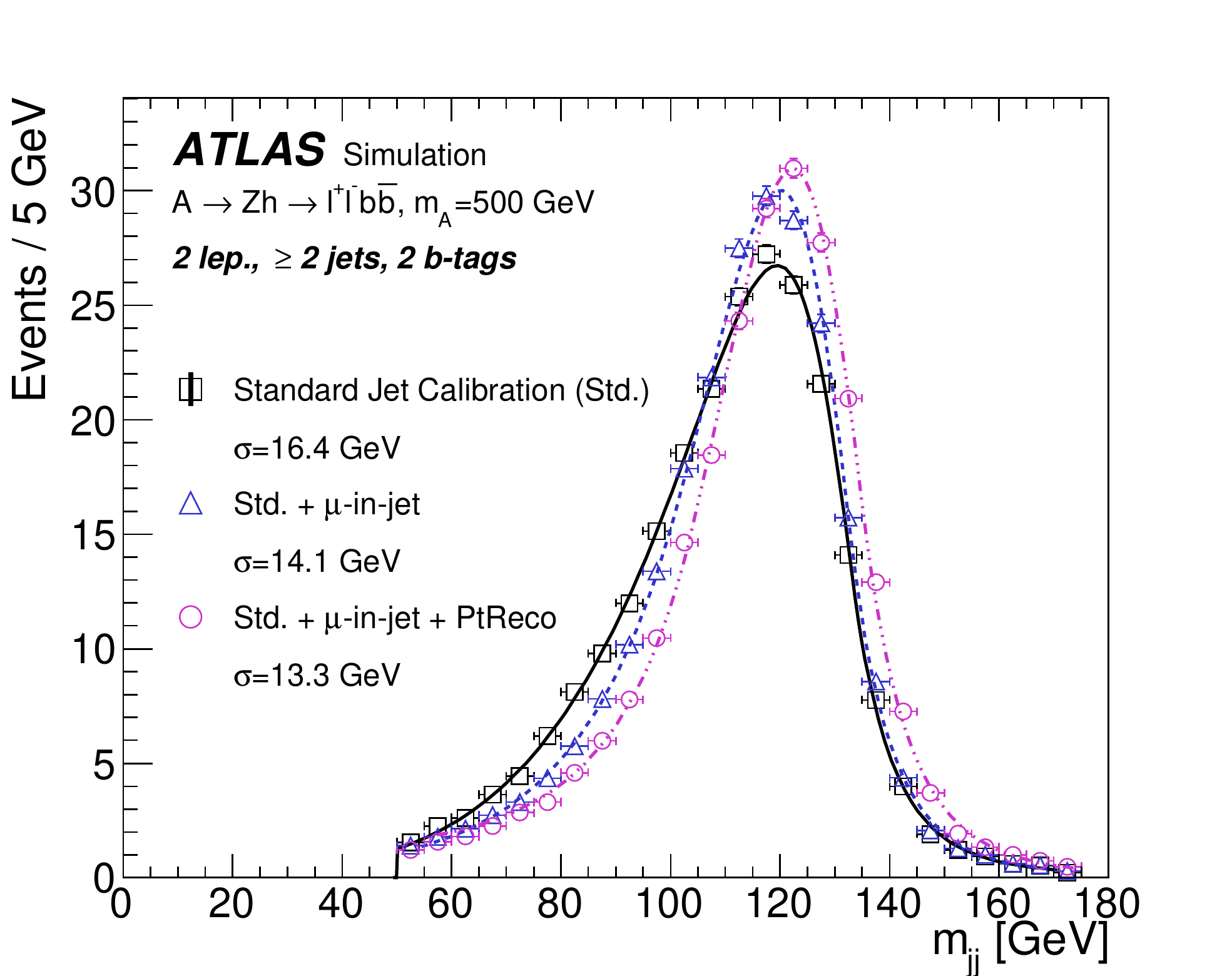}
\includegraphics[width=0.45\textwidth]{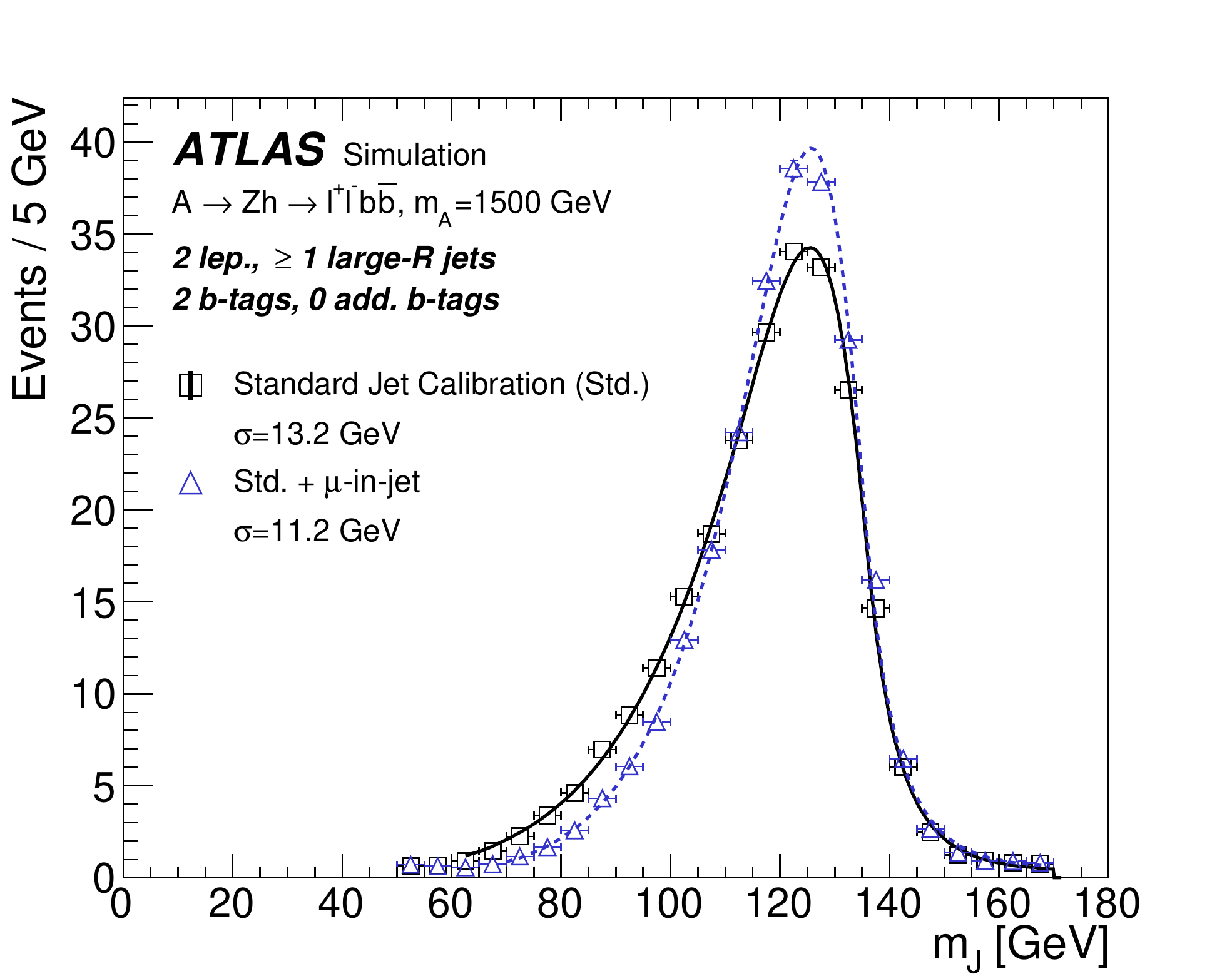}
\caption{ 
Reconstructed mass of the Higgs boson candidates for the (left) resolved and (right) merged event topologies 
in a sample of simulated signal events with $m_A=500$~\GeV\ and $m_A=1500$~\GeV\ respectively.
The different distributions correspond to the different $b$-jet energy corrections applied in each case as described in the text.
The distributions are fit to the asymmetric function described in Ref.~\cite{2007arXiv0711.4449B} and the resolution parameter is shown in the plots. 
}
\label{fig:bJetEnergy_corrections}
\end{figure}

To improve the mass resolution of the Higgs boson candidate, dedicated energy corrections are applied for $b$-tagged \smallR\ and \largeR\ jets 
to account for the semileptonic decays of the $b$-hadrons. The momentum of the closest muon in $\Delta R$ with \pt\ larger than 5~\GeV\ inside 
the jet cone is added to the jet momentum after removing the energy deposited by the muon in the calorimeter (the muon-in-jet 
correction)~\cite{ATLAS-CONF-2016-039}. For this correction, muons are not required to pass the isolation requirements.
For \smallR\ jets only, an additional \pt-dependent correction, denoted ``PtReco'', is applied to the jet four-momentum to account 
for biases in the response of $b$-jets, improving the resolution of the dijet mass. This correction is determined from \V\h($h\rightarrow b\bar{b}$) 
simulated events by calculating the ratio of the \pt\ of the true $b$-jets from the Higgs boson decay to the \pt\ of the reconstructed $b$-tagged jets 
after the muon-in-jet correction. The resolution of the dijet mass, \mdijet\ (\mjet), in this process is improved by 18\% (22\%) for the resolved (merged) 
Higgs boson reconstruction after these corrections, as shown in Figure~\ref{fig:bJetEnergy_corrections}.

Hadronically decaying $\tau$-lepton candidates ($\tau_{\mathrm{had}}$) are identified using \smallR\ jets with $\pT>20$~\GeV\ and $|\eta| < 2.5$, outside the 
transition region between the barrel and endcap calorimeters ($1.37 < |\eta| < 1.52$). These $\tau_{\mathrm{had}}$ candidates must have either one or three 
associated tracks and must satisfy the "Medium" identification criterion~\cite{PERF-2013-06}. 
They are used in the \vvbb\  channel to reject backgrounds with real hadronic $\tau$-leptons.

The presence of neutrinos in the $\nu\bar{\nu}\bbbar$ and $\ell^\pm\nu\bbbar$ final states can be inferred from a momentum imbalance in the transverse plane.
The missing transverse momentum (\metvec) is calculated as the negative vectorial sum of the transverse momenta of all the muons, electrons, \smallR\ jets, and ID tracks 
associated with the primary vertex but not associated with any of those leptons and jets~\cite{PERF-2014-04,ATL-PHYS-PUB-2015-023}. 
To suppress non-collision and multijet backgrounds in the $\nu\bar{\nu}\bbbar$ channel, an additional 
track-based missing transverse momentum estimator, \mptvec, is built independently as the negative vectorial sum of the transverse momenta of all tracks from the primary vertex.

An overlap-removal algorithm is applied to prevent double counting of the leptons and jets used for the resonance reconstruction. A $\tau$-lepton is 
removed if the $\Delta R$ between the $\tau$-lepton and an electron or a muon is below 0.2. In the case of a muon, the $\tau$-lepton is not removed if 
the $\tau$-lepton has \pT\ above 50~\GeV\ and the reconstructed muon is not a combined muon. If a reconstructed muon and electron share 
the same ID track then the electron is removed.  \SmallR\ jets are removed if they are within a cone of size $\Delta R = 0.2$ around an electron or 
muon that has passed the isolation requirements. To account for semi-muonic $b$-jet decays, the jet is only removed if it has fewer than three 
associated tracks, or if more than 70\% of the sum of the \pt\ of its associated tracks comes from the muon and $\pt^{j}/\pt^{\mu}<2$, 
where $\pt^{j}$ ($\pt^{\mu}$) is the \pt\ of the jet (muon). Next, electrons and muons within a cone of size $\Delta R = 0.4$ around a surviving \smallR\ jet 
are discarded if their distance from the jet direction is smaller than $\Delta R = (0.04 + 10\,\GeV/\pt^{\ell})$.  The shrinking cone size ensures a 
high efficiency for boosted topologies. \SmallR\ jets are also removed if they are within  $\Delta R = 0.2$ of the axis of a $\tau_{\mathrm{had}}$ 
candidate. Finally, \largeR\ jets within $\Delta R = 1.2$ of any surviving electron are removed.

\section{Analysis strategy and event selection}
\label{sec:event_selection}

The search for the \Zprime\ and \A\ bosons in the $Z\h \rightarrow \nn\bbbar$ and $Z\h \rightarrow \lplm\bbbar$ decay modes uses 
event samples wherein the number of reconstructed charged leptons is exactly zero or two (0-lepton and 2-lepton channels).
For the \Wprime\ search in the $W\h\rightarrow\lpmnu\bbbar$ channel, events with exactly zero or one charged lepton are used 
(0-lepton or 1-lepton channels). 
The lepton selection requirements described in the previous section  are applied, using the "Loose" identification working point. 
The selections outlined below define regions sensitive to the different models. 

For the 0-lepton channel, an \met\ trigger with a threshold of 70~\GeV\ was used to record the data in 2015 runs; the threshold varied  
between 90 and 110~\GeV\ in 2016 runs due to the increasing instantaneous luminosity. Events are required to have 
$\met>150$~\GeV, where \met\ is reconstructed with fully calibrated leptons and jets. 
The efficiency of the trigger selection exceeds 80\% above 150~\GeV. In the 2-lepton channel, events were recorded using a 
combination of single-lepton triggers with isolation requirements. In 2015, the lowest \pt\ threshold was 24~\GeV; in 2016, it ranged from 24 to 26~\GeV. 
Additional triggers without an isolation requirement are used to recover efficiency for leptons with $\pt>60$~\GeV. 
In the single-electron channel, the same single-electron triggers as in the 2-lepton channel are applied. 
In the single-muon channel, the same \met\ triggers as in the 0-lepton channel are used because they are more 
efficient than the single-muon triggers for this analysis. 
For events selected by the lepton triggers, the lepton that satisfied the trigger is required to match 
a reconstructed electron (muon) with $\pt > 27$~\GeV\ and $|\eta|  < 2.47\ (|\eta|<2.5) $.
%one of the offline-selected leptons in the event. The offline requirement for this lepton is $\pt > 27$~\GeV, and for muons the $|\eta|$ 
%range is restricted to $|\eta|<2.5$.

The wide range of resonance masses probed by this search implies that the resonance decay products can be produced with a wide 
range of transverse momenta. When the Higgs boson has relatively low \pT, the $b$-quarks from its decay can be reconstructed 
as two \smallR\ jets. As the momentum of the Higgs boson increases, the two $b$-quarks become more collimated and a selection 
using a single \largeR\ jet becomes more efficient. Two different methods are used for the reconstruction of the Higgs boson candidate: 
a "resolved" category in which two \smallR\ jets are used to build the Higgs boson candidate, 
and a "merged" category where the highest-\pT\ (``leading'') \largeR\  jet is selected as the Higgs boson candidate. 

For the resolved signal region, two \smallR\ jets are required to have an invariant mass (\mbb) in the range 
110 --140~\GeV\ for the 0- and 1-lepton channels and in the range 100 --145~\GeV\ for the 2-lepton channel. 
The latter selection is relaxed to take advantage of the smaller backgrounds in this channel. 
This dijet candidate is defined by the two leading $b$-tagged  \smallR\ jets when two or more $b$-tagged jets are present in the event. 
In the case where only one $b$-tagged jet is present, the dijet pair is defined by the $b$-tagged jet and the leading \smallR\ jet in the 
remaining set. The leading jet in the pair must have $\pT>45$~\GeV. 
For the merged signal region, a \largeR\ jet is required with mass (\mjet) in the range 75 to 145~\GeV\ and at least one 
associated $b$-tagged track-jet. 

Events which satisfy the selection requirements of both the resolved and merged categories, are assigned to the resolved one,
since its better dijet mass resolution and lower background contamination increases the expected sensitivity.
Events failing to satisfy both the resolved and merged signal region requirements are assigned to control regions 
defined in Section~\ref{sec:background}, with priority given to the resolved category.  %Due to the better Higgs-boson mass resolution for events in the resolved category, 
This procedure provides a higher sensitivity for resonances of mass near 1~\TeV\ compared to a procedure in which the merged 
category is prioritised.

Higgs boson candidates with one or two $b$-tagged jets define the "1 $b$-tag" or "2 $b$-tag" categories, respectively.  
For the merged selection, only one or two leading track-jets associated with the \largeR\ jet are considered in this counting. 
For the 0- and 2-lepton channels, resolved events with more than two $b$-tagged 
jets or merged events with additional $b$-tagged track-jets not associated with the \largeR\ jet are used to define signal regions sensitive to \bbA\ production. 
These are labeled as "3+ $b$-tag" in the resolved category, and "1 $b$-tag additional $b$-tag" or "2 $b$-tag additional $b$-tag" in the merged category. 
In the 2-lepton channel, the latter two are merged and labeled as "1+2 $b$-tag additional $b$-tag''.

%%%%%%%%%%%%%%%%%%%%% Mass reconstruction
The calculation of the reconstructed resonance mass depends on the decay channel.  
In the 0-lepton channel, where it is not possible to reconstruct the $Zh$ system fully due to
the presence of two neutrinos from the $Z$ boson decay, the transverse mass defined as
\begin{linenomath*}
\begin{equation*}
\MTVh = \sqrt{(E_{{\mathrm T}}^{\,\h} + {\met})^2 - (\vec{p}_{\mathrm T}^{\,\h} + \metvec)^2},
\end{equation*}
\end{linenomath*}
is used as the final discriminant. In order to reconstruct the invariant mass of the 
$W\h\rightarrow\lpmnu\bbbar$ system in the 1-lepton channel, the momentum of the neutrino in 
the $z$-direction, $p_{z}$, is obtained by imposing a $W$ boson mass constraint on the lepton--\met\ system. 
In the resulting quadratic equation, the neutrino $p_{z}$ is taken as the real component in the case 
of complex solutions, or as the smaller of the two solutions if both solutions are real. 
The mass resolution of the \V\h\ system is improved in the resolved signal regions of all channels by 
rescaling the four-momentum of the dijet system by 125~\gev / \mdijet. In the 2-lepton channel, the 
four-momentum of the dimuon system is scaled by 91.2 \gev\ / $m_{\mu\mu}$ as well in all signal 
regions. This helps to address the worse momentum resolution of high-momentum muons which 
are measured solely by the tracking detectors.
%For the 2-lepton channel, the four-momentum of the dimuon system is scaled by 91.2 \gev\ / $m_{\mu\mu}$ 
%in the calculation of the mass of the $\Z\h$ system in the signal regions. 
%This helps to address the worse momentum resolution of high-momentum muons which are measured solely by the tracking detectors. 
%Finally, the mass resolution of the \V\h\ system is improved in the resolved signal regions of all channels by rescaling the four-momentum 
%of the dijet system by 125~\gev / \mdijet.

%%%%%%%%%%%%%%%%%%%Additional selections per channel/category
Additional selections are applied for each lepton channel, as outlined below, to reduce the main backgrounds and enhance 
the signal sensitivity.  These selections are summarised in Table~\ref{tab:selKinEvt}. 

%%%%%%%%%%%%%%%%%%  0 Lepton
For the resolved and merged categories in the 0-lepton channel, the following selections are applied to reduce multijet and non-collision backgrounds 
to a negligible level:
\begin{itemize}
\item $\mpt>30$~\GeV\ (not applied in the resolved 2 and 3+ $b$-tag categories);  
\item the azimuthal angle between \metvec\ and \mptvec,  $\Delta \phi(\metvec, \mptvec)<\pi/2$;
\item the azimuthal angle between \metvec\ and the Higgs boson candidate momentum direction, $\Delta \phi (\metvec,\h)>2\pi/3$; 
\item the azimuthal angle between \metvec\ and the nearest \smallR\ jet momentum direction, \newline $\mindphi >\pi/9$ (for the resolved category with four or more jets, $>\pi/6$ is used).
\end{itemize}
For the resolved category, \mindphi\ is calculated using the \smallR\ jets that constitute the Higgs boson candidate and an additional \smallR\ jet, which is the third leading $b$-tagged jet (if the event contains at least three $b$-tagged jets), the leading central jet which is not $b$-tagged (if the event contains only two $b$-tagged jets) and the leading forward jet (if the event contains only two central \smallR\ jets). For the merged category, all central and forward \smallR\ jets are used in the \mindphi\ calculation. For the $\Zprime/A$ search, the \ttbar\ and $W+$jets backgrounds are further reduced  by rejecting events with at least one  identified $\tau_{\mathrm{had}}$ candidate. This veto is not applied when searching for the \Wprime\ boson or in the HVT combined search, because it leads to a loss of signal events in the $W\h \rightarrow \tau^{\pm} \nu b\bar{b}$ final state.

For the 0-lepton resolved category, two additional selections are applied:  
\begin{itemize}
\item the scalar sum of the \pt\ of the three leading central \smallR\ jets, \sumptjet, 
is greater than 150~\gev. In the case where there are only two central \smallR\ jets, the sum of the  \pt\ of these two jets and of the leading forward 
\smallR\ jet, if any, is required to be greater than 120~\gev;
\item the azimuthal angle between the two jets used to reconstruct the Higgs boson candidate, \dphijj, is required to be less than $7\pi/9$. 
\end{itemize}
Finally, for the merged category, the missing transverse momentum must be larger than $200$~\GeV.

{
\def\arraystretch{1.2}
\begin{table}[t]
\caption{\label{tab:selKinEvt} 
Topological and kinematic selections for each channel and category as described in the text. 
$^{(\ast)}$ Applies in the case of only two central jets.
$^{(\ast\ast)}$ Tau veto only applied for the $\Zprime/A$ search.
$^{(\dagger)}$ Tighter threshold (80 \GeV) is used for the single-electron channel.
$^{(\dagger\dagger)}$ Applied only for \MZh\ > 320 \GeV.
$^{(\ddagger)}$ Not applied in the resolved 2 and 3+ $b$-tag categories.
}

\begin{center}

\begin{footnotesize}

\begin{tabular}{l| c| c}

\hline\hline
Variable    & Resolved  & Merged \\
\hline
\multicolumn{3}{c}{Common selection} \\

\hline
\multirow{2}{*}{Number of jets}  & $\geq$2 ${\textrm{small-}R~\mathrm{jets}}$ (0, 2-lep.)  & \multirow{2}{*}{$\geq$1 large-$R$ jet} \\
& 2 or 3 \smallR\ jets (1-lep.) & \\
Leading jet \pt  [\GeV]      &  $> 45$    &  $> 250$ \\ 
$\mbb$, $\mjet$  [\GeV]      & 110--140 (0,1-lep.), 100--145 (2-lep.) & 75--145  \\

\hline
\multicolumn{3}{c}{0-lepton selection} \\
\hline
\met\ [\GeV]                       & $>150$       & $>200$\\ 
\sumptjet\ [\GeV]                  & $>150$ $(120{}^{\ast})$   & -- \\
\dphijj                           & $<7\pi/9$    & -- \\\cline{2-3}
\mpt\ [\GeV]                       & \multicolumn{2}{c}{$>30^{\ddagger}$} \\
$\Delta\phi(\metvec,\mptvec)$     & \multicolumn{2}{c}{$<\pi/2$} \\
$\Delta\phi(\metvec,\h)$          & \multicolumn{2}{c}{$>2\pi/3$}\\
\mindphi                          & \multicolumn{2}{c}{$>\pi/9$ (2 or 3 jets), $>\pi/6$ ($\geq$ 4 jets) } \\
$N_{\tau_{\mathrm{had}}}$         & \multicolumn{2}{c}{0$^{\ast\ast}$ } \\

\hline
\multicolumn{3}{c}{1-lepton selection} \\
\hline
Leading lepton \pt\ [\GeV]               &  $>27$     & $>27$ \\ 
\met\ [\GeV]                     &  $>40\ (80^{\dagger})$ & $>100$ \\
\ptw\ [\GeV]       &  $>$ max[$150,\ 710 - (3.3\times 10^5\ \GeV) / \MVh$]  &  $>$ max[$150,\ 394\cdot\ln(\MVh/(1\ \GeV)) - 2350$] \\\cline{2-3}
\mtw\ [\GeV]   &    \multicolumn{2}{c}{ $< 300$ } \\    

\hline
\multicolumn{3}{c}{2-lepton selection} \\
\hline
Leading lepton \pt\ [\GeV]            &  $>27$         & $>27$ \\ 
Sub-leading lepton \pt\ [\GeV]        &  $>7$          & $>25$ \\ \cline{2-3}
$\met/\sqrt{H_{{\mathrm T}}}$ [$\sqrt{\mathrm{\GeV}}$]  & \multicolumn{2}{c}{$< 1.15\ + 8\times 10^{-3}\cdot \MVh / (1\ \GeV)$}  \\
\ptll\ [\GeV]                         &  \multicolumn{2}{c}{$> 20 + 9\cdot \sqrt{ \MVh/(1\ \GeV) - 320}^{\dagger\dagger}$}  \\ 
\mll\ [\GeV]  & \multicolumn{2}{c}{ [max[$40\ \GeV,\ 87 - 0.030\cdot \MVh / (1\ \GeV) $], $97 + 0.013\cdot\MVh / (1\ \GeV) ]$ } \\

\hline\hline
\end{tabular}
\end{footnotesize}
\end{center}
\end{table}
}

%%%%%%%%%%%%%%%%%%  1 Lepton
For the 1-lepton channel, a selection on the transverse momentum of the $W$ boson candidate (\ptw), which increases as a function of the reconstructed 
resonance mass, is applied to reduce the contribution of \Wjets: 
$\ptw\ > \text{max}[150, 710 - 3.3\times 10^5\ \GeV / \MVh]$~\GeV\ for the resolved category, 
while $\ptw\ > \text{max}[150, 394 \cdot \ln(\MVh/(1\ \GeV)) - 2350]$~\GeV\ for the merged category.
This selection is optimised taking advantage of the larger transverse momentum of  $W$ bosons expected to be produced in the decays of high-mass resonances. 
The \ttbar\ background is reduced in the resolved category by requiring fewer than four central jets in the event 
and in the merged category by rejecting events with additional $b$-tagged track-jets not associated with the \largeR\ jet.
For all categories, the transverse mass of the $W$ candidate (\mtw), calculated from the transverse components
of the lepton and $\met$ momentum vectors, is required to be less than 300 \GeV.

In the 1-lepton channel, a significant contribution of multijet events arises mainly from non-prompt leptons from hadron decays and from jets misidentified as electrons.  
%This background is significantly reduced using requirements on lepton identification and isolation requirements, and on the \met.
This background is significantly reduced by applying tighter selection requirements on the lepton isolation and identification, as well as on \met.
Muons must satisfy the "Medium" identification and electrons must satisfy the "Tight" identification requirements. Stringent lepton isolation requirements are applied: 
the scalar sum of the \pT\ of tracks within a variable-size cone around the lepton (excluding its own track) must be less than 6\% of the lepton \pT. 
In addition, in the case of electrons the sum of the transverse energy of the calorimeter energy clusters in a cone of $\Delta R = 0.2$ around the 
electron  must be less than 6\% of the electron \pT\ \cite{PERF-2015-10,ATLAS-CONF-2016-024}.
Finally, the \met\ value is required to be greater than 100~\GeV\ for the  merged category and greater than 80 (40)~\GeV\ for the resolved category in 
the electron (muon) channel. 

%%%%%%%%%%%%%%%%%%  2 Lepton
In the 2-lepton channel, same-flavour leptons ($ee$ or $\mu\mu$) are used. For both the resolved and merged categories, three kinematic selections 
are optimised as a function of the resonance mass to reduce the \ttbar\ and \Zjets\ backgrounds. Selections on the mass of the dilepton system, 
max[$40\ \GeV,\ 87\ \GeV - 0.030\cdot \MVh$] $< \mll < 97\ \GeV\ + 0.013\cdot\MVh$, 
and on $\met/\sqrt{(1\ \GeV) \cdot\ H_{{\mathrm T}}} < 1.15\ + 8\times 10^{-3}\ \cdot \MVh\ /\ (1\ \GeV)$ 
are relaxed for higher-mass resonances to account for resolution effects and smaller backgrounds. 
The variable $H_{{\mathrm T}}$ is calculated as the scalar sum of the \pt\ of the leptons and \smallR\ jets in the event. 
The momentum of the dilepton system (\ptll) is required to be greater than $20\ \GeV + 9\ \GeV\ \cdot \sqrt{ \MVh/(1\ \GeV) - 320}$ for \MVh\ greater than 320~\GeV.
%In the resolved category, muons are required to have opposite charge to reduce the multijet background and the leading muon is required to have $|\eta|$ less than $2.5$.
In the resolved dimuon category, an opposite-charge requirement is applied since the probability to mis-reconstruct the charge of 
individual muons is extremely low. Additionally, in this category the leading muon is required to have $|\eta|$ less than $2.5$.
Finally, for the merged category, the sub-leading lepton is required to have $\pt>25\ \GeV$ and for muons $|\eta|$ is restricted to be less than $2.5$. 

%The signal selection efficiency multiplied by the acceptance ranges from 7\% to 92\% depending 
%on the resonance mass and the model considered.

%%%%%%%%%%%%%Fit regions
% Spyros: I move this to the results section. 
%The search for the different signal models uses different channels and categories.
%A total of five different fits are performed: \Zprime, \Wprime, HVT, $A$ in gluon fusion, and $A$ with $b$-quark association. 
%The list of signal regions and control regions (defined in the next section) used for the different fits is shown in Table~\ref{tab:regionsFit}.

%\input{sections/fit_regions_table.tex}

\section{Background estimation}
\label{sec:background}

The background contamination in the signal regions is different for each of the three channels studied. 
In the 0-lepton channel, the dominant background sources are \Zjets\ and \ttbar\ events with a significant contribution from \Wjets.  
In the 1-lepton channel, the largest backgrounds are \ttbar, single-top-quark and \Wjets\ production. 
In the 2-lepton channel, \Zjets\ production is the  predominant background followed by the \ttbar\ background. 
%Another background source consists of multijet events with semileptonic  hadron decays or misidentified jets.  This background is found to be negligible in the 0- and 2-lepton channels, after applying the event selections  described in Section~\ref{sec:event_selection}. 
The contribution from diboson, SM $V\h$, $\ttbar\h$, and $\ttbar V$  production is small in all three channels. 
The multijet background, due to semileptonic hadron decays or misidentified jets, is found to be negligible in the 0- and 2-lepton channels after applying the event selections described in Section 5. 
In the 1-lepton channel, the multijet background remains significant only in the resolved 1 $b$-tag category.
All background distribution shapes except those for multijet are estimated from the samples of simulated events with normalisations of the main 
backgrounds estimated from the data; the multijet shape and normalisation is determined using data.

The \WZjets\ simulated event samples are split into different components. In the resolved category, the samples are split according 
to the true flavour of the two \smallR\ jets forming the Higgs boson candidate. In the merged category, they are split according to the 
true flavour of the one or two leading track-jets associated with the \largeR\ jet. 
The true jet flavour is determined by counting true heavy-flavour hadrons with $\pt > 5$~\GeV\ within the cone of 
the reconstructed jet. If a  true $b$-hadron is found, the jet is labelled as a $b$-jet, otherwise if a true $c$-hadron is found the jet is 
labelled as a $c$-jet. If neither a true $b$-hadron nor a true $c$-hadron is associated with the reconstructed jet, it is labelled as a light jet. 
For \largeR\ jets with only one track-jet, the true hadrons are counted within this track-jet. 
Based on this association scheme, the \WZjets\ simulated event samples are split into six 
components:
\WZbb,  \WZbc, \WZbl, \WZcc, \WZcl\ and \WZll; 
in this notation $l$ refers to a light jet.
In the statistical analysis described in Section~\ref{sec:results}, the components \WZbb, \WZbc, and \WZcc\ are treated as a single 
component  denoted by \WZhf. The combination of \WZbl\ and \WZcl\ is denoted by \WZhl. 
For the HVT, \Zprime, and \A\ boson interpretations, the normalisations of the largest components \Zhf\ and \Zhl\ are determined from data.
In the \A\ boson interpretation, the \Zhf\ background normalisation in the 3+ $b$-tag region is determined from this region independently.
The normalisations of \Whf\ and \Whl\ are determined from data for the \Wprime\ and HVT interpretations.

The normalisation of the \ttbar\ background is determined from the fits to data separately for the 0-, 1-, and 2-lepton channels. 
In the 0-lepton channel, only the signal regions are used in the fit. In the 1- and 2-lepton
channels, dedicated control regions enhanced in \ttbar\ events are used in addition to the signal regions. In the 1-lepton channel, 
resolved events in the sidebands of the \mdijet\ distribution between 50~\GeV\ and 200~\GeV\ (excluding the signal region 
with $110 < \mdijet < 140$~\GeV) are primarily composed of \ttbar\ and \Wjets\ events. These control regions are included in the 
fit for the 1 and 2 $b$-tag categories.
In the 2-lepton channel, a \ttbar\ control region is defined using resolved events with different-flavour ($e\mu$), oppositely 
charged leptons, and without the $\met/\sqrt{H_{{\mathrm T}}}$ requirement. The \ttbar\ purity of this selection is greater than 90\%. 
This region combining the 1 and 2 $b$-tag events is used in the $A$, \Zprime, and HVT interpretations; for the $A$ interpretation, a 
control region with $e\mu$ events and 3+ $b$-tags is also included in the fit to provide an independent constraint on \ttbar\ 
production with associated heavy-flavour jets.  

The shape of the multijet background in the 1-lepton channel is estimated from a sample of data events orthogonal to the 
signal regions, the anti-isolated lepton region. In the muon channel, this region is defined by events where the sum of the transverse momentum of tracks in a cone of $\Delta R = 0.2$
around the muon is between 6\% and 15\% of the muon \pt.  In the electron channel, this region is defined by events where 
the sum of the calorimeter energy deposits in a cone of $\Delta R = 0.2$ around the electron is larger than 6\%  of the electron \pt; 
this region is defined after applying the track isolation requirement described in Section~\ref{sec:object_selection}. 
%A "template" shape for this background is extracted after removing from the selected data events the contribution from the simulated \ttbar, single-top, $V+$jets and diboson backgrounds. 
A template shape for the multijet background is extracted from the anti-isolated lepton region after removing the 
contribution from the simulated electroweak and top-quark backgrounds.
In this subtraction, the normalisation of the simulated electroweak and top-quark backgrounds is estimated by fitting them to data
in the region $\met>200$~\gev\ where the contribution from multijet events is negligible. 
In the signal and control regions used in the statistical analysis, the multijet normalisation is determined by fitting the $\met$ multijet template and
the $\met$ combined template of the electroweak and top-quark backgrounds to data in the 1 and 2 $b$-tag categories separately.
Using this method, the multijet contribution is estimated to be less than 6\% in all signal and control regions and is included in the statistical analysis.

\section{Systematic uncertainties}
\label{sec:systematics}

%%%introduction
Two types of systematic uncertainties, experimental and modelling, affect the reconstruction of the \MVh\ and \MTVh\ observables. 
Experimental uncertainties arise due to the trigger selection, the reconstruction, identification, energy/momentum, 
mass, and resolution for the leptons, jets and missing transverse momentum. Modelling uncertainties result in shape and normalisation uncertainties of the different MC 
samples used to model the signal and backgrounds. These stem from uncertainties in the matrix element calculation, the choice of parton shower
and hadronisation models and their free parameters, the PDF set and the choice of renormalisation and factorisation scales.

{
\def\arraystretch{1.2}
\begin{table}[t]
\caption{\label{tab:syst} 
Relative systematic uncertainties in the normalisation, cross-region extrapolation, and shape of the signal and background processes included in the fits described in the text. 
An ``S'' indicates a shape variation is included for the sources listed, 
``/'' indicates a ratio of two regions,
and ``norm.'' is the sum of cross-section and acceptance variations.
A range of values means the value depends on the lepton channel.
Parentheses indicate when the uncertainty applies only to a given fit or a given region.
}

\begin{footnotesize}

\begin{tabular}{cc}

%%%%%%Left table
\begin{minipage}{0.5\linewidth}
\begin{center}
\begin{tabular}{l | c | c }
\hline\hline
Process    & Quantity/source & Value \\
\hline

%%%Signal
\multirow{1}{*}{ Signal } 
& acceptance  & 3--7\% \\
%\hline
%\multirow{1}{*}{ \Wprime,\Zprime } 
%& acceptance  & 1-3\% \\
\hline

%%%%SM 
\multirow{1}{*}{SM \V\h, \ttbar\V, \ttbar\h } 
& norm. & 50\% \\
\hline

%%%%diboson
\multirow{1}{*}{Diboson }  
& norm. & 11\% \\
\hline

%%%%%% QCD
\multirow{2}{*}{Multijet (1-lep.) }  
& norm. & 50\% \\
& template method & S \\
\hline

%%%%single top
\multirow{2}{*}{Single top quark}  
& norm. & 19\% \\
& resolved / merged  & 24\% \\
& \mbb\ SR / \mbb\ CR  (1-lep.) & 7\% \\
\hline

%%%%ttbar
\multirow{2}{*}{\ttbar}  
% & 0 lep. resolved / merged  & 15\% \\
% & 1 lep. resolved / merged  & 26\% \\
% & 2 lep. resolved / merged  & 46\% \\
 & resolved / merged  & 15--46\% \\
 & \mbb\ SR / \mbb\ CR  (1-lep.) & 7\% \\
 & SR / $e\mu$ CR  (2-lep.) & 2\% \\
 & PS, ISR/FSR, ME  & S \\
 & \pT\ reweight & S \\
\hline

%% https://indico.cern.ch/event/631302/contributions/2559224/attachments/1445665/2226949/ModellingUncertainties_180417.pdf
%%%%Z+jets
\multirow{2}{*}{\Zhf}  
 & resolved / merged  & 19\% \\
 & 0-lep. / 2-lep.  & 15\% \\
% & 1-lep. / 2-lep.  & 45\% \\
 & generator, PDF, scale & S \\
\hline

\multirow{2}{*}{\Zhl}  
& resolved / merged  & 28\% \\
& 0-lep. / 2-lep.  & 12\% \\
%& 1-lep. / 2-lep.  & 60\% \\
& generator, PDF, scale & S \\
\hline\hline
\end{tabular}
\end{center}
\end{minipage}

\hfill

%%%%%%Right table
\begin{minipage}{.5\linewidth}
\begin{center}
\begin{tabular}{l | c | c }
\hline\hline
Process    & Quantity/source & Value \\
\hline

\multirow{2}{*}{\Zl}  
& norm.  & 19\% \\
& resolved / merged  & 23\% \\
& 0-lep. / 2-lep.  & 8\% \\
%& 1-lep. / 2-lep.  & 40\% \\
& generator, PDF, scale & S \\
\hline

%%%%W+jets
\multirow{2}{*}{\Whf}  
& norm. ($A$,\Zprime) & 26\% \\
& resolved / merged  & 18--43\% \\
& \mbb\ SR / \mbb\ CR  (1-lep.) & 6\% \\
& 0-lep. / 1-lep.  & 26\% \\
& generator, PDF, scale & S \\
\hline

\multirow{2}{*}{\Whl}  
& norm. ($A$,\Zprime) & 23\% \\
& resolved / merged  & 15--35\% \\
& \mbb\ SR / \mbb\ CR  (1-lep.) & 5\% \\
& 0-lep. / 1-lep.  & 22\% \\
& generator, PDF, scale & S \\
\hline

\multirow{2}{*}{\Wl} 
& norm. & 20--30\% \\ 
& resolved / merged  & 16--20\% \\
& \mbb\ SR / \mbb\ CR  (1-lep.) & 2\% \\
& 0-lep. / 1-lep.  & 19\% \\
& generator, PDF, scale & S \\
\hline\hline
\end{tabular}
\end{center}
\end{minipage}

\\
\end{tabular}

\end{footnotesize}

\end{table}
}

%%%%Experimental
The largest experimental systematic uncertainties are associated with the calibration and resolution of the \smallR\ and \largeR\ jet energy, 
the calibration and resolution of the \largeR\ jet mass, 
and the determination of the jet $b$-tagging efficiency and  misidentification rate. 
The uncertainties in the \smallR\ jet energy scale have contributions from \emph{in situ} calibration studies, from the dependency on the pile-up 
activity and on the flavour composition of jets~\cite{ATL-PHYS-PUB-2015-015,ATLAS-CONF-2015-002}. % 2016 public plots: https://atlas.web.cern.ch/Atlas/GROUPS/PHYSICS/PLOTS/JETM-2016-010/
The \smallR\ jet uncertainties are propagated to the \met\ measurement.
The uncertainty in the scale and resolution of \largeR\ jet energy and mass is estimated by comparing the ratio of calorimeter-based to 
track-based measurements in dijet data and simulation~\cite{ATL-PHYS-PUB-2015-033,ATLAS-CONF-2016-035}. % for 2016 only public plots: https://atlas.web.cern.ch/Atlas/GROUPS/PHYSICS/PLOTS/JETM-2016-009/
The flavour tagging efficiency and its uncertainty for $b$-jets and $c$-jets is estimated in $t\bar{t}$ and $W+c$-jet events, respectively,
while the light-jet misidentification rate and uncertainty is determined using dijet events~\cite{PERF-2012-04,ATL-PHYS-PUB-2016-012,ATLAS-CONF-2014-004,ATLAS-CONF-2014-046}. % a paper on b-jet calibration might be published in time for Moriond, for light jets need to use the old reference
Other experimental systematic uncertainties with a smaller impact are those in the lepton energy and momentum scales, 
in lepton reconstruction and identification efficiency, and in the efficiency of the triggers. 
Finally, a global normalisation uncertainty of \LumiRelUnc\ is assigned due to the luminosity measurement from a preliminary calibration 
of the luminosity scale using $x$--$y$ beam-separation scans performed in August 2015 and May 2016, following a methodology similar to that detailed in Ref.~\cite{DAPR-2013-01}. 
Experimental uncertainties  have an impact  on the shape of the mass distributions and account for possible migration of events across the different regions.

%%%Modelling

%Modelling uncertainties are assigned to each signal and background process and lead to variations in the normalisation and shape of the templates
%in the different regions. 
Modelling uncertainties are assigned to each signal and background process and lead to variations in the normalisation and in the case of main backgrounds 
also in the shape of the templates in the different regions. In addition, for all MC samples, the statistical uncertainty arising from the number of simulated 
events is considered by introducing shape variations determined from the uncertainty in each bin of the \MVh\ or \MTVh\ distributions. %entering the fit.
The modelling uncertainties considered are shown in Table~\ref{tab:syst} and described below.

%For the signal processes, the choice of PDFs, factorisation and renormalisation scales,  parton shower (PS), and underlying event tune have been varied and lead to acceptance variations between 3\% and 8\% at high and low resonance masses, respectively. 

For the signal processes, the uncertainties in the acceptance were derived by considering the following variations: 
the renormalisation and factorisation scales were varied by a factor of two,
the nominal PDF set was replaced by the \MSTW \LO PDF set and the tuned parameters were varied according to the variations derived from the eigentune method~\cite{ATL-PHYS-PUB-2014-021}.
%Variations in the acceptance from the above tests are summed in quadrature.
% and lead to uncertainties  between 3\% and 8\% at high and low resonance masses, respectively. 
For both the $A$ and $V'$ signals, the total variations are less than 3\% at resonance masses above 500~\GeV. The variations increase to 7\% for the $A$ 
boson masses below 500~\GeV.

%%%%% ttbar
%https://twiki.cern.ch/twiki/bin/viewauth/AtlasProtected/ReferenceCONF13TeVMC15
% description from modeling note: https://cds.cern.ch/record/2111370/files/ATL-COM-PHYS-2015-1474.pdf 

The modelling uncertainties affecting \ttbar\ and single-top-quark processes are derived as follows~\cite{ATL-PHYS-PUB-2016-004}.  
A variation of the parton shower, hadronisation, and the underlying-event model is obtained by replacing \PYTHIA 6.428 by \Herwigpp (version 2.7.1) ~\cite{Bahr:2008pv} with the 
UE-EE-5 tune and the \CTEQ PDF set~\cite{Pumplin:2002vw}. 
To assess potential differences in the matrix element calculation, a comparison is made to a sample where \POWHEG is replaced by \MGMCatNLO~\cite{Alwall:2014hca}.
A comparison is also made to samples with smaller and larger amount of initial- and final-state radiation (ISR/FSR) by changing the renormalisation and factorisation scales by a factor of 
two and switching to the corresponding low- and high-radiation Perugia 2012 tunes. 
Finally, the difference between the nominal and corrected distributions due to the top-quark and \ttbar\  \pT\ reweighting described in Section~\ref{sec:data_MC_samples} is included as a symmetrised shape uncertainty. 

%%%%%%%V+jes
%https://twiki.cern.ch/twiki/bin/viewauth/AtlasProtected/ReferenceCONF13TeVMC15
Similarly, for  \WZjets\ backgrounds, the following comparisons have been performed. 
The PDF set in the nominal samples was replaced by the alternative PDF sets: the hundred \NNPDF 3.0 \NNLO replicas, including the sets resulting 
from variations of \alphas~\cite{Ball:2014uwa}, the MMHT2014 NNLO set, and the CT14 NNLO PDF set~\cite{Butterworth:2015oua}.
The scale uncertainties are estimated by comparing samples where the renormalisation and factorisation scales were modified by a factor of two.
Finally, a comparison was made to a sample generated using \MGMCatNLO~v2.2.2 interfaced to \PYTHIA~8.186 and using the 
A14 tune together with the \NNPDF 2.3 \LO PDF set~\cite{Sjostrand:2014zea,Alwall:2014uj,Ball:2012cx}.

%%%%%%common procedures top and V+jets
For the \ttbar, single-top-quark, and \WZjets\ backgrounds, the acceptance differences that affect the relative normalisation across 
regions with a common background normalisation are estimated by summing in quadrature the relative yield variations between 
the different regions. These uncertainties are assigned to all regions used in the fit as shown in Table~\ref{tab:regionsFit} and 
across the different lepton channels as shown in Table~\ref{tab:syst}. 

%%%%% multijet
For the multijet background included in the 1-lepton channel, a 50\% uncertainty in the normalisation is estimated from the fit to the \met\ distribution  described in Section~\ref{sec:background}.  
Also, a shape variation is included to account for uncertainties in the determination of the template in the anti-isolated lepton region, arising from differences in the trigger scheme between isolated and anti-isolated regions and uncertainties in the normalisation of the top-quark and electroweak backgrounds in this region.

%%%%% diboson, SM Vh, ttV, ttH , multijet
Finally, for the remaining small backgrounds only a normalisation uncertainty is assigned.
For the diboson backgrounds a normalisation uncertainty of 11\% is applied~\cite{DibosonUnc}.
For the SM $Vh$, $ttV$, and $tth$ production, a 50\% uncertainty is assigned  which covers the uncertainty in the cross-sections.

\section{Results}
\label{sec:results}

%%%% fit discussion

In order to test for the presence of a massive resonance, the $\MTVh$ and $\MVh$ templates obtained from the signal and background simulated event samples
are fit to data using a binned maximum-likelihood approach  based on the \textsc{RooStats} framework ~\cite{Moneta:1289965,Verkerke:2003ir,Baak2015}.
A total of five different fits are performed according to the signal interpretation: \Zprime, \Wprime, HVT, $A$ in gluon--gluon fusion, and $A$ in $b$-quark associated production. 
The list of channels and regions used for the different fits is shown in Table~\ref{tab:regionsFit}.

{
\def\arraystretch{1.2}
\begin{table}[t]
\caption{\label{tab:regionsFit} 
A list of the signal and control regions (separated by commas below) included in the statistical analysis of the $A$ and HVT model hypotheses. The notation 
1+2 $b$-tag indicates the 1 and 2 $b$-tag regions are combined, and add. $b$-tag indicates the regions with additional $b$-tags not associated with the \largeR\ jet.
}

\begin{center}

\begin{footnotesize}

\begin{tabular}{l c| c | c| c | c}
\hline\hline

\multicolumn{2}{c|}{\multirow{2}{*}{Fit}} & \multirow{2}{*}{Channel}  & Resolved        & Merged          & Resolved  \\
& &         & signal regions  & signal regions  & control regions \\

\hline
\multicolumn{2}{c|}{\multirow{2}{*}{$A$}}
& 0-lepton & 1, 2, 3+ $b$-tag & 1, 2 $b$-tag, and 1, 2 $b$-tag add. $b$-tag   &  --\\
& & 2-lepton & 1, 2, 3+ $b$-tag & 1, 2 $b$-tag, and 1+2 $b$-tag add. $b$-tag  &  1+2 $b$-tag , 3+ $b$-tag $e\mu$\\
\hline
\multirow{3}{*}{HVT} & \Zprime, \Wprime &  0-lepton  & 1, 2 $b$-tag &  1, 2 $b$-tag & -- \\
& \Wprime &  1-lepton & 1, 2 $b$-tag &  1, 2 $b$-tag & 1, 2 $b$-tag  \mdijet\ sideband \\
& \Zprime & 2-lepton & 1, 2 $b$-tag &  1, 2 $b$-tag & 1+2 $b$-tag $e\mu$\\

\hline\hline
\end{tabular}
\end{footnotesize}
\end{center}
\end{table}
}

The fits are performed on the $\MTVh$ distribution in the 0-lepton channel and the $\MVh$ distribution in the 1- and 2-lepton channels 
using a binning of the distributions  chosen to optimise the search sensitivity while minimising statistical fluctuations.
%As described in Section~\ref{sec:background}, for all fits the \ttbar, \Zhf, and \Zhl\ backgrounds are left freely floating, the \Whf\ and \Whl\ are also floated in the \Wprime\ and HVT fits.  
As described in Section~\ref{sec:background}, the normalisations of the \ttbar, \Zhf, and \Zhl\ backgrounds are free parameters in all fits, 
as are the normalisations of \Whf\ and \Whl\  in the \Wprime\ and HVT fits.  
The systematic uncertainties described in Section~\ref{sec:systematics} are incorporated 
in the fit as nuisance parameters with correlations across regions and processes taken into account. 
The signal normalisation is a free parameter in the fit.
In order to account for migrations of signal events across different channels due to lepton reconstruction and selection inefficiencies, the 
$Z\h\rightarrow \lplm\bbbar$ ($W\h\rightarrow \lpmnu\bbbar$) signal samples are included in the 1(0)-lepton categories.  

The total uncertainty in the signal yield is dominated by different sources of systematic uncertainty depending on the mass of the resonance used in the fit.
The uncertainties in the \WZhf\ shape and normalisation, \ttbar\ normalisation, and in the flavour tagging efficiencies constitute the dominant sources of 
systematic uncertainty for low-mass resonances.
%For high-mass resonances, above approximately 1~\TeV, the statistical uncertainty dominates.
For all interpretations, the statistical uncertainty dominates for resonances above 1~\TeV.
The uncertainties in the \largeR\ jet mass resolution and in the track-jet $b$-tagging efficiency constitute the dominant systematic uncertainties at high masses.

\begin{figure}[tb]
\centering
\includegraphics[width=0.43\textwidth]{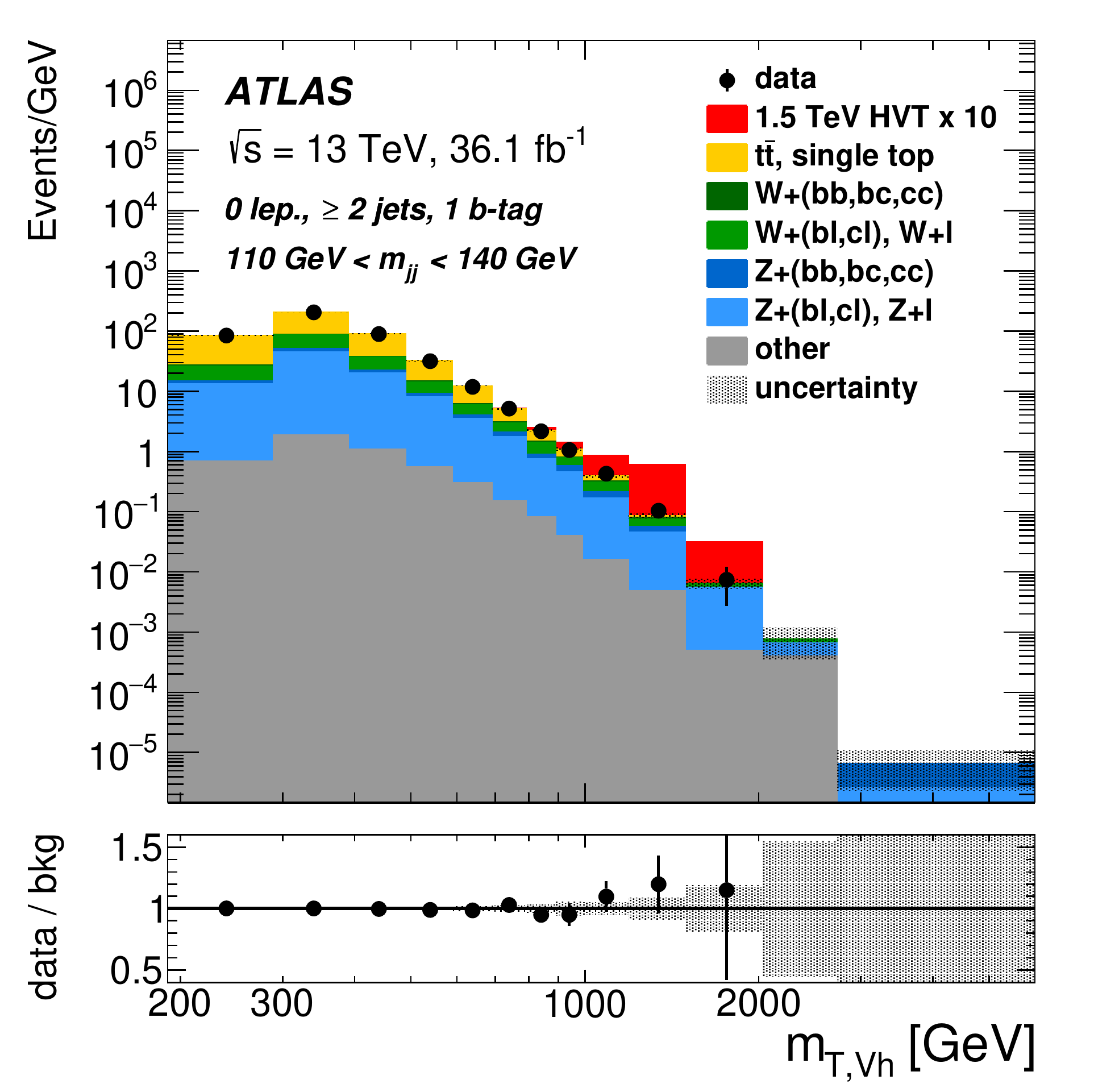}
\includegraphics[width=0.43\textwidth]{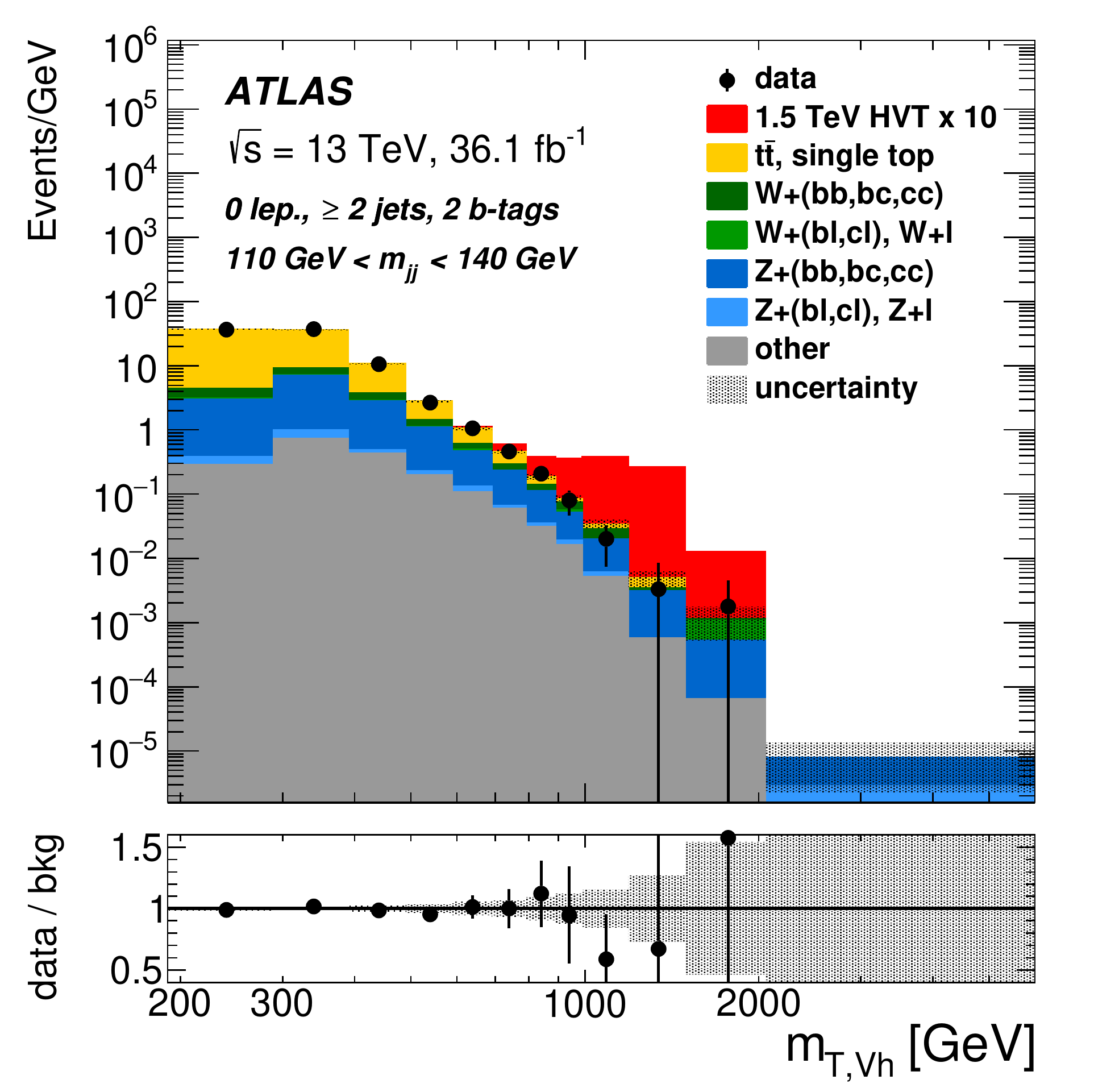}
\\
\includegraphics[width=0.43\textwidth]{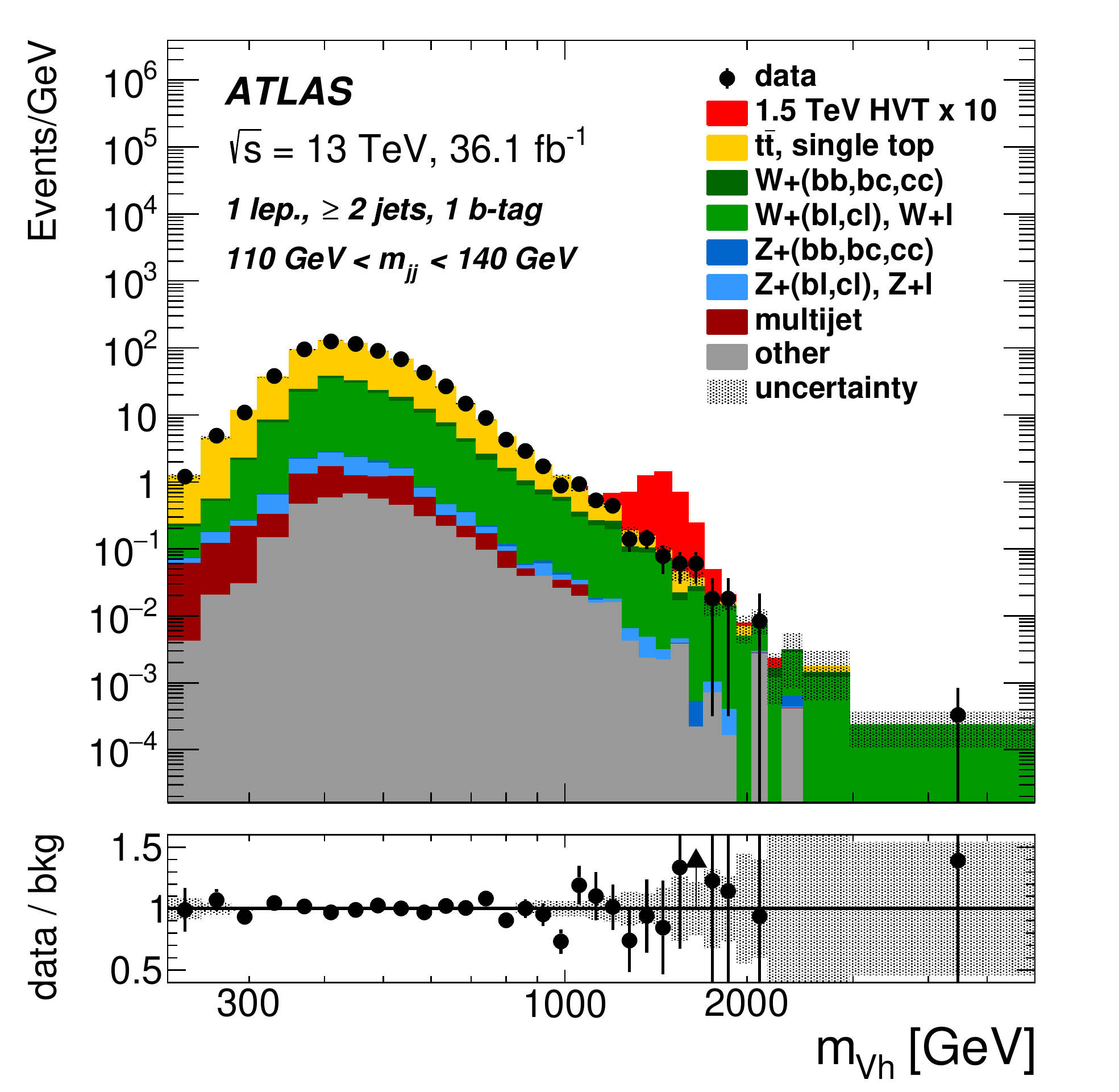}
\includegraphics[width=0.43\textwidth]{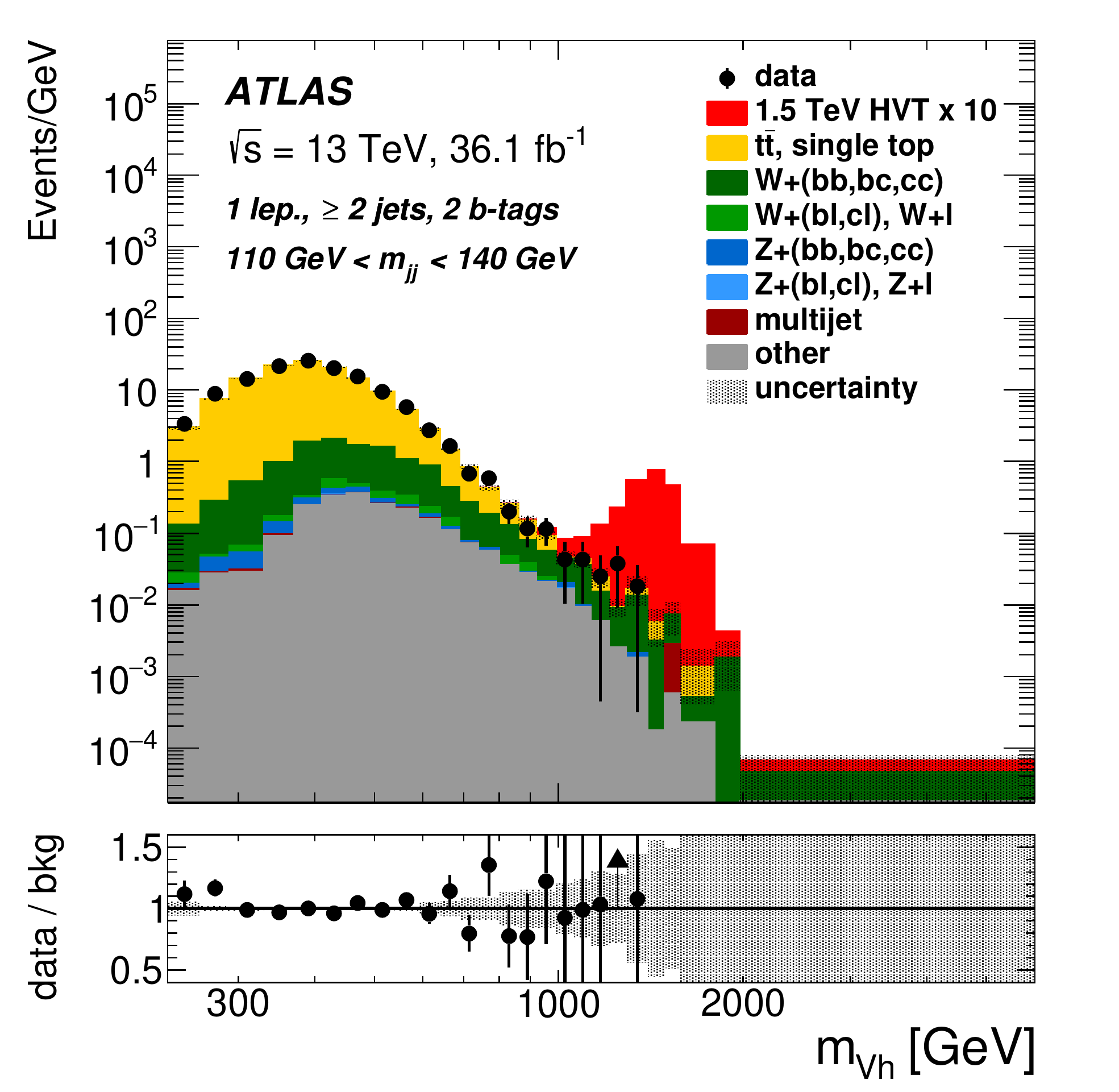}
\\
\includegraphics[width=0.43\textwidth]{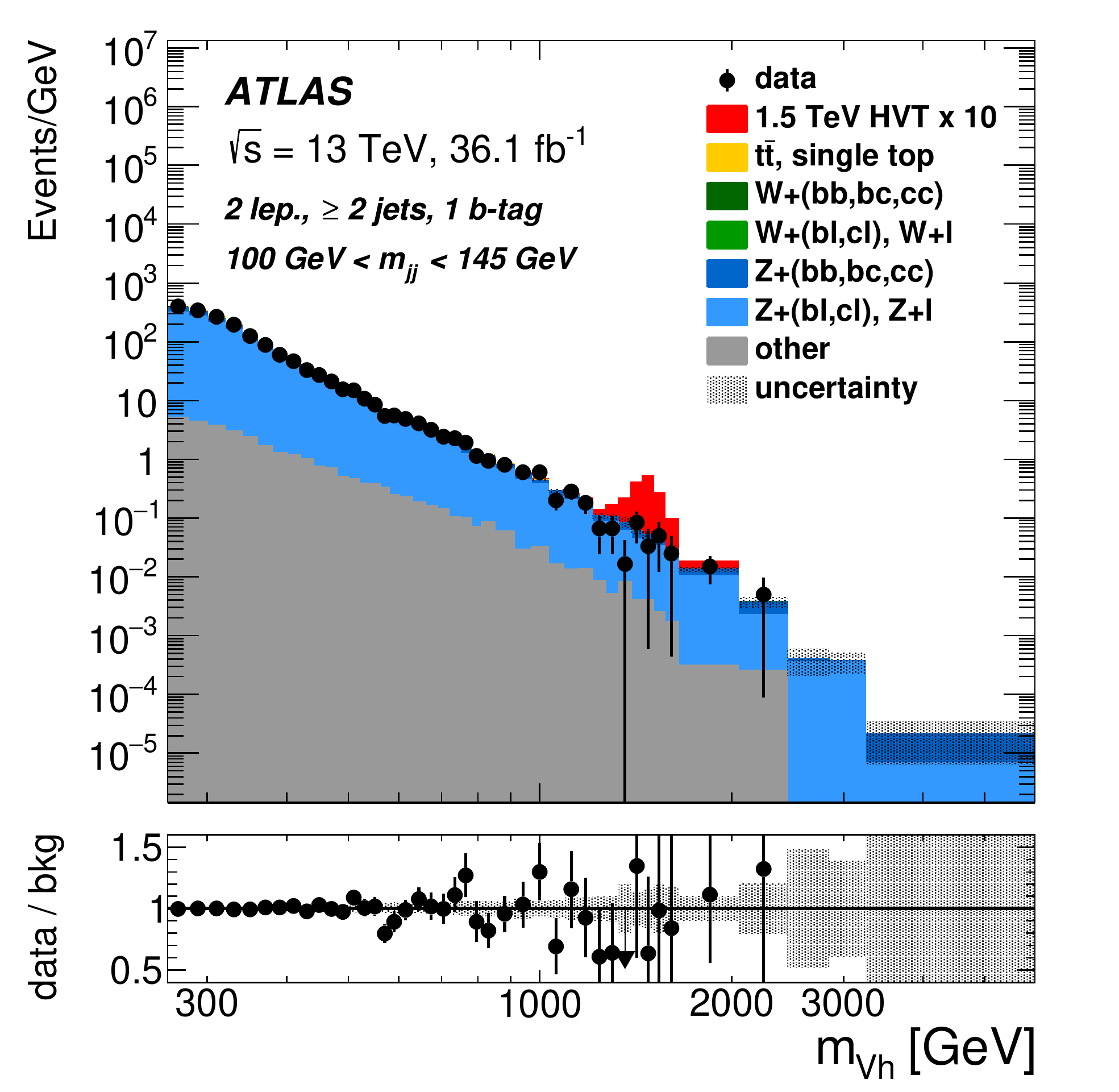}
\includegraphics[width=0.43\textwidth]{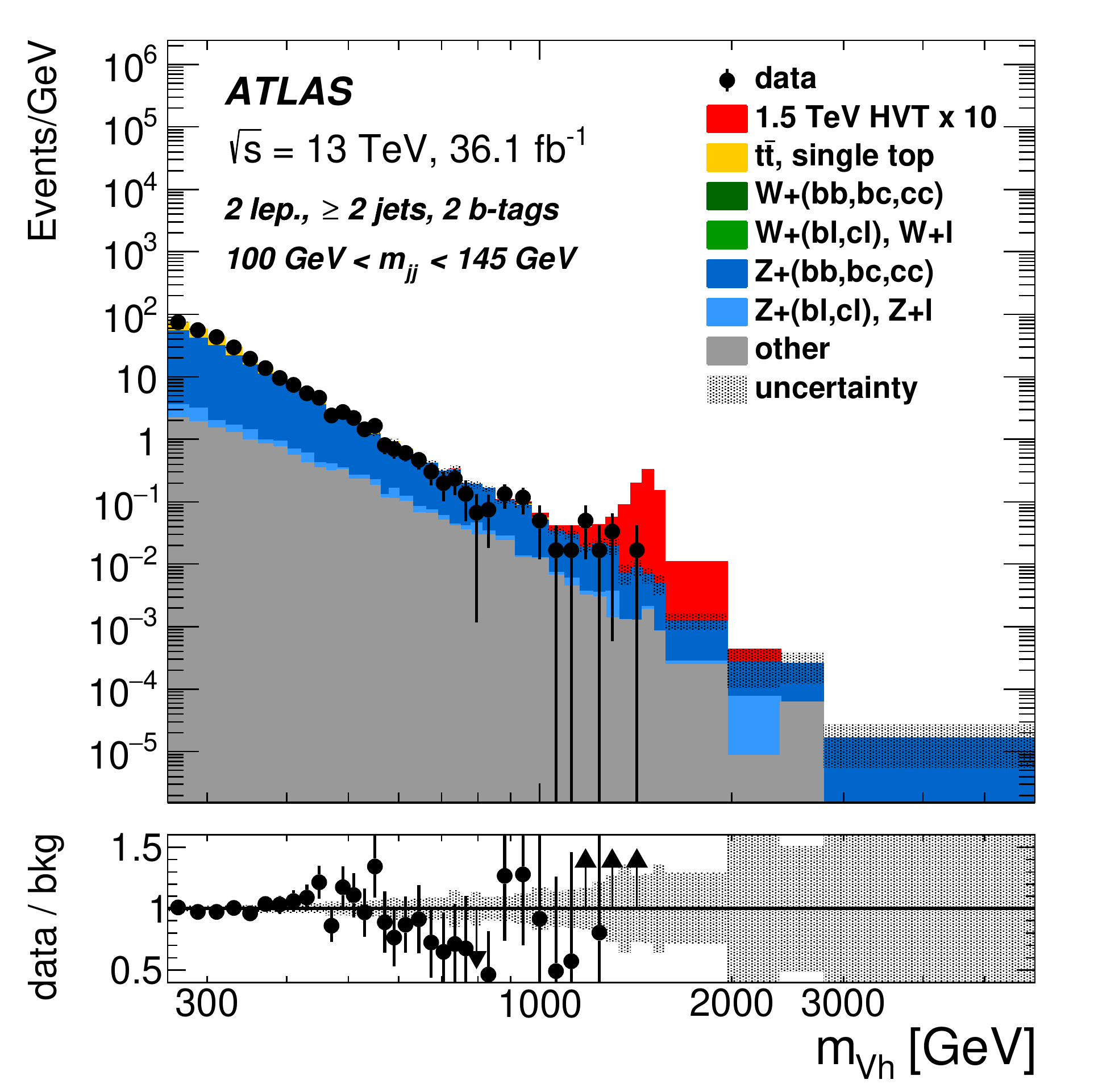}
\caption{
Event distributions of $m_{\mathrm{T}, V\h}$  for the 0-lepton channel, and  $m_{V\h}$ for the 1-lepton  and 2-lepton channels in the resolved categories.
The quantity on the vertical axis is the number of data events divided by the bin width in \GeV.
The background prediction is shown after a background-only maximum-likelihood fit to the data.
The signal for the benchmark HVT \textit{Model A} with $m_{V'} = 1.5 $~\TeV\ is normalised to 10 times the theoretical cross-section.
The background uncertainty band shown includes both the statistical and systematic uncertainties after the fit added in quadrature. The lower panels show the ratio of the observed data to the estimated SM background.
} \label{fig:mVH_HVTpostfit_resolved}
\end{figure}

\begin{figure}[tb]
\centering
\includegraphics[width=0.43\textwidth]{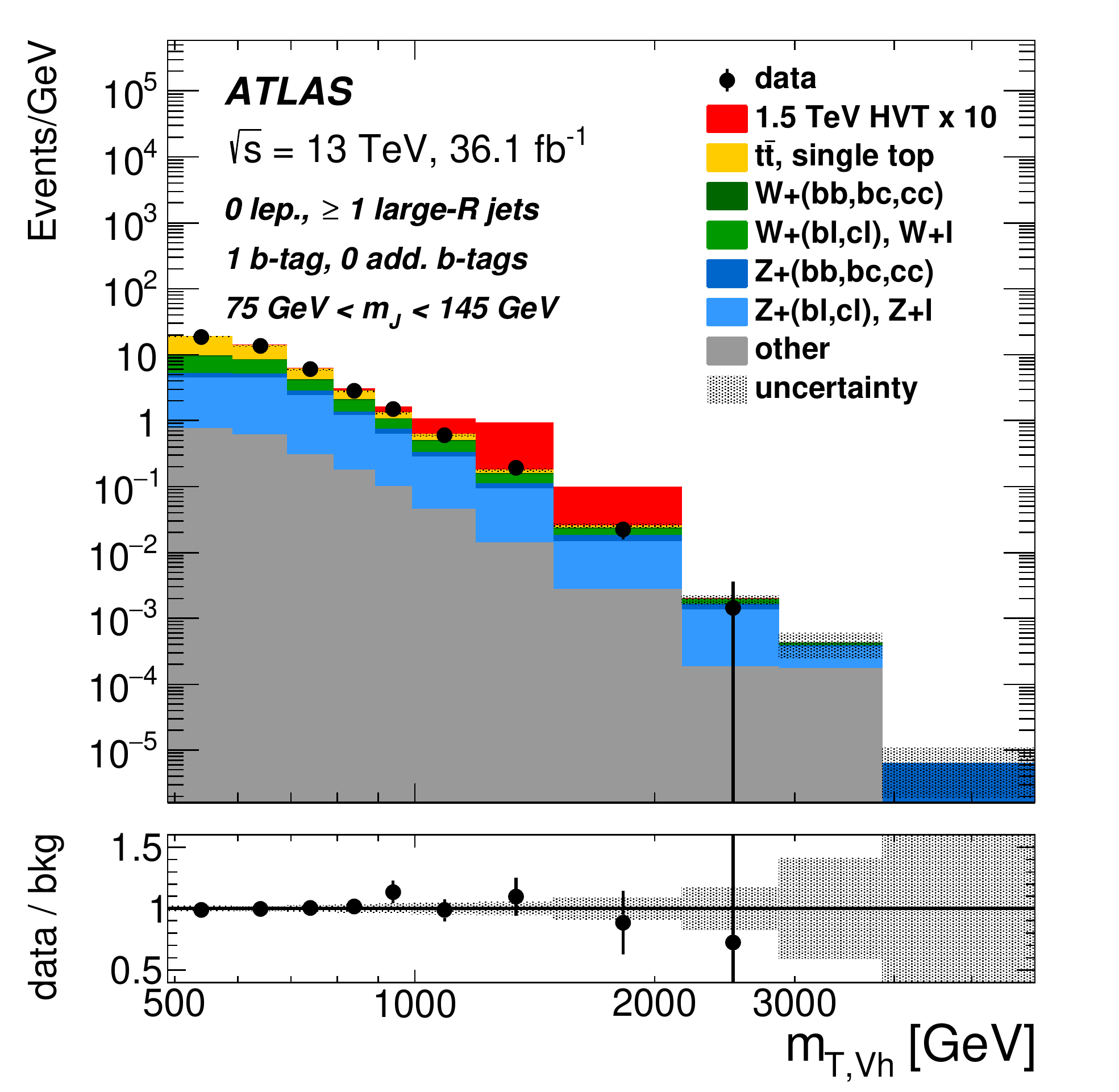}
\includegraphics[width=0.43\textwidth]{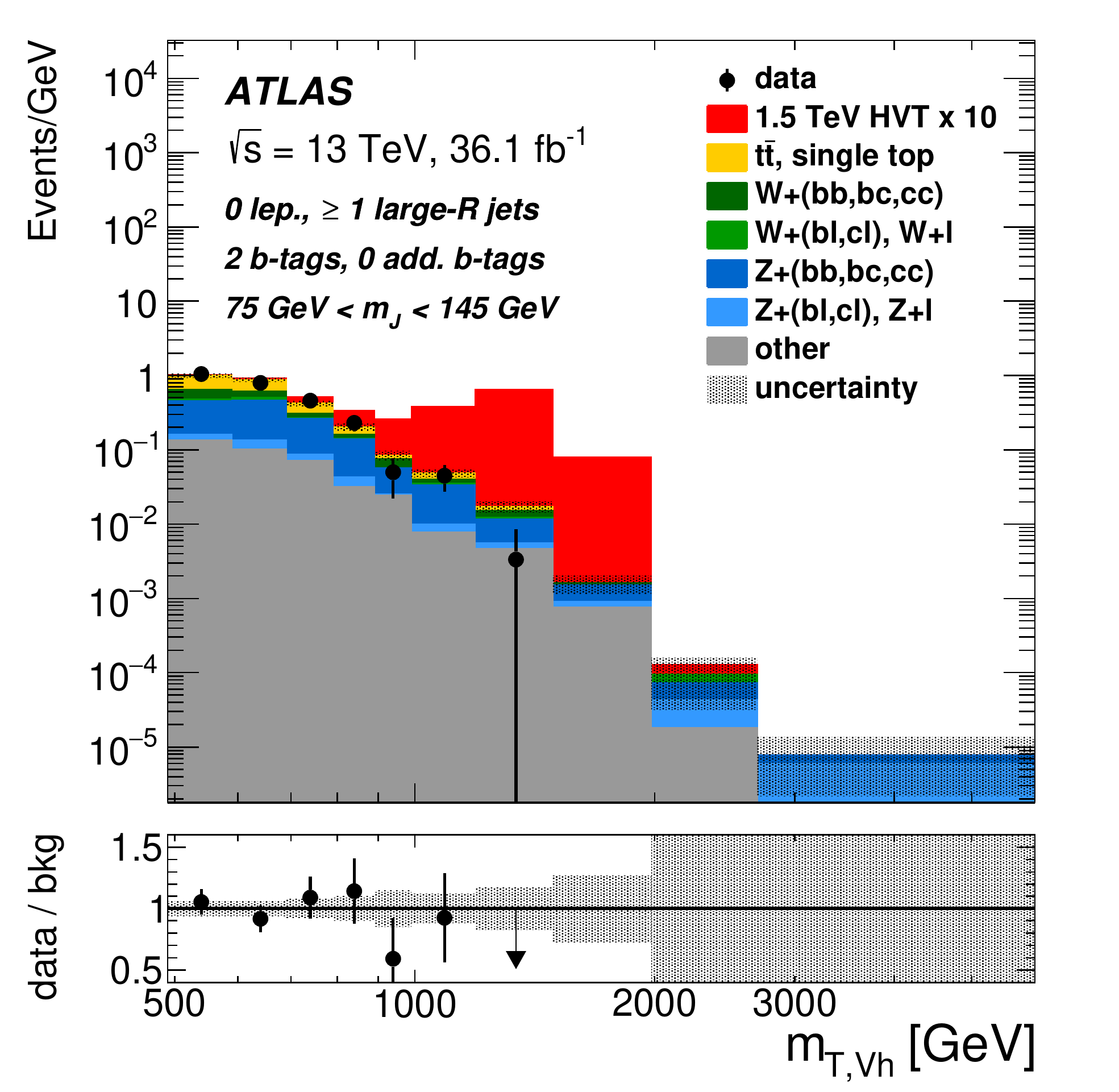}
\\
\includegraphics[width=0.43\textwidth]{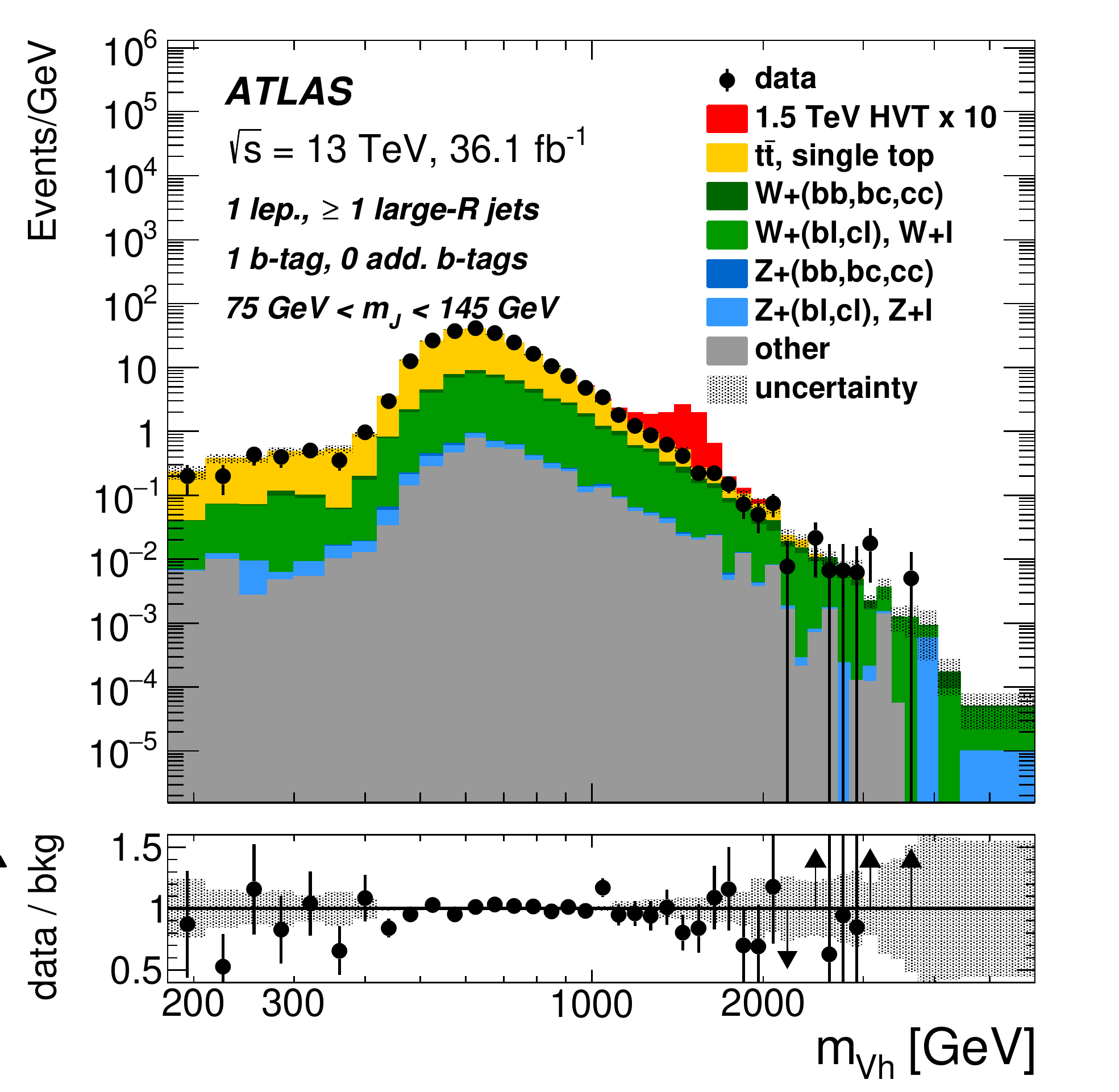}
\includegraphics[width=0.43\textwidth]{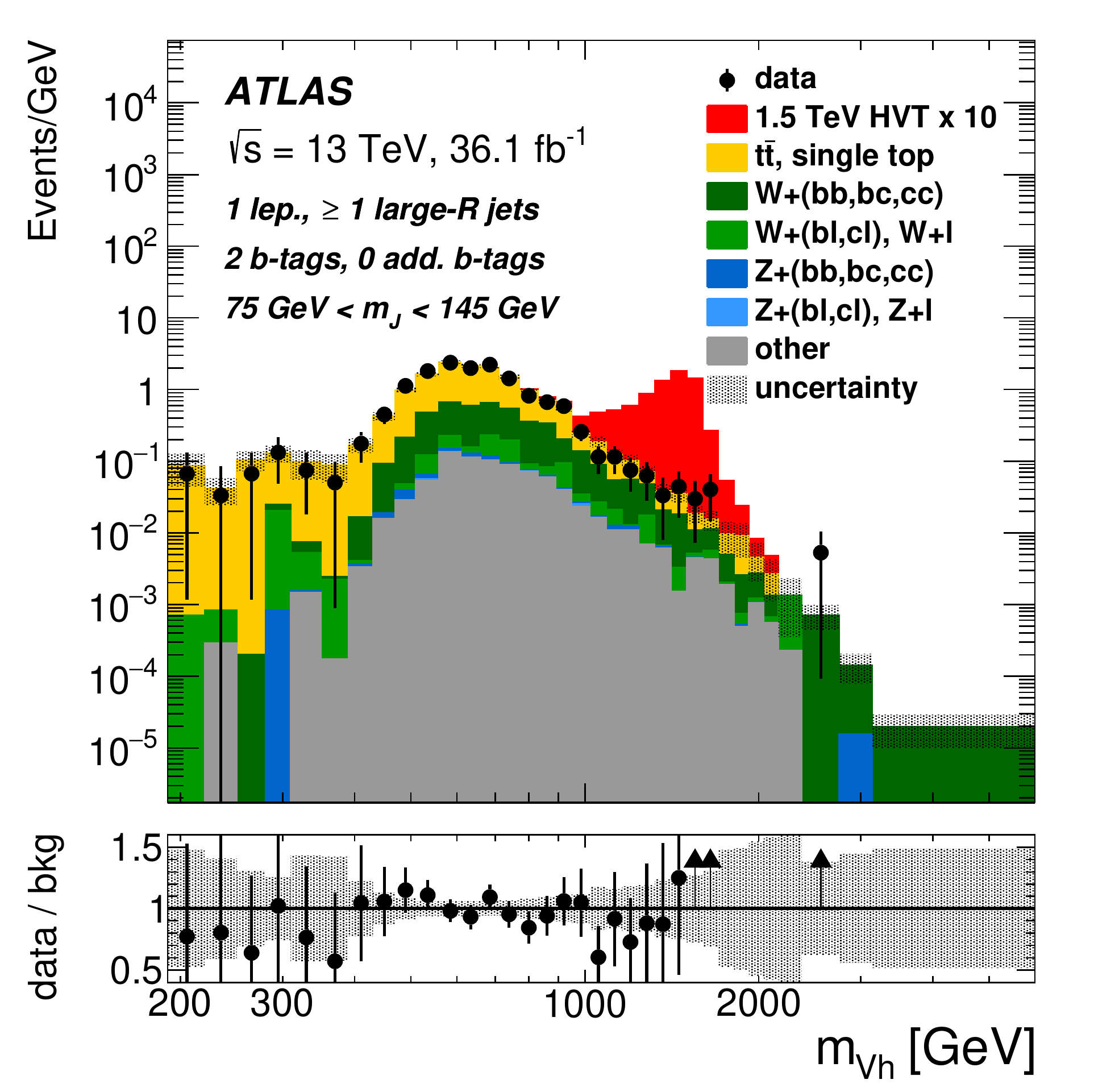}
\\
\includegraphics[width=0.43\textwidth]{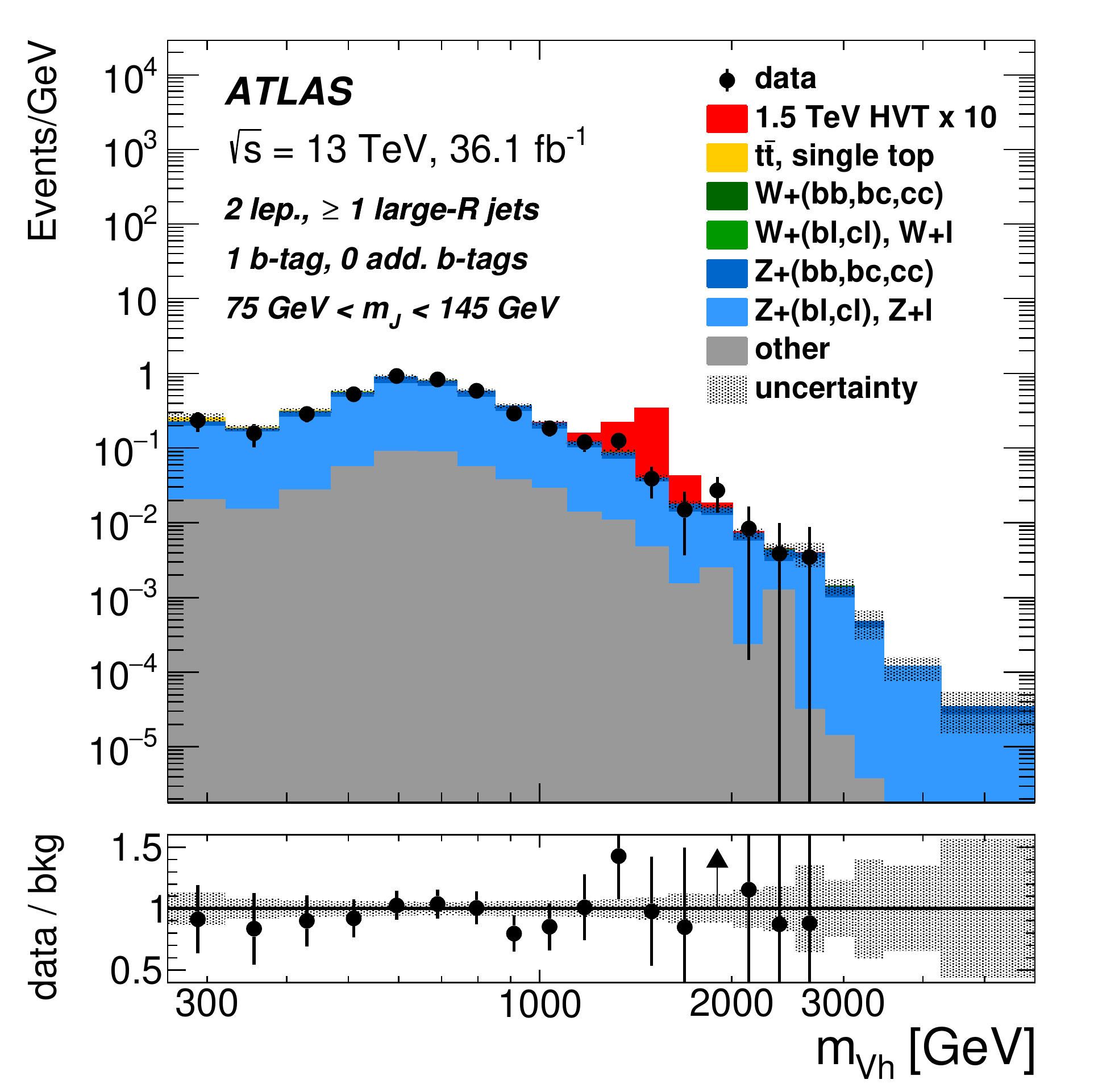}
\includegraphics[width=0.43\textwidth]{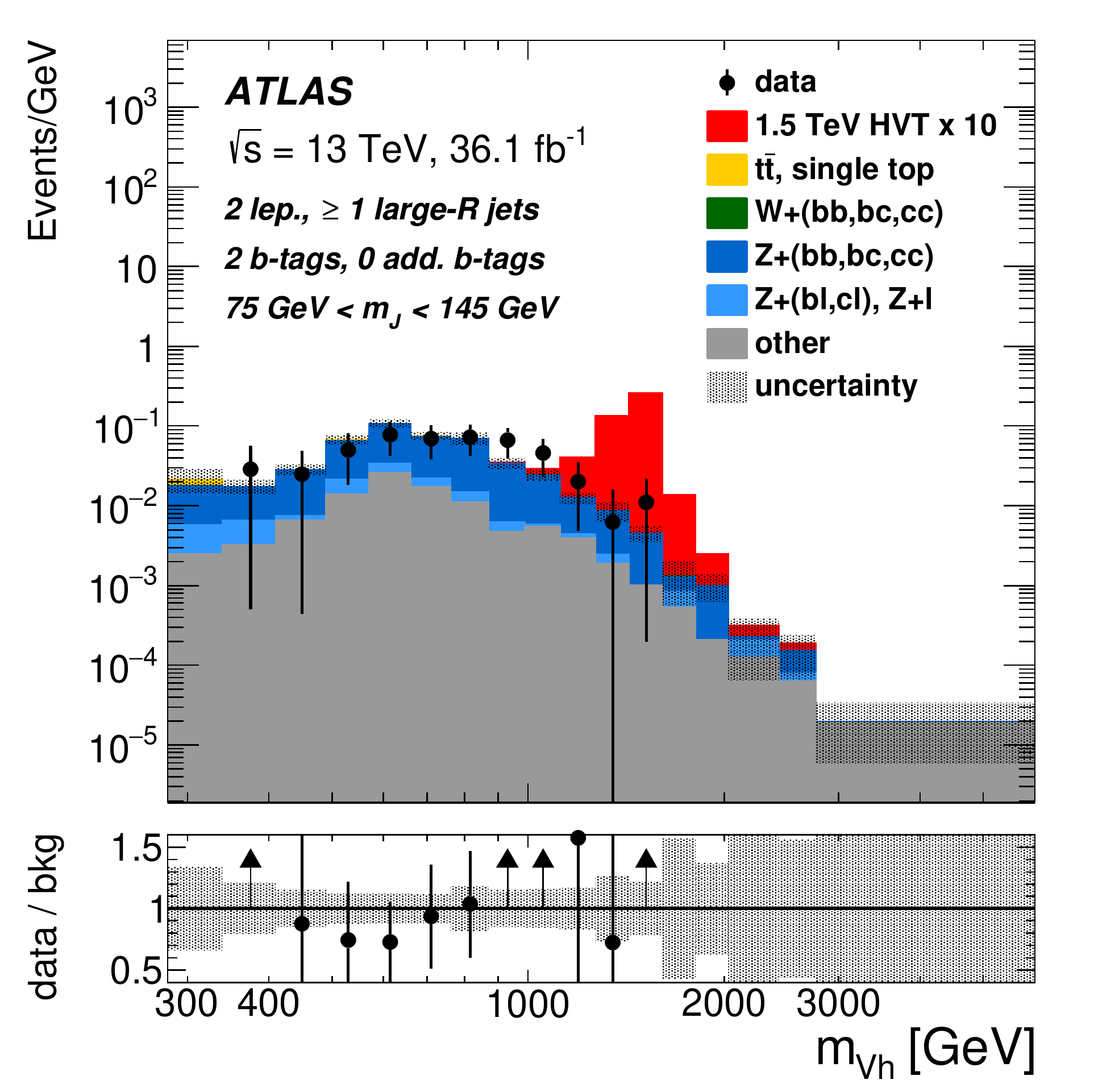}
\caption{
Event distributions of $m_{\mathrm{T}, V\h}$  for the 0-lepton channel, and  $m_{V\h}$ for the 1-lepton  and 2-lepton channels in the merged categories.
The quantity on the vertical axis is the number of data events divided by the bin width in \GeV.
The background prediction is shown after a background-only  maximum-likelihood fit to the data.
The signal for the benchmark HVT \textit{Model A} with $m_{V'} = 1.5 $~\TeV\ is normalised to 10 times the theoretical cross-section.
The background uncertainty band shown includes both the statistical and systematic uncertainties after the fit added in quadrature. The lower panels show the ratio of the observed data to the estimated SM background.
    } \label{fig:mVH_HVTpostfit_merged}
\end{figure}

\begin{figure}[htb]
\centering
\includegraphics[width=0.44\textwidth]{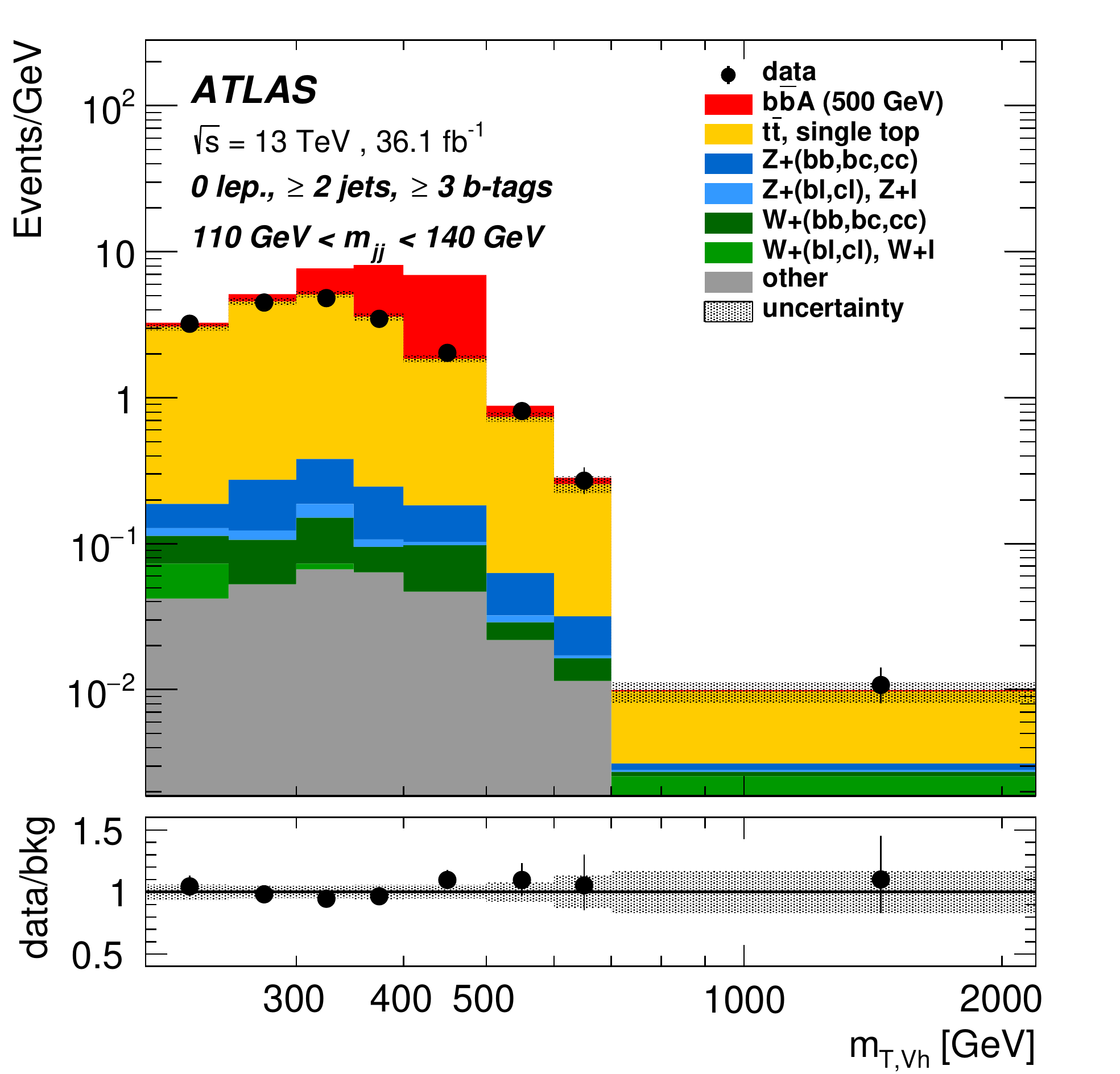}
\includegraphics[width=0.44\textwidth]{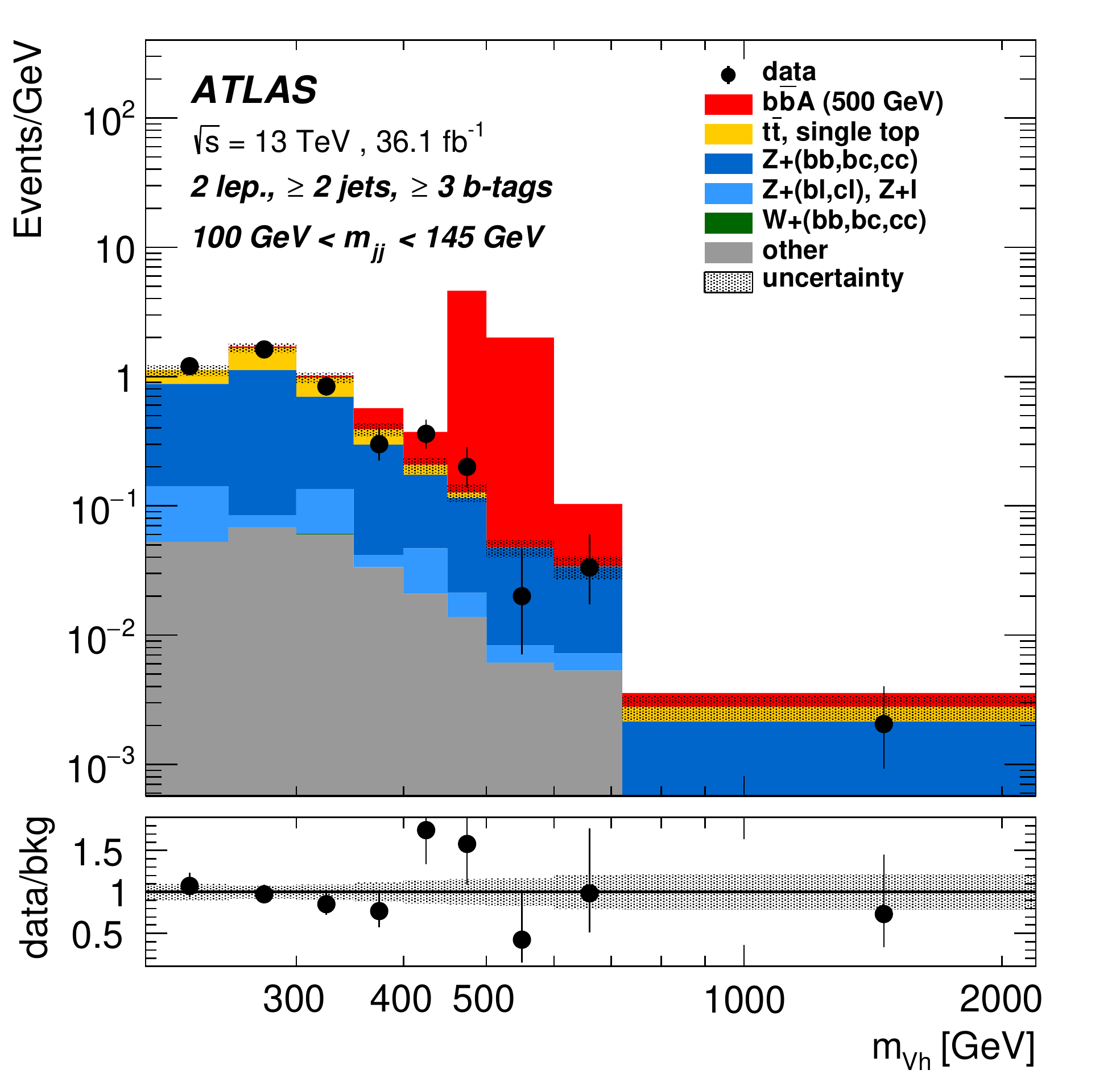}
\\
\includegraphics[width=0.44\textwidth]{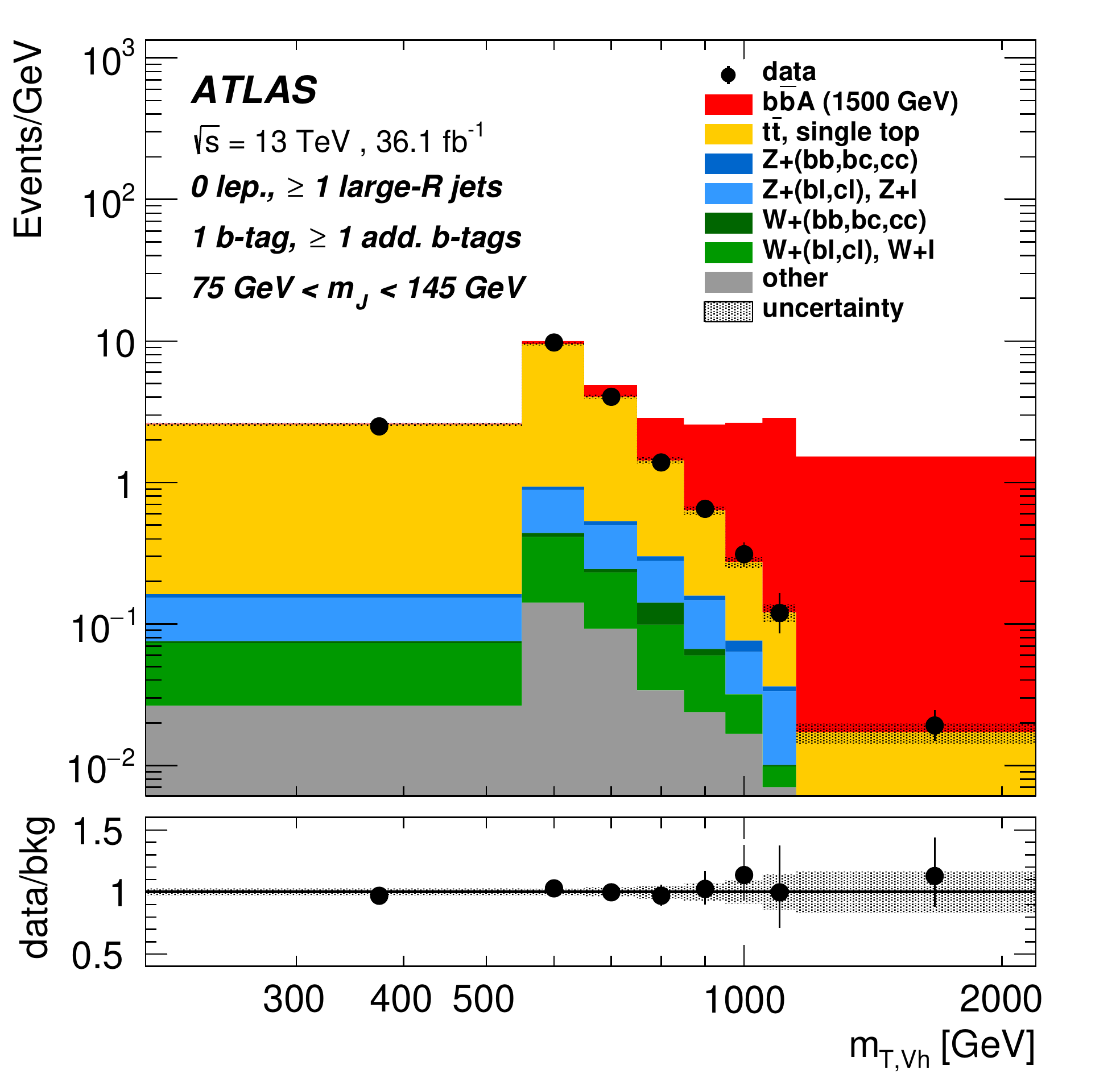}
\includegraphics[width=0.44\textwidth]{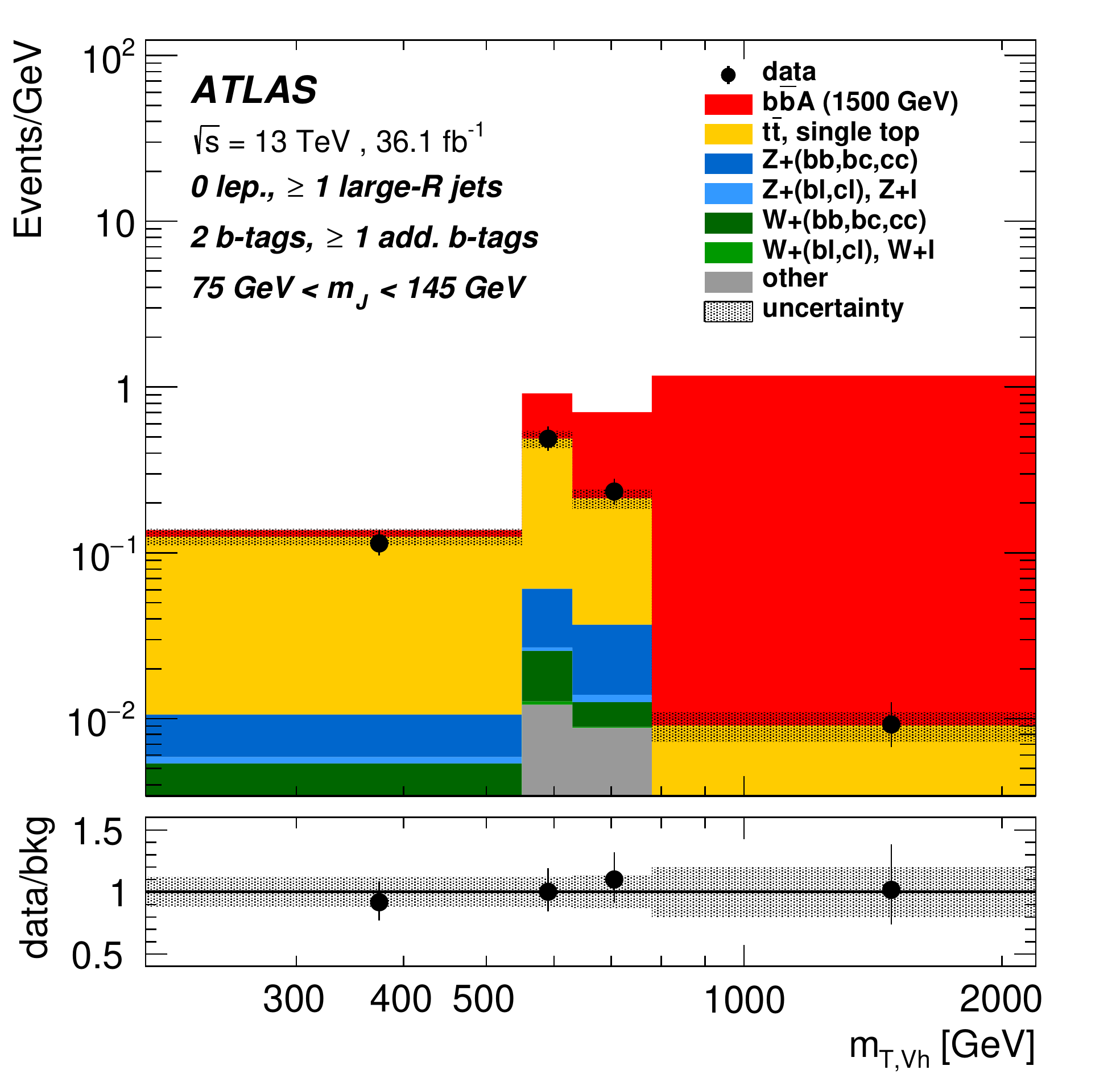}
\\
\includegraphics[width=0.44\textwidth]{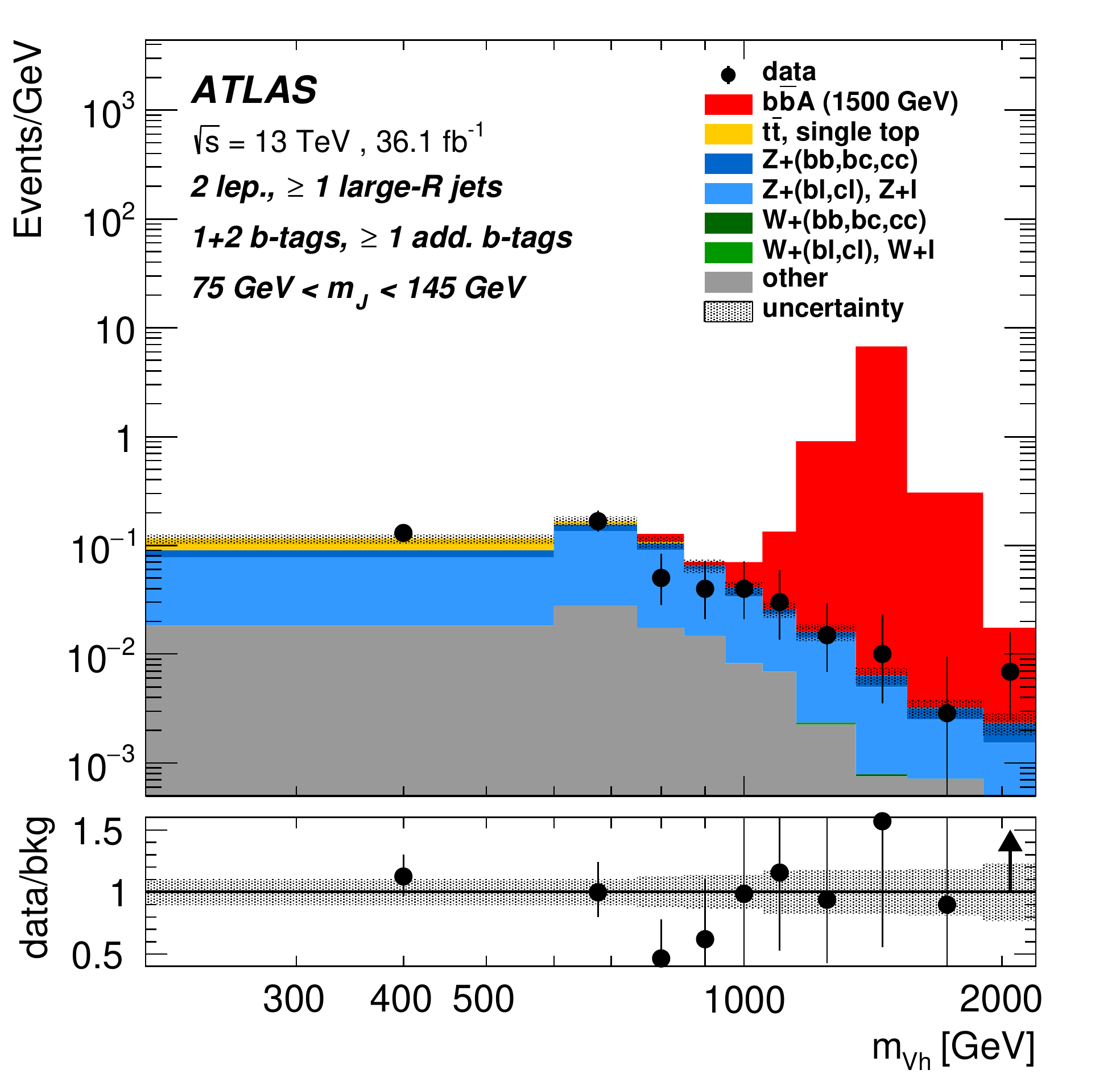}
\caption{ 
Event distributions of  $\MTVh$ for the 0-lepton channel and $\MVh$ for the 2-lepton channel after a background-only fit to the categories used in the search for the $A$ boson produced in association with $b$-quarks.  
The quantity on the vertical axis is the number of data events divided by the bin width in \GeV.
The distribution for an $A$ boson with mass of 500~\GeV\ (1.5~\TeV) is shown for illustration in the resolved (merged) regions normalised using a cross-section of 5~pb.
The background uncertainty band shown includes both the statistical and systematic uncertainties after the fit added in quadrature. The lower panels show the ratio of the observed data to the estimated SM background.
}
\label{fig:mVH_AZhpostfit_3ptagmerged}  
\end{figure}

The expected and observed event yields after the HVT fit are shown in Table~\ref{tab:postFitYields}. 
%The yields are given separately in each region, but the fit is performed simultaneously as described above.
%No significant excess beyond the expectation from the background prediction is observed in the data. 
The $\MTVh$ and $\MVh$ distributions after the HVT fit are shown in 
Figures~\ref{fig:mVH_HVTpostfit_resolved} and~\ref{fig:mVH_HVTpostfit_merged}.  
Similar distributions are obtained from the \Wprime, \Zprime\ and $A$ fits with background yields consistent within the uncertainties.
The mass distributions for the resolved 3+ $b$-tag category and the merged categories with additional $b$-tagged jets, used in the $A$ boson fits, are shown in Figure~\ref{fig:mVH_AZhpostfit_3ptagmerged}.

%%%%Limits discussion

As no significant excess over the background prediction is observed, upper limits at the 95\% CL are set on the
production cross-section times the branching fraction for each model. 
The limits are evaluated using a modified frequentist method known as CL$_\text{s}$~\cite{Read:2002hq} and the profile-likelihood-ratio
test statistic~\cite{Cowan:2010js} using the asymptotic approximation. 

\begin{figure*}[ht!]
  \centering
  \subfloat[]{
    \includegraphics[width=0.45\textwidth]{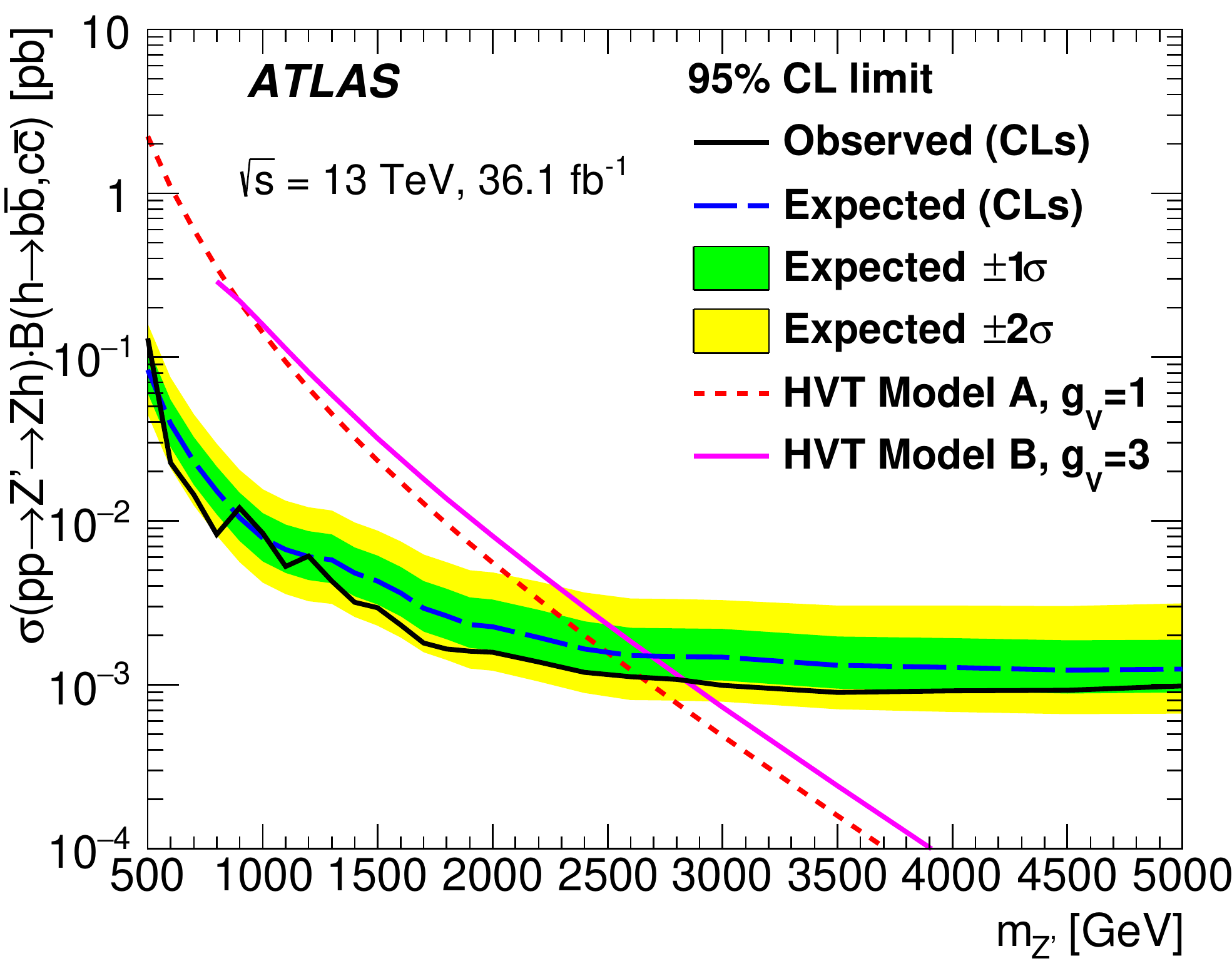}
  }
  \subfloat[]{
    \includegraphics[width=0.45\textwidth]{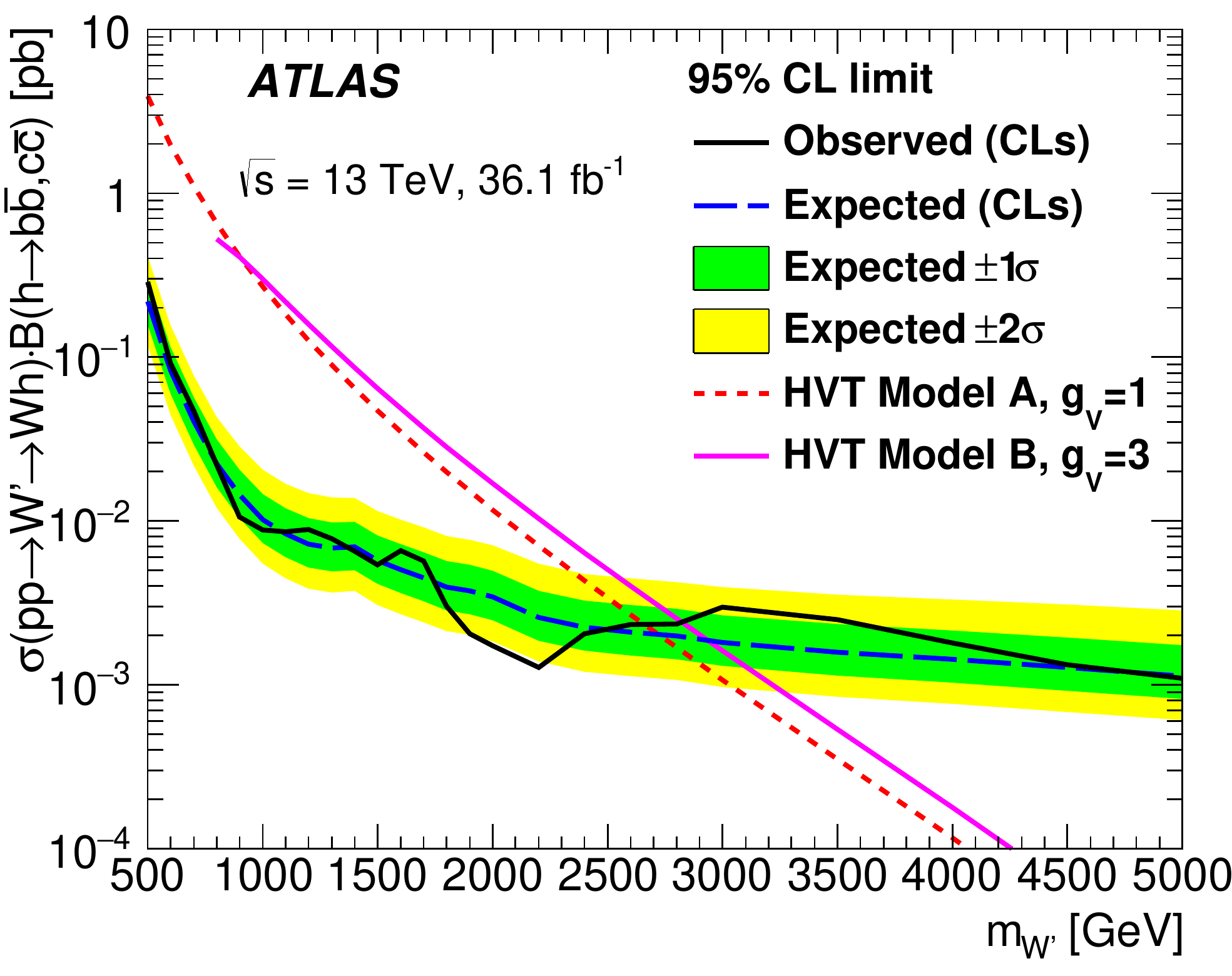}
  }\\
  \subfloat[] {
    \includegraphics[width=0.45\textwidth]{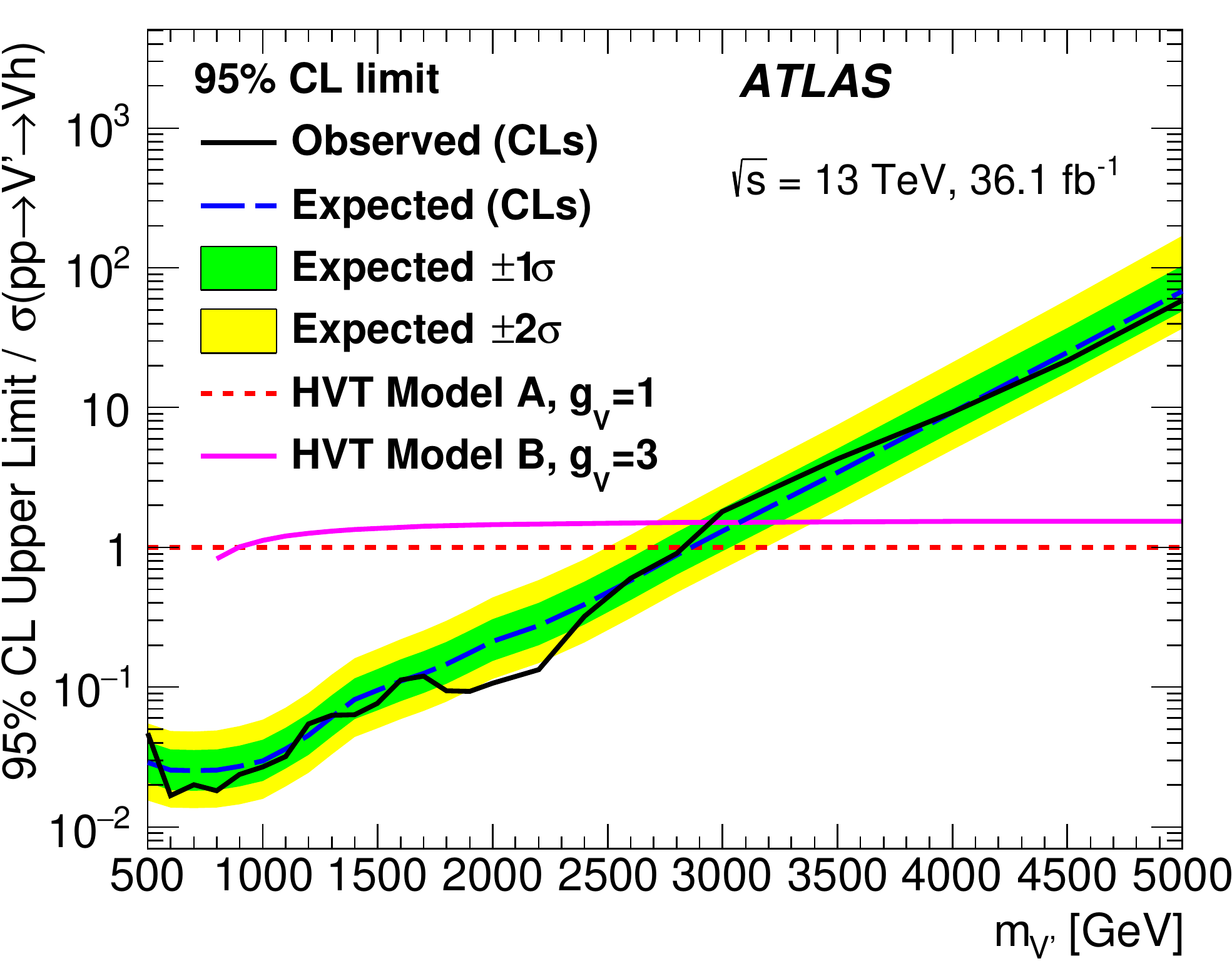}
  }
  \subfloat[] { 
    \includegraphics[width=0.45\textwidth]{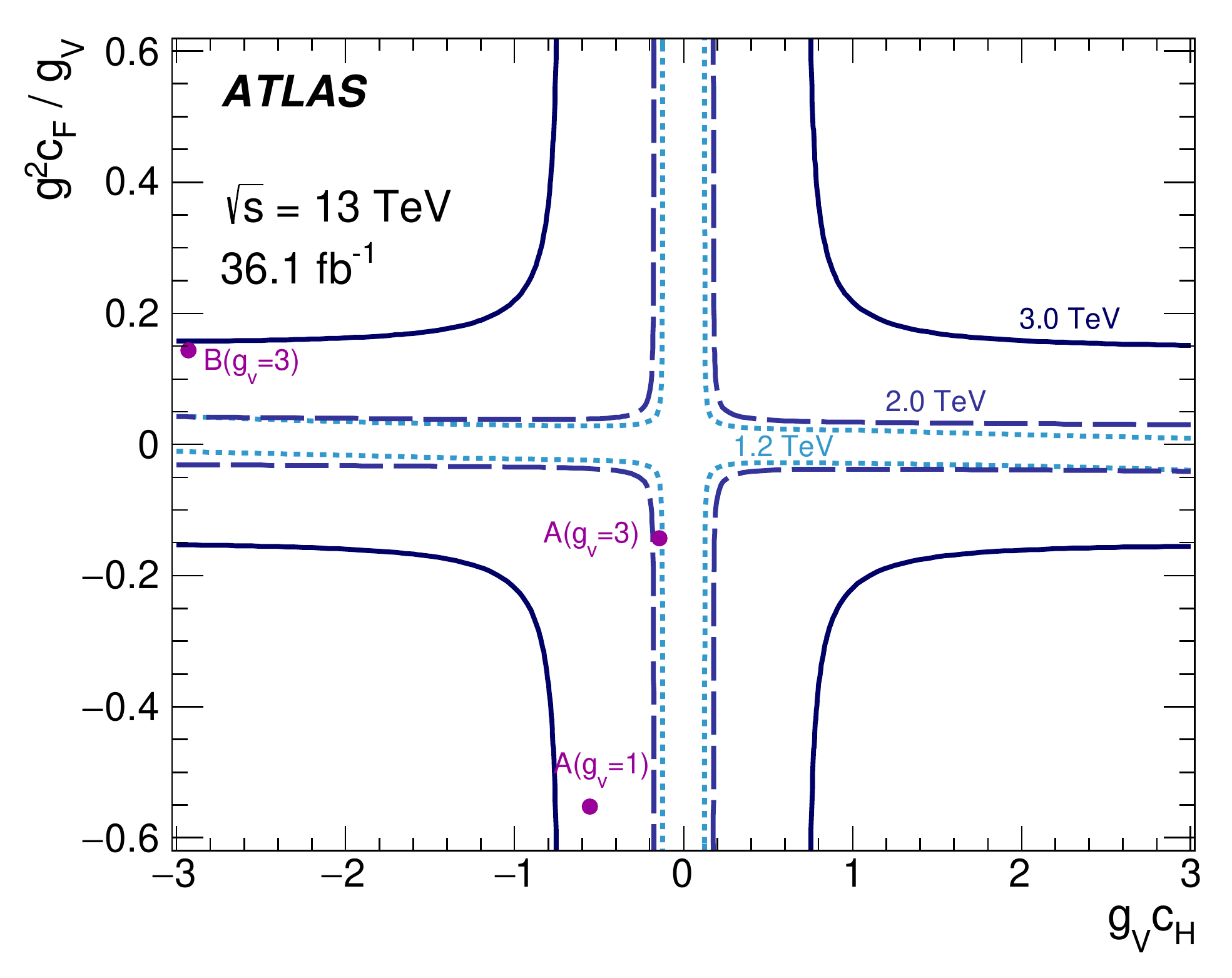}    
  }
  \caption{
    Upper limits as a function of the resonance mass at the 95\% CL for (a) the production
    cross-section of $Z^{\prime}$ times its branching fraction to
    $Z\h$ and the branching fraction $B(\h\rightarrow
    b\bar{b},c\bar{c})$ and (b) the production cross-section of
    $W^{\prime}$ times its branching fraction to $W\h$ and the
    branching fraction $B(\h \rightarrow b\bar{b},c\bar{c})$.    
    (c) Upper limits at the 95\% CL for the scaling factor of the
    production cross-section for $V'$ times its branching fraction
    to $W\h/Z\h$ in \textit{Model A}. 
    The production cross-sections predicted by \textit{Model A}
    and \textit{Model B} are shown for comparison in (a)--(c).
    (d) Observed 95\% CL exclusion contours in the HVT parameter
    space \{$g_\text{V}c_\text{H}$, ($g^2/g_\text{V}$)$c_\text{F}$\} for resonances of mass
    1.2~\TeV, 2.0~\TeV\ and 3.0~\TeV. The areas outside the curves are
    excluded.  
    Also shown are the benchmark model parameters A($g_\text{V}=1$), A($g_\text{V}=3$) and B($g_\text{V}=3$). 
  } \label{fig:hvtlimit}
\end{figure*}

The 95\% CL upper limits on the production cross-section multiplied by the
branching fraction for $W^\prime\rightarrow W\h$ and $Z^\prime\rightarrow Z\h$ and the sum of branching fractions
$B(\h\rightarrow\bbbar) + B(\h\rightarrow\ccbar)$, which is fixed to 60.6\%~\cite{Heinemeyer:2013tqa},
are shown in Figure~\ref{fig:hvtlimit}(a) and Figure~\ref{fig:hvtlimit}(b) as a function of the resonance mass.
The existence of $W^{\prime}$ and $Z^{\prime}$ bosons with masses $m_{W^{\prime}} < \WpExclusionModelA$~\TeV\ and 
$m_{Z^{\prime}} < \ZpExclusionModelA$~\TeV, respectively, are excluded for the HVT benchmark \textit{Model A} with coupling constant
$g_\text{V}=1$~\cite{Pappadopulo:2014qza}. For \textit{Model B} with coupling constant $g_\text{V}=3$~\cite{Pappadopulo:2014qza}, the corresponding excluded masses are 
$m_{W^{\prime}} < \WpExclusionModelB$~\TeV\ and $m_{Z^{\prime}} < \ZpExclusionModelB$~\TeV.

To study the scenario in which the masses of charged and neutral
resonances are degenerate, a likelihood fit over all the
signal regions and control regions is performed.
The 95\% CL upper limit on the combined signal strength for the processes 
$W' \rightarrow W\h$ and $Z' \rightarrow Z\h$, assuming $m_{W'}=m_{Z'}$, 
relative to the HVT model predictions, is shown in Figure~\ref{fig:hvtlimit}(c). 
For \textit{Model A} (\textit{Model B}), $m_{V'} < \HVTExclusionModelA$~\TeV\ (\HVTExclusionModelB~\TeV) is excluded.

The exclusion contours in the HVT parameter space \{$g_\text{V}c_\text{H}$,
($g^2/g_\text{V}$)$c_\text{F}$\} for resonances of mass 1.2~\TeV, 2.0~\TeV\ and
3.0~\TeV\ are shown in Figure~\ref{fig:hvtlimit}(d) where all three
lepton channels are combined, taking into account the branching fractions to
$W\h$ and $Z\h$ from the HVT model prediction.
Here, the parameter $c_\text{F}$ is assumed to be the same for quarks and
leptons, including third-generation fermions, and other parameters
involving more than one heavy vector boson, $g_\text{V}c_\text{VVV}$,  $g_\text{V}^2c_\text{VVhh}$ 
and $c_\text{VVW}$, are assumed to have negligible contributions to the overall
cross-sections for the processes of interest.

\begin{figure*}[ht!]
\centering
\subfloat[][Pure gluon--gluon fusion production]{
\includegraphics[width=0.45\textwidth]{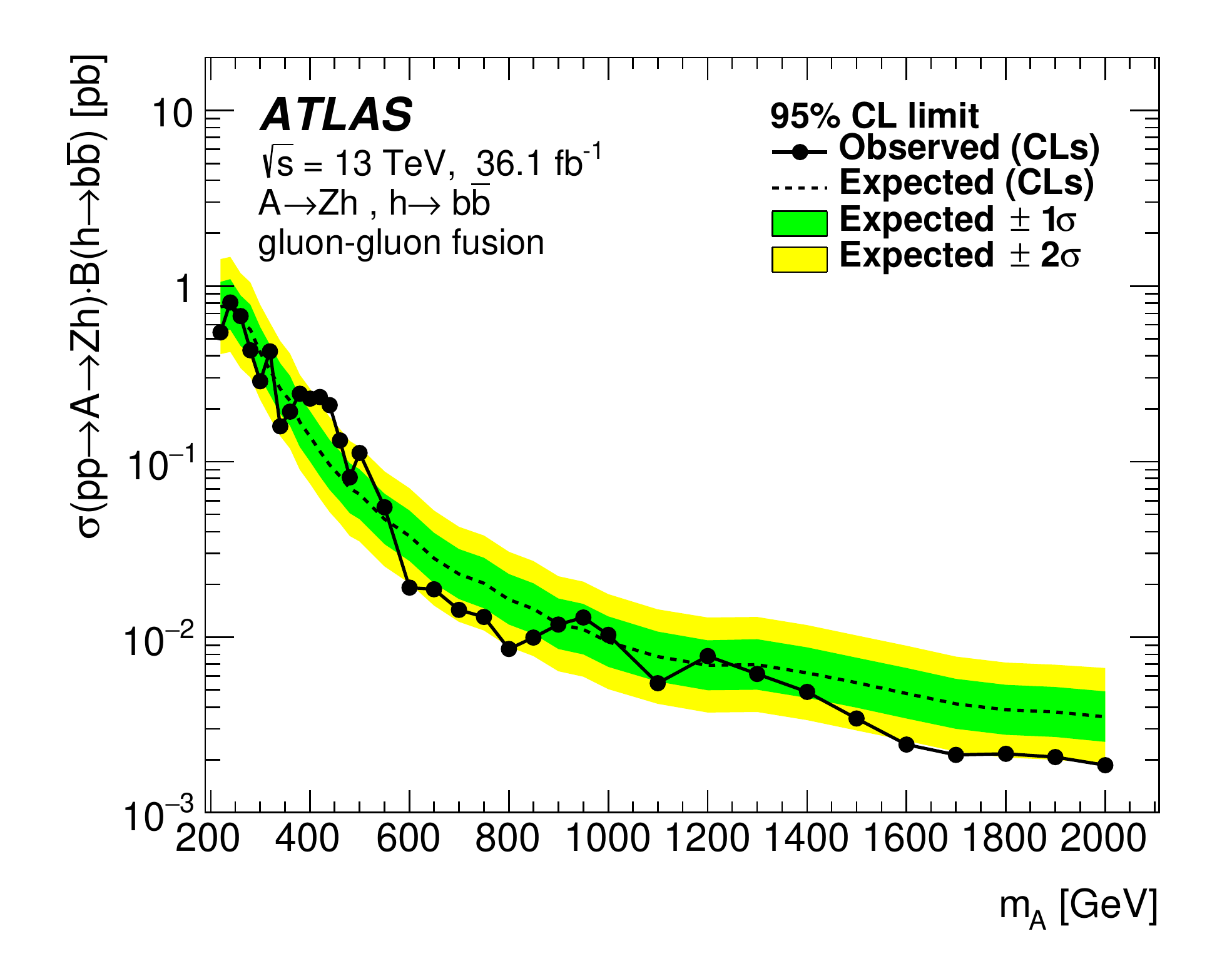}
}
\subfloat[][Pure $b$-quark associated production]{
\includegraphics[width=0.45\textwidth]{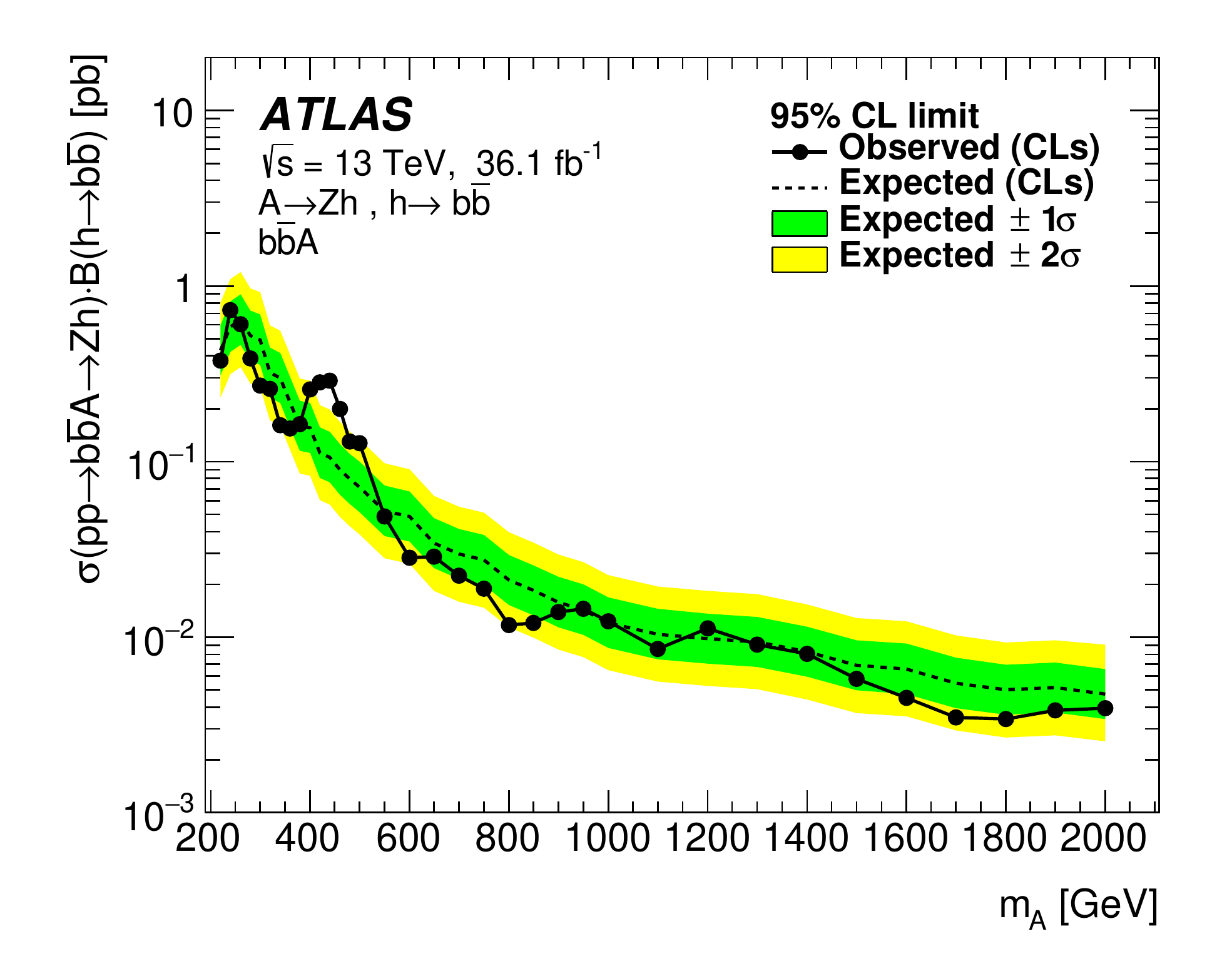}
}
\caption{
Upper limits at the 95\% CL on the product of the production cross-section 
        for $\pp \to  \A$  and the branching fractions for $\AZh$ and $\hbb$ evaluated by combining the 0-lepton and 2-lepton channels. 
        The possible signal components of the data are interpreted assuming (a) pure gluon--gluon fusion production, 
        and (b) pure $b$-quark associated production.
}
\label{fig:azhlimit}
\end{figure*}

Figures~\ref{fig:azhlimit}(a) and~\ref{fig:azhlimit}(b) show the 95\% CL upper limits on the production cross-section of 
the $A$ boson times its branching fraction to $Z\h$ and the branching fraction of $\h\rightarrow\bbbar$
as a function of the resonance mass. Upper limits are placed separately for a signal arising from pure gluon--gluon fusion 
production (Figure~\ref{fig:azhlimit}(a)) and from pure $b$-quark associated production (Figure~\ref{fig:azhlimit}(b)). In the search for the $A$ boson with $b$-quark associated production, 
a mild excess of events is observed around 440~\GeV, mainly driven by the dimuon channel in the resolved 
category with 3+ $b$-tags. The local significance of this excess with respect to the background-only hypothesis 
is estimated to be $3.6\ \sigma$, and the global significance, accounting for the look-elsewhere 
effect~\cite{Gross2010} is estimated to be $2.4\ \sigma$.

The data are also interpreted in terms of limits at 95\%\ CL on the 2HDM parameters $\tan(\beta)$ and $\cos(\beta-\alpha)$.
The admixture of gluon--gluon fusion and $b$-quark associated production, and the variation of the \A\ and \h\ boson widths and branching fractions are taken into account according to the predictions of the different models. 
In this interpretation, the $\MTVh$ and $\MVh$ distributions of the simulated signal events are smeared according to a Breit--Wigner function with a width predicted by the parameters of the model. This procedure has been verified to produce the same line-shape as the one including non-resonant and interference effects for widths $\Gamma_A/m_A<10\%$.

\begin{figure*}[ht!]
\centering
\subfloat[][2HDM Type-I]{\includegraphics[height=0.4\textheight]{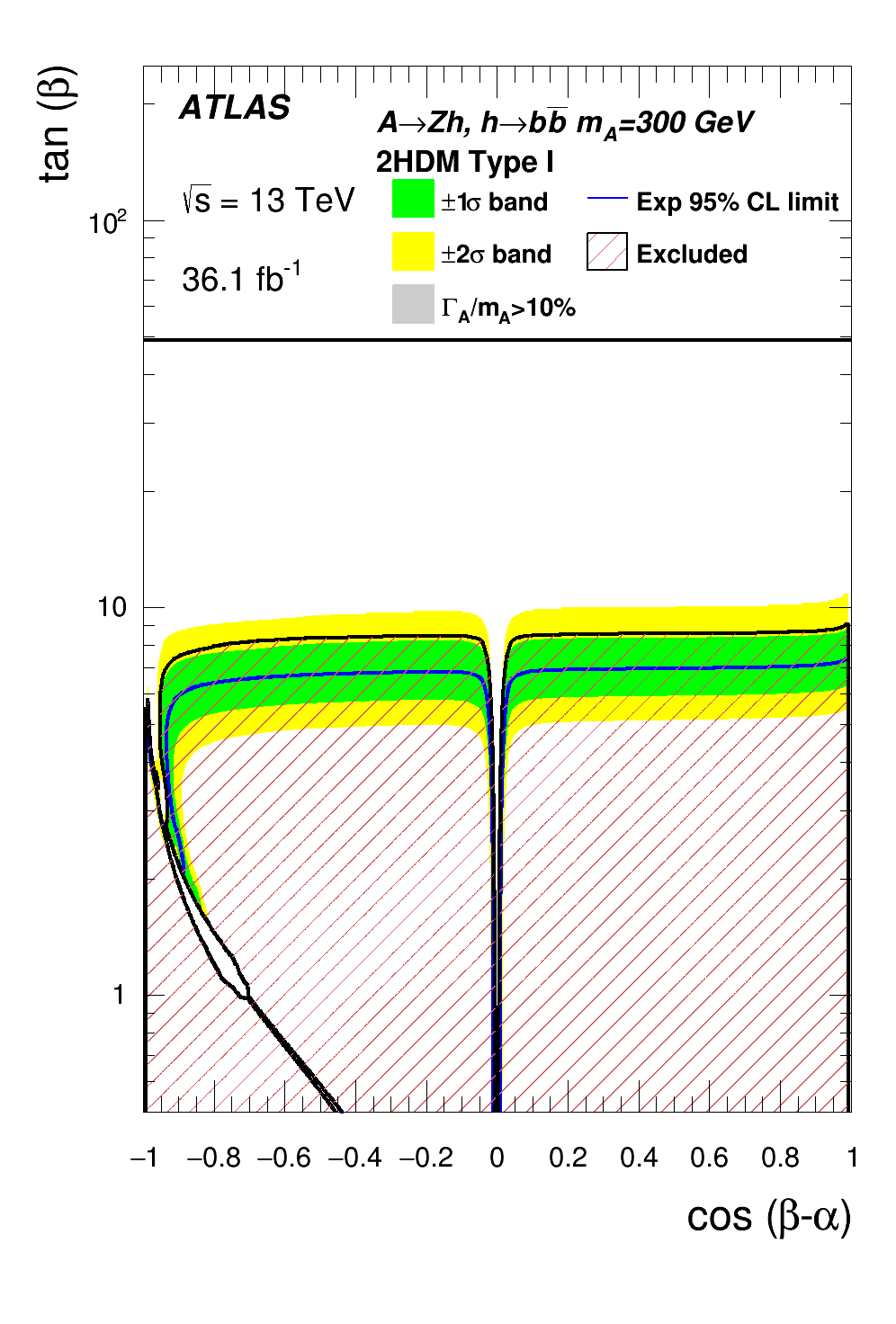}}
\subfloat[][2HDM Type-II]{\includegraphics[height=0.4\textheight]{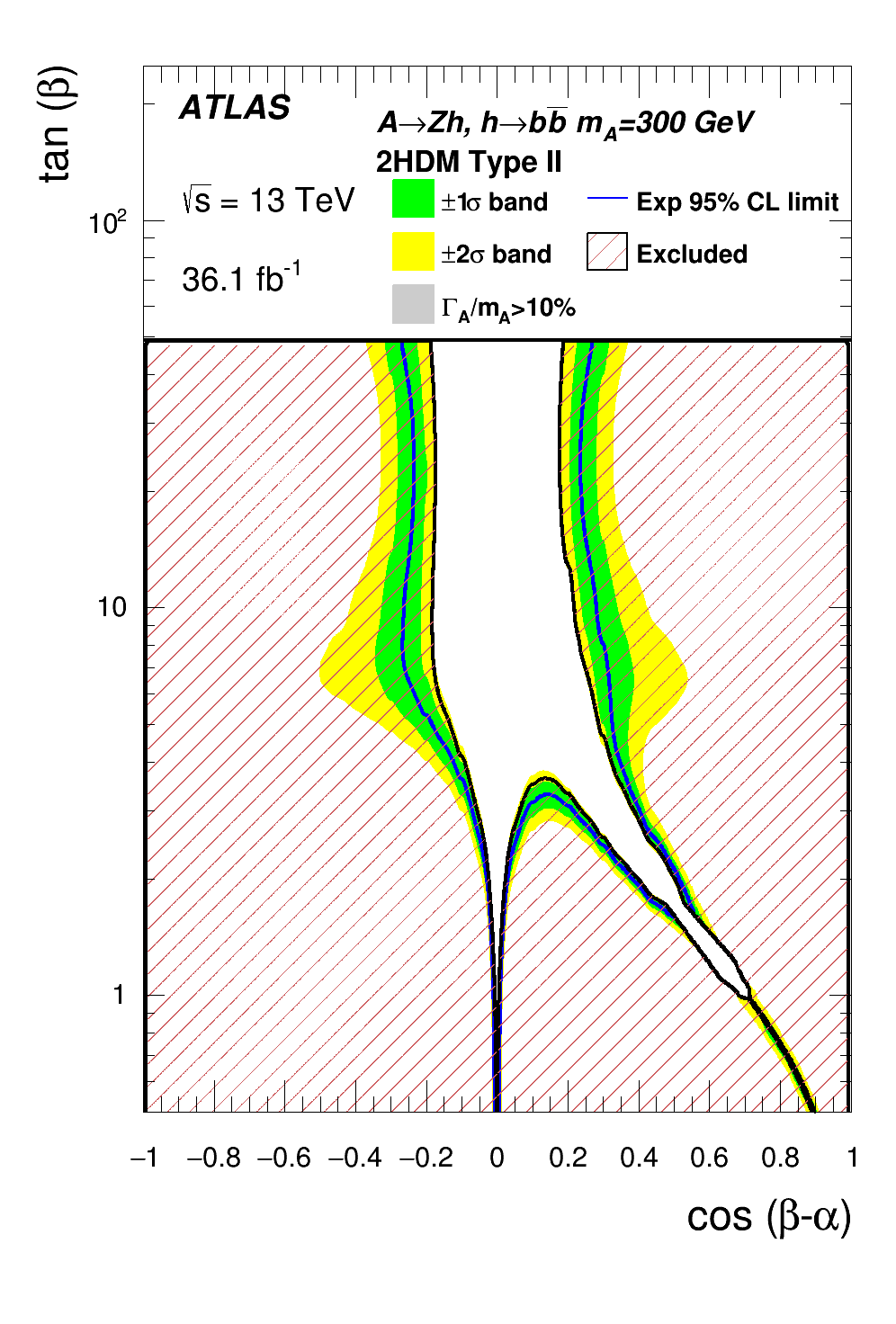}}\\
\subfloat[][2HDM Lepton-specific]{\includegraphics[height=0.4\textheight]{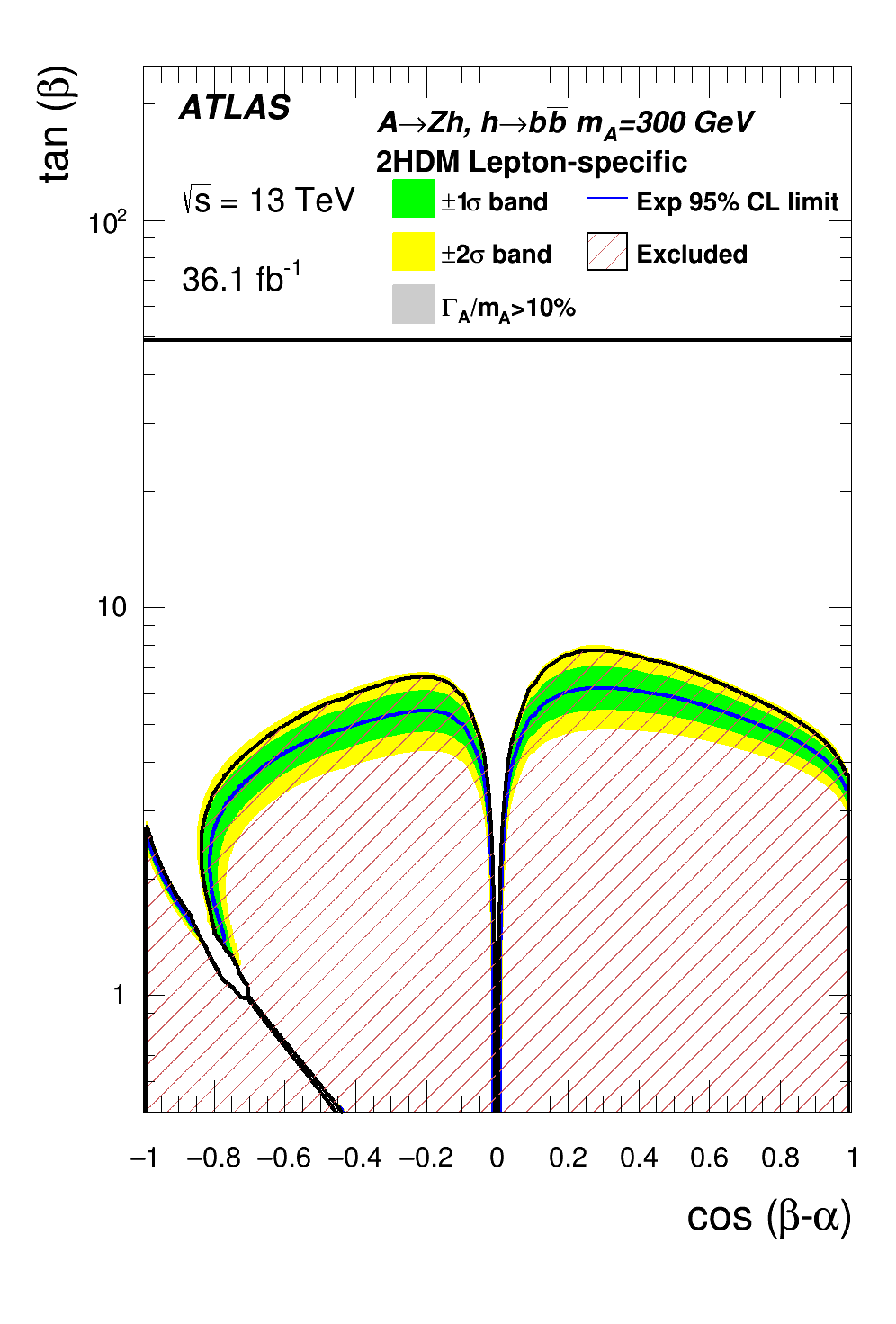}}
\subfloat[][2HDM Flipped]{\includegraphics[height=0.4\textheight]{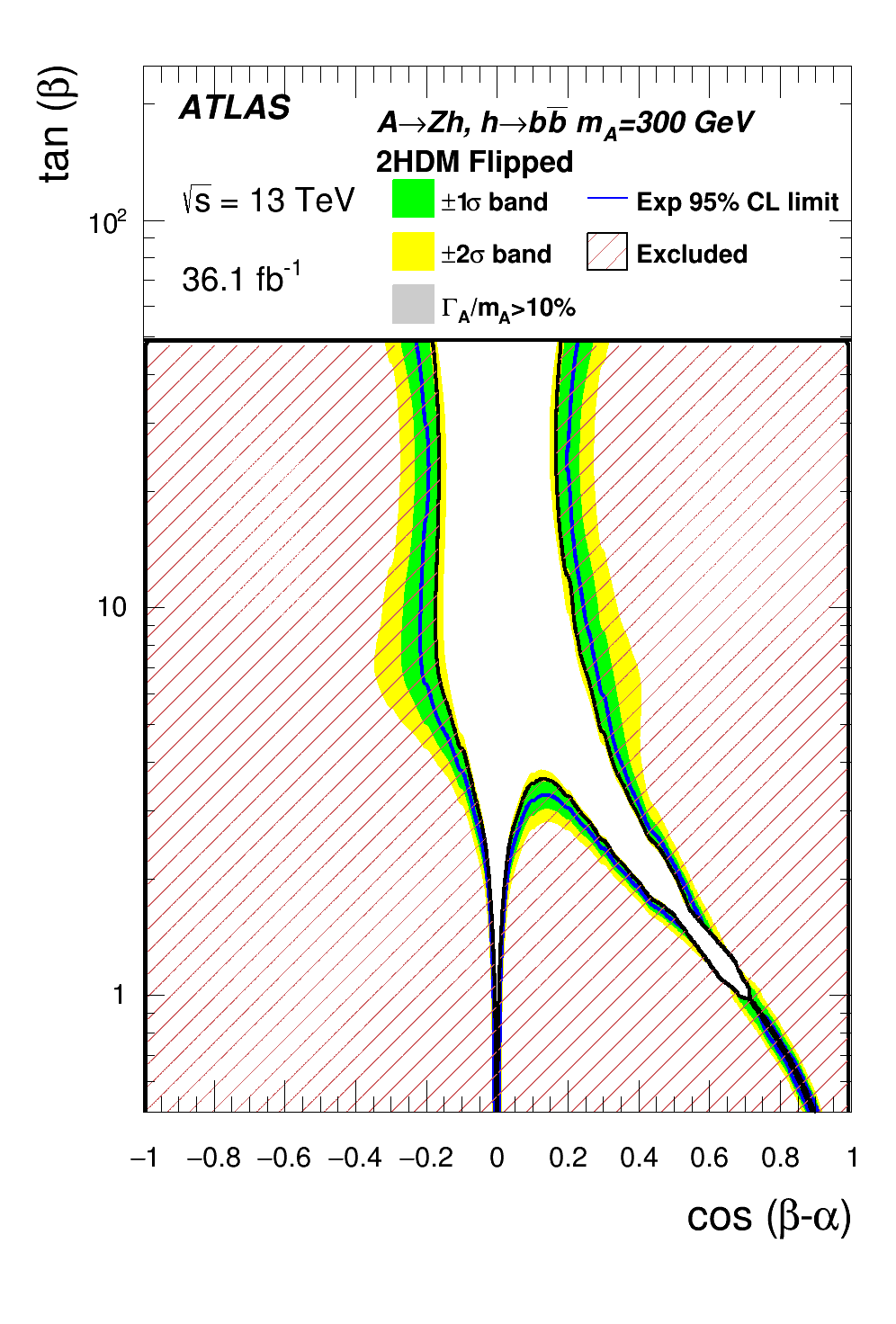}}
\caption{
The interpretation of the cross-section limits in the context of the various 2HDM types as a function of the parameters $\tan(\beta)$ and $\cos(\beta - \alpha)$ for $m_A$ = 300~\GeV: (a) Type-I , (b) Type-II, (c) Lepton-specific, and (d) Flipped. 
Variations of the natural width up to $\Gamma_A/m_A$=10\% have been taken into account. 
For the interpretation in Type-II and Flipped 2HDM, the $b$-quark associated production is included in addition to the gluon--gluon fusion production. 
}
\label{fig:2hdmlimitmA300}
\end{figure*}

Figure~\ref{fig:2hdmlimitmA300} shows the excluded parameter space for a resonance mass of $m_A=300$~\GeV\ in four 2HDM types:
I, II, Lepton-Specific, and Flipped. Greater sensitivity is observed at high $\tan(\beta)$ for the Type-II and Flipped models, due to an increased cross-section for $b$-quark associated production.
The narrow regions with no exclusion power in Type-I and Type-II at low $\tan(\beta)$ that are far from $\cos(\beta - \alpha)$=0 are caused by the vanishing branching fraction of \hbb.

\begin{figure*}[ht!]
\centering
\subfloat[][2HDM Type-I]{\includegraphics[height=0.4\textheight]{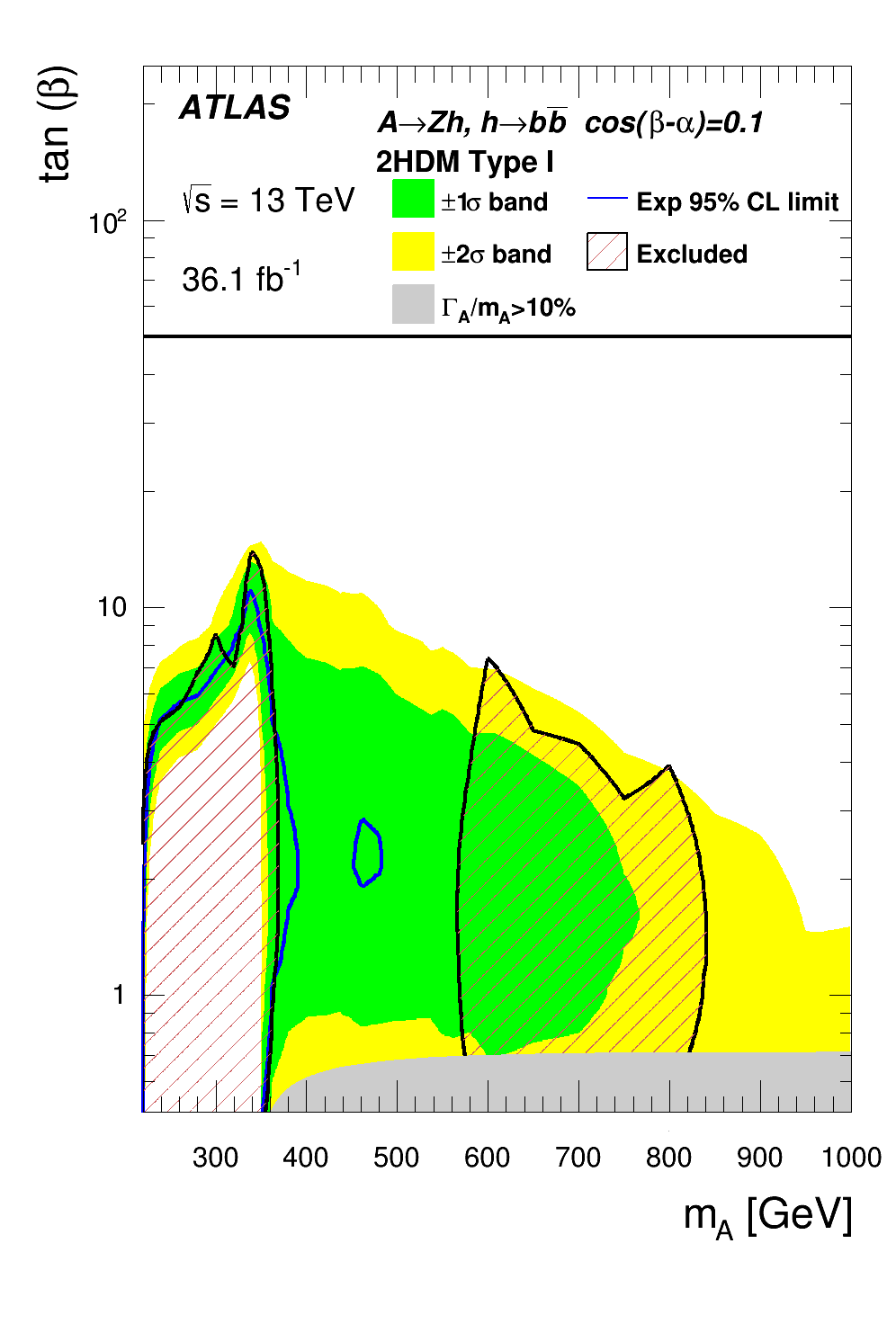}}
\subfloat[][2HDM Type-II]{\includegraphics[height=0.4\textheight]{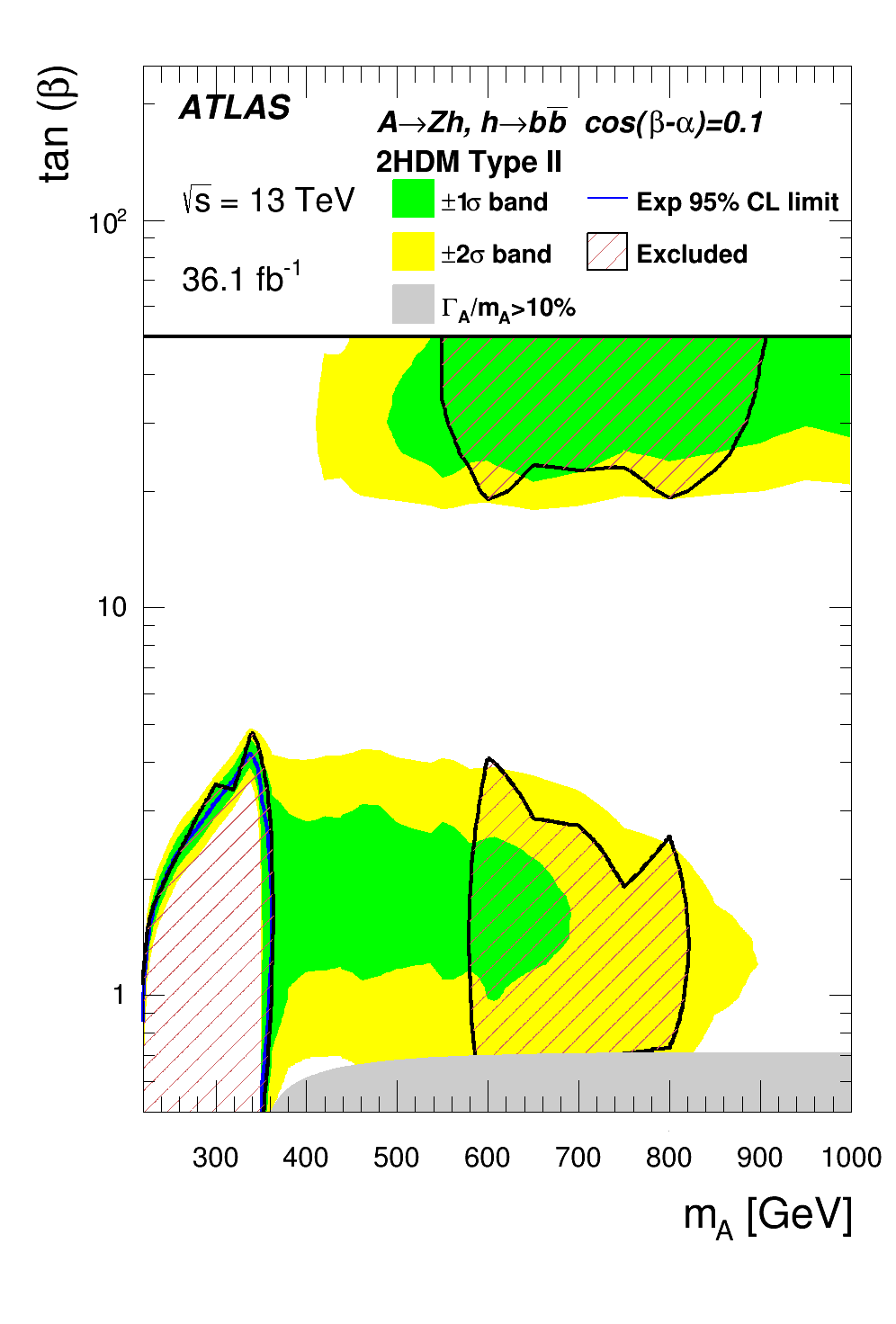}}\\
\subfloat[][2HDM Lepton-specific]{\includegraphics[height=0.4\textheight]{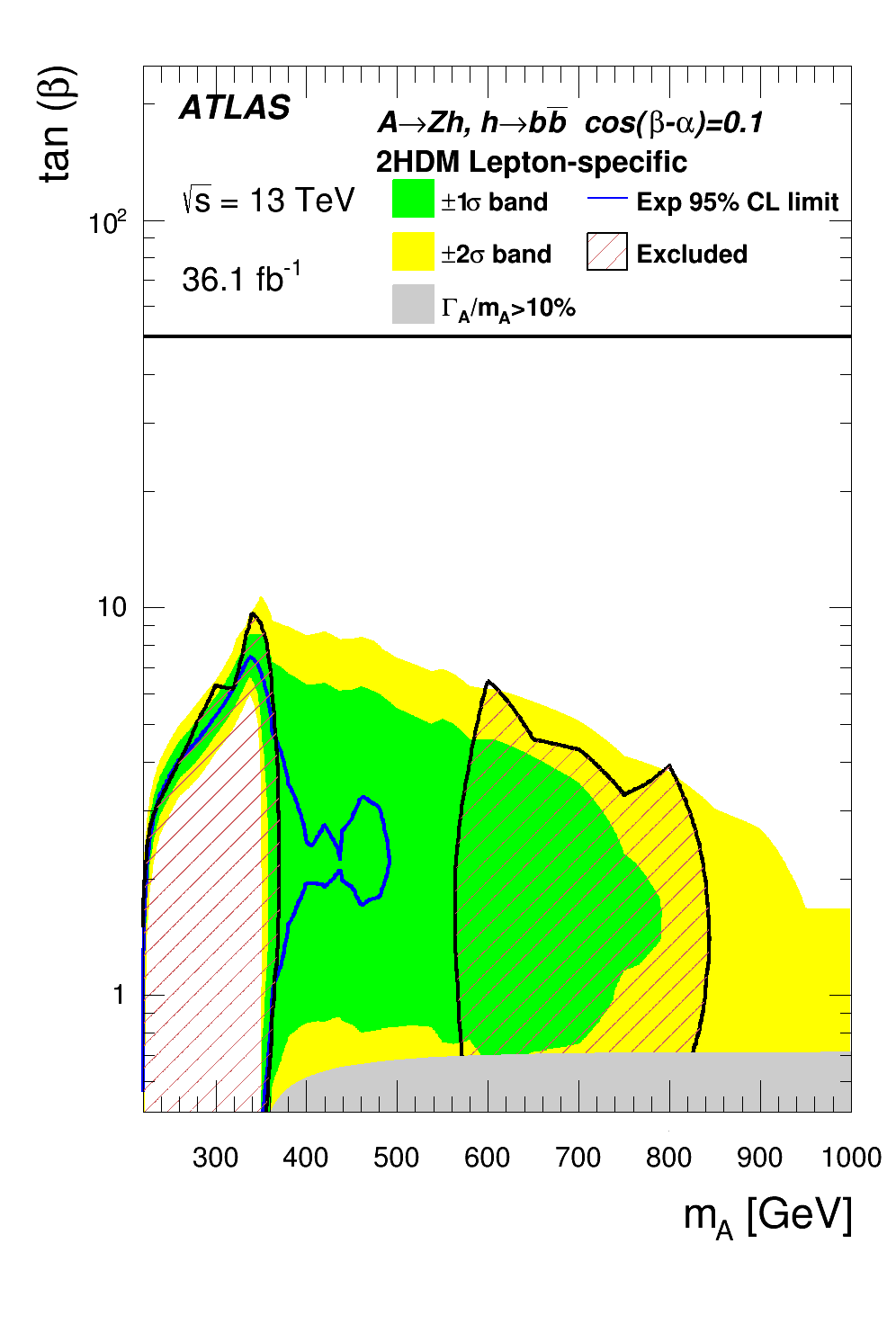}}
\subfloat[][2HDM Flipped]{\includegraphics[height=0.4\textheight]{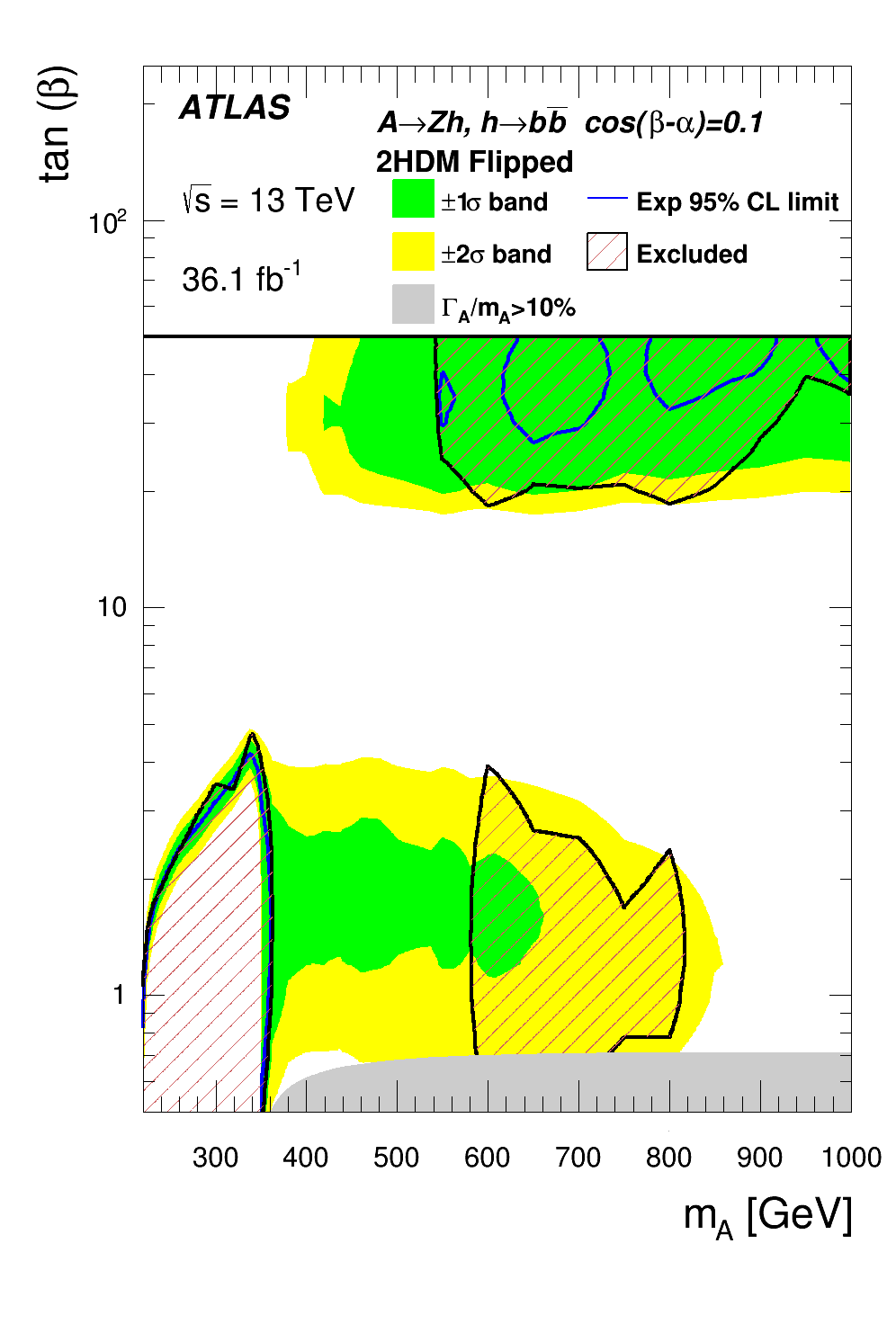}}
\caption{
The interpretation of the cross-section limits in the context of 2HDMs of Type  (a) I , (b) II, (c) Lepton-specific, and (d) Flipped, as a function of the parameters $\tan(\beta)$ and $m_A$ for $\cos(\beta - \alpha)$= 0.1. 
Variations of the natural width up to $\Gamma_A/m_A$=10\% have been taken into account.
}
\label{fig:2hdmlimitCos01}
\end{figure*}

Figure~\ref{fig:2hdmlimitCos01} shows the parameter exclusion for the four models in the $\tan(\beta)$--$m_A$ plane for $\cos(\beta-\alpha)=0.1$.  
For the interpretation in Type-II and Flipped 2HDMs, the $b$-quark associated production is included in addition to the gluon--gluon fusion production.
The shape of the expected exclusions is determined by the interplay of the expected cross-section limit, which decreases as a function of $m_A$, and the signal production cross-section times the $A\rightarrow Zh$ branching fraction at a given $m_A$ and $\tan(\beta)$. This branching fraction decreases significantly at $m_A=350$ \GeV\ due to the opening of the $A\rightarrow t\bar{t}$ channel, but increases again at higher $m_A$, maintaining similar sensitivity into this $m_A$ region. The variable $\tan(\beta)$ controls the admixture of the gluon--gluon fusion and $b$-quark associated production thereby affecting the rate at which the signal cross-section falls as a function of $m_A$, which leads to a varying sensitivity as a function of  $\tan(\beta)$. The excesses or deficits in the data visible in Figure~\ref{fig:azhlimit} are also reflected in Figure~\ref{fig:2hdmlimitCos01}.

{
\def\arraystretch{1.1}
\begin{table}[t]
\caption{\label{tab:postFitYields} 
The predicted and observed event yields in the signal regions defined in the text. The yields in the 1 and 2 $b$-tag regions 
correspond to the HVT fit for a signal of mass 1.5~\TeV. In the 3+ $b$-tag and 1 and 2 $b$-tag with additional $b$-tags 
regions, the yields are from the fit using the $A$ boson produced in association with $b$-quarks as signal with a mass of 1.5~\TeV.
The quoted uncertainties are the statistical and systematic uncertainties combined in quadrature after the fit.
The uncertainties in the individual background predictions are larger than the total background uncertainty due to correlations in the normalisation parameters in the fit. 
}
\begin{center}
\begin{footnotesize}

\begin{tabular}{l|ccc|cccc}

\hline\hline
\multirow{2}{*}{0-lepton} & \multicolumn{3}{c|}{ Resolved }  & \multicolumn{4}{c}{ Merged }  \\
%& \multirow{2}{*}{1 $b$-tag} & \multirow{2}{*}{2 $b$-tag} & \multirow{2}{*}{3+ $b$-tag} &  \multirow{2}{*}{1 $b$-tag} & \multirow{2}{*}{2 $b$-tag} & 1 $b$-tag & 2 $b$-tag \\
%&  &  &  &   &  & add. $b$-tag  &  add. $b$-tag\\

& \multirow{2}{*}{1 $b$-tag} & \multirow{2}{*}{2 $b$-tag} & \multirow{2}{*}{3+ $b$-tag} & \multirow{2}{*}{1 $b$-tag} & \multirow{2}{*}{2 $b$-tag} & 1 $b$-tag & 2 $b$-tag \\
& & & & && add. $b$-tag & add. $b$-tag \\
\hline

\ttbar & 22900 $\pm$ 890\phantom{0} & 6640 $\pm$ 180 & 1000 $\pm$ 34\phantom{00} & 1650 $\pm$ 160 & 68 $\pm$ 12 & 2110 $\pm$ 70\phantom{0} & 105 $\pm$ 11\phantom{0} \\
Single top quark & 2440 $\pm$ 330 & 552 $\pm$ 76 & 25.8 $\pm$ 5.6\phantom{.} & 217 $\pm$ 52 & 15.4 $\pm$ 4.1\phantom{0} & 136 $\pm$ 50 & 5.6 $\pm$ 2.4 \\
Diboson & 317 $\pm$ 41 & 41.2 $\pm$ 5.8 & \phantom{.}4.5 $\pm$ 1.1 & 188 $\pm$ 30& 34.8 $\pm$ 4.8\phantom{0} & 12.9 $\pm$ 2.3 & 1.6 $\pm$ 0.4 \\
\Zl & \phantom{0}580 $\pm$ 210& \phantom{0}1.3 $\pm$ 1.3 &  \phantom{0}--  & \phantom{0}310 $\pm$ 130 & 0.38 $\pm$ 0.29 & 11.8 $\pm$ 8.2 & 0.1 $\pm$ 0.1 \\
\Zhl & 8240 $\pm$ 840 & \phantom{0}50 $\pm$ 17 & \phantom{0.}5.4 $\pm$ 1.8\phantom{0} & \phantom{0}910 $\pm$ 160 & 10.1 $\pm$ 3.7\phantom{0} & 118 $\pm$ 27 & 0.6 $\pm$ 0.4 \\
\Zhf & 1280 $\pm$ 170 & 1270 $\pm$ 150 & 41 $\pm$ 8 & 238 $\pm$ 45& 101 $\pm$ 16\phantom{0} & 16.8 $\pm$ 4.2 & 8.6 $\pm$ 2.3 \\
\Wl & \phantom{0}960 $\pm$ 300 & \phantom{0}3 $\pm$ 2 &  \phantom{0}--  & 227 $\pm$ 95 & 1.0 $\pm$ 0.6 & \phantom{0}5.4 $\pm$ 3.9 & 0.02 $\pm$ 0.02 \\
\Whl & \phantom{0}5960 $\pm$ 1100 & 56 $\pm$ 17\phantom{.} & \phantom{.}3.7 $\pm$ 2.3 & \phantom{0}770 $\pm$ 230 & 6.6 $\pm$ 3.2 & \phantom{0}65 $\pm$ 21 & 0.1 $\pm$ 0.1 \\
\Whf & \phantom{0}530 $\pm$ 150 & \phantom{0}470 $\pm$ 130 & 16.5 $\pm$ 4.7\phantom{.} & 112 $\pm$ 44 & 40 $\pm$ 16 & 10.2 $\pm$ 5.1 & 3 $\pm$ 2 \\
SM $Vh$ & \phantom{0}55 $\pm$ 21&102 $\pm$ 39& 1.04 $\pm$ 0.57 & \phantom{0}7.4 $\pm$ 2.9 & 4.7 $\pm$ 1.8 & \phantom{0}0.4 $\pm$ 0.2 & 0.06 $\pm$ 0.04 \\
\ttbar $h$ & 10.4 $\pm$ 5.3 & \phantom{0}7.8 $\pm$ 3.9 & \phantom{0}6 $\pm$ 3 & \phantom{0}1.4 $\pm$ 0.7 & 0.2 $\pm$ 0.1 & \phantom{0}4 $\pm$ 2 & 0.6 $\pm$ 0.3 \\
\ttbar $V$ & 102 $\pm$ 54& \phantom{0}41 $\pm$ 22\ & \phantom{0.}8.7 $\pm$ 4.5 & 17.7 $\pm$ 9.5\phantom{0} & 1.4 $\pm$ 0.8 & \phantom{0}24 $\pm$ 12 & 1.8 $\pm$ 1.0 \\
\hline
Total & 43400 $\pm$ 200\phantom{0} & 9240 $\pm$ 95\phantom{0} & 1110 $\pm$ 30 & 4650 $\pm$ 79\phantom{0} & 282 $\pm$ 14\phantom{0} & 2510 $\pm$ 50\phantom{0} & 127 $\pm$ 11\phantom{0} \\
Data & 43387 & 9236 & 1125 & 4657 & 283 & 2516 & 127 \\

\hline\hline
1-lepton &1 $b$-tag &2 $b$-tag & & 1 $b$-tag &2 $b$-tag && \\ 
\hline

\ttbar & 16300 $\pm$ 600\phantom{0} & 3900 $\pm$ 120\phantom{0} &  & 8100 $\pm$ 300 & 400 $\pm$ 50\phantom{0} & \multicolumn{2}{c}{} \\
Single top quark & 4100 $\pm$ 600 & 860 $\pm$ 130 &  & 1100 $\pm$ 300 & 120 $\pm$ 30\phantom{0} & \multicolumn{2}{c}{} \\
Diboson & 110 $\pm$ 20 & 12 $\pm$ 2\phantom{0} &  & 220 $\pm$ 30 & 34 $\pm$ 5\phantom{0} & \multicolumn{2}{c}{} \\
\Zl & \phantom{0}40 $\pm$ 10 & 0.09 $\pm$ 0.05 &  & 14 $\pm$ 6 & 0.2 $\pm$ 0.1 & \multicolumn{2}{c}{} \\
\Zhl & 170 $\pm$ 10 & 0.7 $\pm$ 0.5 &  & 38 $\pm$ 6 & 0.4 $\pm$ 0.2 & \multicolumn{2}{c}{} \\
\Zhf & 27 $\pm$ 4 & 17 $\pm$ 2\phantom{0} &  & 11 $\pm$ 2 & 4.5 $\pm$ 0.6 & \multicolumn{2}{c}{} \\
\Wl & \phantom{0}550 $\pm$ 180 & 3 $\pm$ 3 &  & \phantom{0}590 $\pm$ 230 & 0.2 $\pm$ 0.2 & \multicolumn{2}{c}{} \\
\Whl & 5700 $\pm$ 440 & 24 $\pm$ 8\phantom{0} &  & 1800 $\pm$ 300 & 30 $\pm$ 10 & \multicolumn{2}{c}{} \\
\Whf & \phantom{0}820 $\pm$ 140 & 420 $\pm$ 70\phantom{0} &  & 350 $\pm$ 80 & 180 $\pm$ 40\phantom{0} & \multicolumn{2}{c}{} \\
SM $Vh$ & \phantom{0}60 $\pm$ 20 & 90 $\pm$ 30 &  & 14 $\pm$ 6 & 11 $\pm$ 4\phantom{0} & \multicolumn{2}{c}{} \\
Multijet & \phantom{0}200 $\pm$ 100 & 1.7 $\pm$ 0.9 &  &  \phantom{0}--  &  --  & \multicolumn{2}{c}{} \\
\hline
Total & 28100 $\pm$ 170\phantom{0} & 5320 $\pm$ 70\phantom{00} &  & 12200 $\pm$ 120 & 780 $\pm$ 30\phantom{0} & \multicolumn{2}{c}{} \\
Data & 28073 & 5348 &  & 12224 & 775 & \multicolumn{2}{c}{} \\

\hline\hline
2-lepton &1 $b$-tag &2 $b$-tag &3+ $b$-tag & 1 $b$-tag &2 $b$-tag &\multicolumn{2}{c}{1+2 $b$-tag add. $b$-tag} \\ 
\hline

\ttbar & 2570 $\pm$ 80\phantom{0} & 1940 $\pm$ 110\phantom{0} & 58 $\pm$ 9\phantom{0} & \phantom{0}5.3 $\pm$ 2.6 & 0.4 $\pm$ 0.2 & \multicolumn{2}{c}{11 $\pm$ 5\phantom{0}} \\
Single top quark & 185 $\pm$ 25 & 58 $\pm$ 9\phantom{0} & 1.5 $\pm$ 0.4 & \phantom{0}0.7 $\pm$ 0.1 & 0.2 $\pm$ 0.2 & \multicolumn{2}{c}{0.5 $\pm$ 0.3} \\
Diboson & 570 $\pm$ 80 & 159 $\pm$ 24\phantom{0} & 5.2 $\pm$ 1.3 & 35 $\pm$ 5 & 8.5 $\pm$ 1.3 & \multicolumn{2}{c}{4.6 $\pm$ 0.8} \\
\Zl & 2210 $\pm$ 950 & 2 $\pm$ 3 &  --  & \phantom{0}85 $\pm$ 34 & 1.0 $\pm$ 0.5 & \multicolumn{2}{c}{6 $\pm$ 4} \\
\Zhl & 37200 $\pm$ 1100 & 130 $\pm$ 50\phantom{0} & 12 $\pm$ 5\phantom{0} & 240 $\pm$ 40 & 2.3 $\pm$ 0.8 & \multicolumn{2}{c}{55 $\pm$ 11} \\
\Zhf & 7840 $\pm$ 690 & 6320 $\pm$ 170\phantom{0} & 150 $\pm$ 20\phantom{0} & \phantom{0}74 $\pm$ 12 & 34 $\pm$ 5\phantom{0} & \multicolumn{2}{c}{12 $\pm$ 3\phantom{0}} \\
\Wl & \phantom{0}1.9 $\pm$ 0.7 &  --  &  --  & \phantom{0}0.03 $\pm$ 0.01 &  --  & \multicolumn{2}{c}{0.01 $\pm$ 0.01} \\
\Whl & 37 $\pm$ 9 & 0.9 $\pm$ 0.7 &  --  & \phantom{0}0.4 $\pm$ 0.1 &  --  & \multicolumn{2}{c}{0.01 $\pm$ 0.01} \\
\Whf & \phantom{0}5.4 $\pm$ 1.4 & 1.9 $\pm$ 0.3 & 0.03 $\pm$ 0.01 & \phantom{0}0.17 $\pm$ 0.06 & 0.02 $\pm$ 0.01 & \multicolumn{2}{c}{0.06 $\pm$ 0.05} \\
SM $Vh$ & 105 $\pm$ 40 & 140 $\pm$ 60\phantom{0} & 1.3 $\pm$ 0.7 & \phantom{0}1.6 $\pm$ 0.6 & 0.8 $\pm$ 0.3 & \multicolumn{2}{c}{0.2 $\pm$ 0.1} \\
\ttbar $h$ & \phantom{0}0.9 $\pm$ 0.5 & 1.6 $\pm$ 0.8 & 1.1 $\pm$ 0.5 & \phantom{0}0.05 $\pm$ 0.02 & 0.01 $\pm$ 0.01 & \multicolumn{2}{c}{0.15 $\pm$ 0.07} \\
\ttbar $V$ & 140 $\pm$ 80 & 60 $\pm$ 30 & 6 $\pm$ 3 & 10 $\pm$ 5 & 0.6 $\pm$ 0.3 & \multicolumn{2}{c}{12 $\pm$ 6\phantom{0}} \\
\hline
Total & 50900 $\pm$ 230\phantom{0} & 8810 $\pm$ 90\phantom{00} & 240 $\pm$ 20\phantom{0} & \phantom{0}450 $\pm$ 20 & 47 $\pm$ 5\phantom{0} & \multicolumn{2}{c}{101 $\pm$ 9\phantom{0}} \\
Data & 50876 & 8798 & 235 & 439 & 50 & \multicolumn{2}{c}{101} \\

\hline\hline
\end{tabular}

\end{footnotesize}
\end{center}
\end{table}
}

\clearpage

\clearpage

\clearpage

%\clearpage

% All figures and tables should appear before the summary and conclusion.
% The package placeins provides the macro \FloatBarrier to achieve this.
\FloatBarrier
\section{Conclusion}
\label{sec:conclusion}

A search for \Wprime\ and \Zprime\ bosons and for a CP-odd Higgs boson $A$ in the \vvbb, \lvbb\ and \llbb\ final states is performed 
using 36.1~fb$^{-1}$ of 13~\TeV\ $pp$ collision data collected with the ATLAS detector at the LHC.
%It uses techniques capable of reconstructing these resonances over a large range of masses up to 5~\TeV. 
No significant excess of events is observed above the SM predictions in all three channels.

Upper limits are placed at the 95\% CL on the cross-section times branching fraction,
$\sigma(pp\rightarrow V^\prime\rightarrow Vh)\times B(h\rightarrow\bbbar,\ccbar)$,
ranging between $1.1\times 10^{-3}$ and $2.8\times 10^{-1}$~pb for the \Wprime\
boson and between $9.0\times 10^{-4}$ and $1.3\times 10^{-1}$~pb for the \Zprime\ boson 
in the mass range  of 500~\GeV\ to 5~\TeV. 
The \Wprime\ and \Zprime\ bosons with masses $m_{W^{\prime}} < \WpExclusionModelA$~\TeV\ (\WpExclusionModelB~\TeV)
and $m_{Z^{\prime}} <\ZpExclusionModelA$~\TeV\ (\ZpExclusionModelB~\TeV) are excluded for 
the benchmark HVT \textit{Model A (Model B)}, while for the combined HVT search  masses up to 
\HVTExclusionModelA~\TeV\ (\HVTExclusionModelB~\TeV) are excluded.

For an $A$ boson, upper limits are placed on $\sigma(pp\rightarrow A \rightarrow Zh)\times B(h\rightarrow\bbbar)$ 
between $1.9\times 10^{-3}$ and $8.1\times 10^{-1}$~pb for gluon--gluon fusion production and 
between $3.4\times 10^{-3}$ and $7.3\times 10^{-1}$~pb for production with associated $b$-quarks 
in the mass range 220~\GeV\ to 2~\TeV. 
%A mild excess of events, coming mainly from the 2-lepton channel with  3 or more $b$-tagged jets, is observed at about $m_{A}=440$~\GeV\ with a local significance of $3.6 \sigma$ and a global significance of $2.4 \sigma$.

%-------------------------------------------------------------------------------
\section*{Acknowledgements}
\label{sec:acknowledgemnts}
% Acknowledgements for papers with collision data
% Version 24-Oct-2018

% Standard acknowledgements start here
%----------------------------------------------
We thank CERN for the very successful operation of the LHC, as well as the
support staff from our institutions without whom ATLAS could not be
operated efficiently.

We acknowledge the support of ANPCyT, Argentina; YerPhI, Armenia; ARC, Australia; BMWFW and FWF, Austria; ANAS, Azerbaijan; SSTC, Belarus; CNPq and FAPESP, Brazil; NSERC, NRC and CFI, Canada; CERN; CONICYT, Chile; CAS, MOST and NSFC, China; COLCIENCIAS, Colombia; MSMT CR, MPO CR and VSC CR, Czech Republic; DNRF and DNSRC, Denmark; IN2P3-CNRS, CEA-DRF/IRFU, France; SRNSFG, Georgia; BMBF, HGF, and MPG, Germany; GSRT, Greece; RGC, Hong Kong SAR, China; ISF and Benoziyo Center, Israel; INFN, Italy; MEXT and JSPS, Japan; CNRST, Morocco; NWO, Netherlands; RCN, Norway; MNiSW and NCN, Poland; FCT, Portugal; MNE/IFA, Romania; MES of Russia and NRC KI, Russian Federation; JINR; MESTD, Serbia; MSSR, Slovakia; ARRS and MIZ\v{S}, Slovenia; DST/NRF, South Africa; MINECO, Spain; SRC and Wallenberg Foundation, Sweden; SERI, SNSF and Cantons of Bern and Geneva, Switzerland; MOST, Taiwan; TAEK, Turkey; STFC, United Kingdom; DOE and NSF, United States of America. In addition, individual groups and members have received support from BCKDF, CANARIE, CRC and Compute Canada, Canada; COST, ERC, ERDF, Horizon 2020, and Marie Sk{\l}odowska-Curie Actions, European Union; Investissements d' Avenir Labex and Idex, ANR, France; DFG and AvH Foundation, Germany; Herakleitos, Thales and Aristeia programmes co-financed by EU-ESF and the Greek NSRF, Greece; BSF-NSF and GIF, Israel; CERCA Programme Generalitat de Catalunya, Spain; The Royal Society and Leverhulme Trust, United Kingdom. 

The crucial computing support from all WLCG partners is acknowledged gratefully, in particular from CERN, the ATLAS Tier-1 facilities at TRIUMF (Canada), NDGF (Denmark, Norway, Sweden), CC-IN2P3 (France), KIT/GridKA (Germany), INFN-CNAF (Italy), NL-T1 (Netherlands), PIC (Spain), ASGC (Taiwan), RAL (UK) and BNL (USA), the Tier-2 facilities worldwide and large non-WLCG resource providers. Major contributors of computing resources are listed in Ref.~\cite{ATL-GEN-PUB-2016-002}.
%----------------------------------------------

%-------------------------------------------------------------------------------
% If you use biblatex and either biber or bibtex to process the bibliography
% just say \printbibliography here
\printbibliography
% If you want to use the traditional BibTeX you need to use the syntax below.
%\bibliographystyle{bib/bst/atlasBibStyleWithTitle}
%\bibliography{mydocument,bib/ATLAS,bib/CMS,bib/ConfNotes,bib/PubNotes}
%-------------------------------------------------------------------------------
\clearpage
% ATLAS Collaboration author list
% Reference date of EXOT-2016-10 is 2017-06-17
% Author list last updated on date 15-NOV-18
% Data extracted on 15-Nov-2018 for paper reference EXOT-2016-10
% at 12:49pm
 
\begin{flushleft}
{\Large The ATLAS Collaboration}

\bigskip

M.~Aaboud$^\textrm{\scriptsize 34d}$,    
G.~Aad$^\textrm{\scriptsize 99}$,    
B.~Abbott$^\textrm{\scriptsize 125}$,    
O.~Abdinov$^\textrm{\scriptsize 13,*}$,    
B.~Abeloos$^\textrm{\scriptsize 129}$,    
S.H.~Abidi$^\textrm{\scriptsize 165}$,    
O.S.~AbouZeid$^\textrm{\scriptsize 143}$,    
N.L.~Abraham$^\textrm{\scriptsize 153}$,    
H.~Abramowicz$^\textrm{\scriptsize 159}$,    
H.~Abreu$^\textrm{\scriptsize 158}$
% ...
\bigskip
\\

% ...
$^{13}$Institute of Physics, Azerbaijan Academy of Sciences, Baku; Azerbaijan.\\
% ...
$^{34}$$^{(a)}$Facult\'e des Sciences Ain Chock, R\'eseau Universitaire de Physique des Hautes Energies - Universit\'e Hassan II, Casablanca;$^{(b)}$Centre National de l'Energie des Sciences Techniques Nucleaires (CNESTEN), Rabat;$^{(c)}$Facult\'e des Sciences Semlalia, Universit\'e Cadi Ayyad, LPHEA-Marrakech;$^{(d)}$Facult\'e des Sciences, Universit\'e Mohamed Premier and LPTPM, Oujda;$^{(e)}$Facult\'e des sciences, Universit\'e Mohammed V, Rabat; Morocco.\\
% ...
$^{99}$CPPM, Aix-Marseille Universit\'e, CNRS/IN2P3, Marseille; France.\\
% ...
$^{125}$Homer L. Dodge Department of Physics and Astronomy, University of Oklahoma, Norman OK; United States of America.\\
% ...
$^{129}$LAL, Universit\'e Paris-Sud, CNRS/IN2P3, Universit\'e Paris-Saclay, Orsay; France.\\
% ...
$^{143}$Santa Cruz Institute for Particle Physics, University of California Santa Cruz, Santa Cruz CA; United States of America.\\
% ...
$^{153}$Department of Physics and Astronomy, University of Sussex, Brighton; United Kingdom.\\
% ...
$^{158}$Department of Physics, Technion, Israel Institute of Technology, Haifa; Israel.\\
$^{159}$Raymond and Beverly Sackler School of Physics and Astronomy, Tel Aviv University, Tel Aviv; Israel.\\
% ...
$^{165}$Department of Physics, University of Toronto, Toronto ON; Canada.\\
% ...
$^{a}$ Also at  Department of Physics, University of Malaya, Kuala Lumpur; Malaysia.\\
$^{b}$ Also at Borough of Manhattan Community College, City University of New York, NY; United States of America.\\
$^{c}$ Also at Centre for High Performance Computing, CSIR Campus, Rosebank, Cape Town; South Africa.\\
$^{d}$ Also at CERN, Geneva; Switzerland.\\
$^{e}$ Also at CPPM, Aix-Marseille Universit\'e, CNRS/IN2P3, Marseille; France.\\
% ...
$^{*}$ Deceased

\end{flushleft}

% Created with Glance <Atlas.Glance@cern.ch>

%-------------------------------------------------------------------------------
% Print the list of contributors to the analysis
% The argument gives the fraction of the text width used for the names
%-------------------------------------------------------------------------------
%\clearpage
%The supporting notes for the analysis should also contain a list of contributors.
%This information should usually be included in \texttt{mydocument-metadata.tex}.
%The list should be printed either here or before the Table of Contents.
%\PrintAtlasContribute{0.30}

%-------------------------------------------------------------------------------
%\clearpage
%\part*{Auxiliary material}
%\addcontentsline{toc}{part}{Auxiliary material}
%\input{sections/auxiliary.tex}
%-------------------------------------------------------------------------------

\end{document}